%% file: main.tex
\documentclass[format=acmsmall, review=false]{acmart}

\renewcommand\footnotetextcopyrightpermission[1]{} % removes footnote with conference information in first column 
\usepackage{booktabs} % For formal tables

\usepackage[ruled]{algorithm2e} % For algorithms

\SetAlFnt{\small}
\SetAlCapFnt{\small}
\SetAlCapNameFnt{\small}
\SetAlCapHSkip{0pt}
\IncMargin{-\parindent}

\setcopyright{rightsretained}

\acmDOI{0000001.0000001}

\received{August 08, 2017}
\received[revised]{January 08, 2017}
\received[accepted]{(Under Review)}
\usepackage{booktabs} % For formal tables
\usepackage{amsmath}
\usepackage{amsfonts}
\usepackage{graphicx}
\usepackage{color}
\usepackage{tikz}
\usepackage{listings}
\usepackage{pifont}% http://ctan.org/pkg/pifont
\usepackage{multirow}
\usepackage{xspace}
\usepackage{array}
\usepackage{arydshln}
\usepackage[framemethod=tikz]{mdframed}
\usepackage{enumitem}
\usepackage{diagbox}
\usepackage{relsize}
\usepackage{colortbl}
\usepackage{rotating}
\usepackage{tcolorbox}% http://ctan.org/pkg/tcolorbox
\usepackage{color,soul}
\usepackage{epigraph}
\usepackage{wrapfig}

\usetikzlibrary{arrows}

\definecolor{mycolor}{rgb}{0.122, 0.435, 0.698}% Rule colour
\definecolor{gray1}{gray}{0.3}

\definecolor{darkgreen}{rgb}{0.0, 0.5, 0.0}
\definecolor{darkred}{rgb}{0.82, 0.1, 0.26}

\newcommand{\todo}[1]{}%
\renewcommand{\todo}[1]{{\color{red} TODO: {#1}}}%
\newcommand{\pythia}{\textsc{Pythia}\xspace}

\newcommand{\rpm}{\sbox0{$1$}\sbox2{$\scriptstyle\pm$}
  \raise\dimexpr(\ht0-\ht2)/2\relax\box2 }

\newmdenv[innerlinewidth=0.5pt, roundcorner=4pt,linecolor=gray,innerleftmargin=0pt,
innerrightmargin=1pt,innertopmargin=2pt,innerbottommargin=0pt]{quotes}

\newif\ifblinded

\tikzset{
  treenode/.style = {align=center, inner sep=0pt, text centered,
    font=\sffamily},
  arn_n/.style = {treenode, circle, black, draw=black,
    text width=1.5em, minimum size=1.1cm},
  arn_r/.style = {treenode, circle, black, draw=gray1, 
    text width=1.5em, minimum size=1.1cm},
  arn_x/.style = {treenode, rectangle, gray1, draw=gray1,
    minimum width=0.9cm, minimum height=0.9cm}
}

\renewcommand{\hl}{}

\newtheorem*{hypo*}{Main Hypothesis}

\begin{document}
\title{STADS: Software Testing as Species Discovery}
\subtitle{Spatial and Temporal Extrapolation from Tested Program Behaviors}

\author{Marcel B\"{o}hme} 
\orcid{0000-0002-4470-1824}
\affiliation{%
  \institution{{National University of Singapore \emph{and} Monash University, Australia}}
}
\email{marcel.boehme@acm.org}
\authornote{Dr. B\"{o}hme conducted this research at the National University of Singapore and has since moved to Monash University.}

\renewcommand{\shortauthors}{B\"{o}hme}

\begin{abstract}
\hl{A fundamental challenge of software testing is the statistically well-grounded \emph{extrapolation} from program behaviors observed during testing. For instance, a security researcher who has run the fuzzer for a week has currently \emph{no} means (i)~to estimate the total number of \emph{feasible} program branches, given that only a fraction has been covered so far, (ii)~to estimate the additional time required to cover 10\% more branches (or to estimate the coverage achieved in one more day, resp.), or (iii)~to assess the residual risk that a vulnerability exists when no vulnerability has been discovered. Failing to discover a vulnerability, does not mean that none exists---even if the fuzzer was run for  a week (or a year). Hence, testing provides \emph{no formal correctness guarantees}. 

In this article, I establish an unexpected connection with the otherwise unrelated scientific field of \emph{ecology}, and introduce a statistical framework that models Software Testing and Analysis as Discovery of Species (STADS). For instance, in order to study the species diversity of arthropods in a tropical rain forest, ecologists would first sample a large number of individuals from that forest, determine their species, and extrapolate from the properties observed in the sample to properties of the whole forest. The estimation (i)~of the total number of species, (ii)~of the additional sampling effort required to discover 10\% more species, or (iii)~of the probability to discover a new species are classical problems in ecology. The STADS framework draws from over three decades of research in ecological biostatistics to address the fundamental extrapolation challenge for automated test generation. Our preliminary empirical study demonstrates a good estimator performance even for a fuzzer with adaptive sampling bias---AFL, a state-of-the-art vulnerability detection tool. %We also show that estimator performance improves with testing time. 
 The STADS framework provides \emph{statistical correctness guarantees} with quantifiable accuracy.
}

\end{abstract}

\begin{CCSXML}
<ccs2012>
<concept>
<concept_id>10002978.10002986.10002989</concept_id>
<concept_desc>Security and privacy~Formal security models</concept_desc>
<concept_significance>500</concept_significance>
</concept>
<concept>
<concept_id>10011007.10011074.10011099</concept_id>
<concept_desc>Software and its engineering~Software verification and validation</concept_desc>
<concept_significance>300</concept_significance>
</concept>
<concept>
<concept_id>10011007.10011074.10011099.10011102.10011103</concept_id>
<concept_desc>Software and its engineering~Software testing and debugging</concept_desc>
<concept_significance>500</concept_significance>
</concept>
<concept>
<concept_id>10011007.10010940</concept_id>
<concept_desc>Software and its engineering~Software organization and properties</concept_desc>
<concept_significance>300</concept_significance>
</concept>
</ccs2012>
\end{CCSXML}

\ccsdesc[500]{Security and privacy~Penetration testing}
\ccsdesc[500]{Software and its engineering~Software testing and debugging}
%\keywords{Statistical guarantees, extrapolation, fuzzing, stopping rule, code coverage, species coverage, discovery probability, security, reliability, measure of confidence, measure of progress}

\maketitle

\input{intro}
\input{sections/examples}

\input{sections/data} 
\input{sections/progress}
\input{sections/extrapolation}
\input{sections/empirical}
\input{sections/overlapping}

\input{sections/related}

\input{sections/future}
\input{conclusion}

\section*{Acknowledgements}
I would like to thank Anne Chao from the Institute of Statistics at National Tsing Hua University for her interesting comments about our model of software testing and analysis as discovery of species (i.e., the STADS model) and her suggestion to view testing objectives where each input can be assigned to multiple species as producing incidence rather than abundance data. I would also like to thank David Clark from the University College London and the attendants of the 41st CREST Open Workshop on ``Software Engineering And Computer Science Using Information'' for the interesting discussions about the role of entropy in automated software testing. In this article, the Shannon-entropy quantifies a program's difficulty to being automatically tested by a fuzzer. Finally, I am grateful for the permission to publish a picture taken from an exhibit at the Lee Kong Chian Natural History Museum in Singapore (\autoref{fig:arthropods}).

This research was partially supported by a grant from the National Research Foundation, Prime Minister's Office, Singapore under its National Cybersecurity R\&D Program (TSUNAMi project, No. NRF2014NCR-NCR001-21) and administered by the National Cybersecurity R\&D Directorate.

\bibliographystyle{ACM-Reference-Format}
%\bibliography{sigproc} 
\input{bibtex.bbl}

\end{document}

%% file: intro.tex
\section{Introduction}
The development of automated and practical approaches to vulnerability detection has never been more important. The recent world-wide WannaCry cyber-epidemic clearly demonstrates the vulnerability of our well-connected software systems. WannaCry exploits a \emph{software vulnerability} on Windows machines to gain root access on a huge number of computers all over the world. The ransomware uses the root access to encrypt all private data which would be released only if a ransom is paid. Hospitals had to shut down because life-saving medical devices were infected \cite{guardian}.

\hl{In 2017, a company's cost of cyber attacks world-wide was on average US\$ 11.7 \emph{million}, which is a 22.7\% increase from the preceeding year} \cite{ponemon1}.
\hl{In February 2017, a bug was discovered in the HTML parser of Cloudflare, a company which offers performance and security services to about six million customer websites (incl. OKCupid and Uber). The bug leaked information, including private keys and passwords }\cite{cloudbleed}\hl{. In July 2017, a hacker stole 31 \emph{million} USD from Ethereum, a blockchain-based platform, exploiting a vulnerability in the implementation of a protocol that was formally verified to be cryptographically sound }\cite{ether}. 
To discover software vulnerabilities \emph{at scale}, we need automated testing tools that can be used in practice, that work by the push of a button.\footnote{We concretely position this work within the software security domain and leverage the appropriate terminology. We take this position due to the practical impact and the recent, considerable traction of automated testing in the security domain. The security domain also provides a more compact terminology: ``Fuzzing'' instead of ``automated software testing'', ``fuzzer'' instead of ``testing tool'', ``fuzzing campaign'' instead of ``execution of the testing tool'', etc. Nevertheless, the central concepts that we present in this article apply to automated software testing and analysis, in general.}

Automated software testing (or fuzzing) has been an extremely successful automated vulnerability detection technique in practice. Our own fuzzers \cite{aflfast,aflgo,MWF,cie} discovered 100+ bugs and more than 40 vulnerabilities in large security-critical software systems. Fuzzers, such as AFL \cite{afl}, Libfuzzer \cite{libfuzzer}, syzkaller \cite{syzkaller}, Peach \cite{peach}, Monkey \cite{monkey}, and Sapienz \cite{sapienz} are now routinely used as automated testing and vulnerability detection techniques in large companies, such as Google \cite{oss}, Microsoft \cite{springfield}, Mozilla \cite{mozillaFuzzing}, and Facebook \cite{harmanFacebook}. The 2004 DARPA Grand Challenge inspired substantial research in self-driving cars that are now a reality. The 2016 DARPA \emph{Cyber} Grand Challenge \cite{cgc}, the world's first machine-only hacking tournament with \$3.75 million in prize money, will arguably provide a similar push of research in advanced automated vulnerability detection. A \emph{fuzzer} generates and executes program inputs, while a \emph{dynamic analysis} (e.g., injected program assertions \cite{asan,msan}) identifies test executions that expose a vulnerability. 

\subsection{Extrapolation: A Fundamental Challenge of Automated Testing}
\hl{A fundamental challenge of software testing is the statistically well-grounded \emph{extrapolation} from program behaviours observed during testing }\cite{whalen}\hl{. Unlike automated verification, fuzzing does not allow to make universal statements over program properties }\cite{djikstra}.

\hl{\textbf{No formal guarantees}.
If a verifier terminates without a counter-example, it formally guarantees the absence of vulnerabilities for \emph{all} inputs. In contrast, a fuzzer perpetually generates random inputs and checks whether any of those exposes a vulnerability. Clearly, if the fuzzer generates a vulnerability-exposing input, a vulnerability exists. Yet, failing to expose a vulnerability does \emph{not} mean that none exists. In fact, Hamlet and Taylor }\cite{tsEffectiv4}\hl{ argue that no matter how long the fuzzer is run (e.g., a year)---if no vulnerability is discovered, we cannot report with \emph{any} degree of confidence that none exists. So then, \emph{what is the utility of a fuzzing campaign that exposes no vulnerabilities?}}

\hl{\textbf{No cost-effectiveness analysis}. 
Suppose, a security researcher has run the fuzzer for one week and exercised 60\% of all program branches. Today, she has no means to estimate how much longer it would take to achieve, say 70\% coverage, or how much coverage would be achieved after, say one more week. 
Perhaps the program is just very difficult to fuzz. However, there exists no formal measure of fuzzability, either, that would allow to estimate the resources needed to achieve an acceptable progress during a fuzzing campaign.
In fact, our security researcher has no means to determine whether the fuzzer \emph{can} even achieve 70\% branch coverage, at all. Some branches may just not be feasible. Perhaps 100\% of feasible branches have already been covered. In that case, how should a security researcher judge the \emph{campaign's progress towards completion}? In practice, exactly when to abort a fuzzing campaign is mostly a judgement call that requires experience and guesswork.}

\hl{\textbf{No smart scheduling}.
The lack of oversight has consequences not only for individual security researchers but for large multi-national companies as well. For instance, Google Security has heavily invested into a large-scale fuzzing infrastructure called OSS-Fuzz which is now generating some 10 \emph{trillion} test inputs per day for more than 50 security-critical open-source software projects }\cite{oss}\hl{. Each project is assigned roughly the same time budget. This is a waste of resources since fuzzing campaigns for certain programs stop making any progress after only a few hours while campaigns for other programs continue to make progress for days on end.  
For now, there is no automated mechanism to measure how far a fuzzing campaign has progressed towards completion. Hence, no \emph{smart} scheduling strategies for fuzzing campaigns have been developed, yet.}

\hl{A security researcher has no means to estimate the progress of the current fuzzing campaign towards completion or the confidence that the campaign inspires in the program's correctness. At any time into the campaign, the researcher has no means to gauge (let alone predict) the expected return on investment: How much more would she learn if she continued the campaign?} 

\subsection{An Unexpected Connection With Ecology}\label{sec:ecology}

\begin{figure}\centering
\tcbox[size=fbox,colframe=black,colback=black]{%
\includegraphics[width=0.98\textwidth]{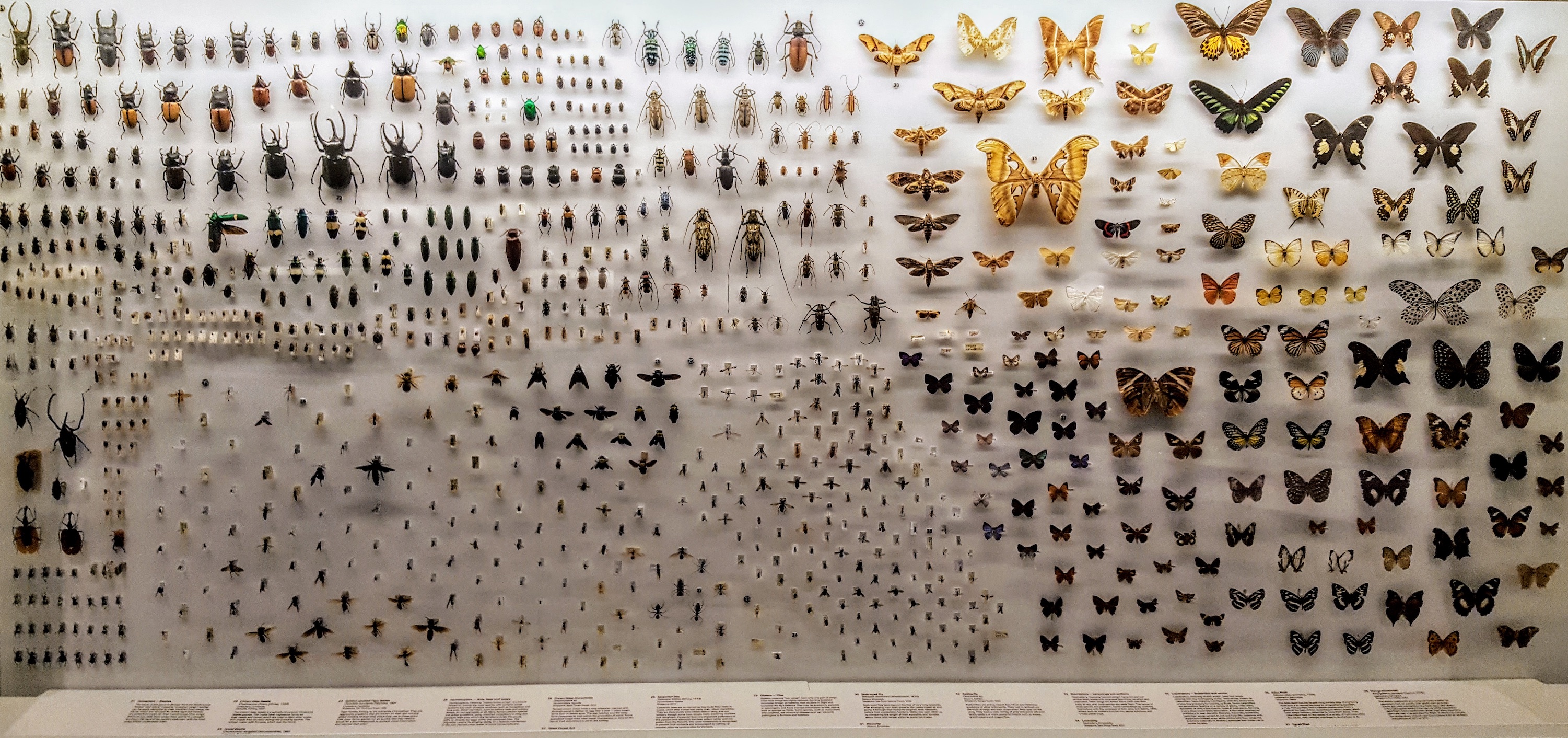}% 
}
\caption{\hl{Species of arthropods (i.e., ``bugs'') discovered during ecological surveys in Singapore and Malaysia. 
The diversity and richness of arthropod species in tropical rain forests are notoriously difficult to assess due to the immense sampling effort that is required. According to a recent estimate }\cite{tropical}\hl{, there are 6.1 \emph{million} tropical arthropod species (high richness), most of which are rare (high diversity).
\emph{Photo Credit}: Marcel B\"{o}hme with the permission from Lee Kong Chian Natural History Museum, Singapore.
}\vspace{-0.2cm}
}
\label{fig:arthropods}
\end{figure}

\hl{In this article, I establish an unexpected connection with the scientific field of \emph{ecology}, a branch of biology that deals with the relations of organisms to one another and to their physical surroundings. I argue that methodologies to estimate the number of species in an assemblage\footnote{An \emph{assemblage} is a group of individuals belonging to a number of different species that occur together in space and time. For\,example, all\,birds\,that live\,on\,an\,island\,today\,form\,an\,assemblage; \mbox{all\,plants\,currently\,on\,Earth\,form\,an\,assemblage;\,etc.}} provide an ideal statistical framework within which one can assess and extrapolate the progress of a fuzzing campaign towards completion and the confidence it inspires in the program's correctness. I conduct a preliminary empirically evaluation and outline future research directions to tailor and improve these methodologies for the requirements of automated software engineering and security.}

\hl{\textbf{Discovery in testing}. My \emph{key observation} is that automated software testing and analysis are about \emph{discovery}.
A fuzzer generates test inputs by sampling from the program's input space, and thus discovers properties about the program's behavior. Depending on the concrete objective, discovery means to find new bugs or vulnerabilities }\cite{efficiency}\hl{, to exercise interesting program paths }\cite{afl}\hl{, to cover new coverage goals, to kill stubborn mutants }\cite{mutation}\hl{, to explore new program states }\cite{peso,prv}\hl{, to report unexpected information flows }\cite{ase17}\hl{, or to explore new event sequences }\cite{monkey}. 

\hl{\textbf{Discovery in ecology}. Similarly, ecologists are concerned with the discovery of species in an assemblage. For instance, in order to study the biodiversity of arthropods in a tropical rain forest (}\autoref{fig:arthropods}\hl{), ecologists would first sample a large number of individuals from that forest and determine their species. However, since sampling effort is necessarily limited, the sample is usually incomplete. The sample may contain several abundant species and miss many rare species. Biostatisticians spent the last three decades constructing a well-grounded statistical framework within which they can extrapolate, with quantifiable accuracy, from properties of the sample to properties of the complete assemblage (e.g., arthropod diversity in the tropical rain forest) }\cite{incidenceSurvey,collwell20,speciesReview}.

\hl{\textbf{STADS framework}. My key observation allows us to model software testing and analysis as discovery of species (STADS). Consequently, STADS provides direct access to a rich statistical framework in ecology. Within the STADS framework, security researchers can leverage methodologies to accurately estimate the degree to which a software has been tested and to extrapolate, with quantifiable accuracy, from the behavior observed during testing to the complete program behavior. We show that an estimate of the probability to discover a new species provides a \emph{statistical correctness guarantee}. Moreover, we present novel methodologies to assess \emph{campaign completeness} (i.e., the progress of an ongoing campaign towards completion), \emph{cost effectiveness} (e.g., the additional resources required to achieve an acceptable completeness), and \emph{residual risk} that a vulnerability exists when none has been discovered.}

\hl{\textbf{Terminology}. A \emph{fuzzer} generates test inputs for a program. In STADS, a \emph{test input} corresponds to an individual or sampling unit. A \emph{dynamic analysis} identifies the \emph{species} for an input. For instance, the AFL }\cite{afl}\hl{ instrumentation identifies the path exercised by an input; AddressSanitizer }\cite{asan}\hl{ identifies the memory error exposed by an input (if at all). A species is \emph{rare} if only a few generated test inputs belong to that species while a species is \emph{abundant} if a large number of test inputs belong to that species. The \emph{relative abundance} of a species describes the probability to generate a test input that belongs to that species. The program's \emph{input space} represents the assemblage. The set of test inputs generated throughout a fuzzing campaign corresponds to the survey sample.%The \emph{total number of species} in the program's input space is called species richness.
} We refer to Chao and Collwell (2017) \cite{incidenceSurvey}, Chao and Lou \cite{coverageSurvey}, and Collwell et al. \cite{sampleSurvey} for recent reviews of the literature on the pertinent models and estimators spanning three decades of research in ecology.

\hl{\textbf{Hypothesis}. I hypothesize that within STADS rare species which have been discovered explain the species within the fuzzer's search space that remain undiscovered. Intuitively, it is the ``difficulty'' to discover a rare species---measured by the total number of test inputs that needed to be generated before discovering the rare species---that provides insights on the discovery of undetected species which are evidently much rarer. A similar hypothesis is underpinning the nonparametric biostatistics in ecology }\cite{hypothesis}\hl{. In order to test this central hypothesis, we need to establish the accuracy of existing estimators and extrapolators from ecology within the STADS framework.}

\textbf{Species richness} $S$. Estimating the total number of species $S$ in the assemblage is a classical problem in ecology.
If an ecologist samples $n$ individuals and discovers $S(n)$ species, then $(S-S(n))$ species remain \emph{undetected}.
In order to quantify the species richness of the complete assemblage, nonparametric estimators $\hat S$ have been developed that become more accurate as sampling effort $n$ increases \cite{chao1,chao2}. For instance, recently ecologists estimated the total number of species on Earth as \emph{8.7 million} \cite{species} while only 14\% have been discovered despite two centuries of taxonomic classification. In STADS, an estimate $\hat S$ of the asymptotic total number of species  allows us to estimate the proportion $\hat G=S(n)/\hat S$ of all $\hat S$ species that have been discovered. \hl{For instance, we could estimate the \emph{feasible} branch coverage, i.e., the proportion of actually feasible branches covered so far. The species coverage $G$ can be used to assess \emph{campaign completeness}, i.e., how much progress has been made towards completion. It could also be used to devise \emph{smart scheduling strategies} for fuzzing campaigns that automatically abort a campaign that has reached a certain degree of completeness $\hat G$, and schedule the next one.}

\hl{\textbf{Discovery probability} $U(n)$. In ecology, the discovery probability $U(n)$ measures the probability to discover a new species with the $n+1$th generated test input. The discovery probability can be estimated accurately and efficiently from the sample alone }\cite{good}\hl{. In the STADS framework, if the dynamic analysis is able to identify vulnerabilities, then the discovery probability $U$ provides a \emph{statistical guarantee} that no detectable vulnerability exists if none has been discovered. In other words, security researchers can use the STADS statistical framework for residual risk assessment. In ecology, the sample coverage $C=1-U$ quantifies the completeness of the sample, i.e., the proportion of individuals in the assemblage whose species is represented in the sample. Sample coverage is routinely used to choose the most accurate estimator for other quantities, such as species richness $S$ }\cite{brose}\hl{ and to compare attributes of species across different assemblages }\cite{coverageSurvey}. 

\hl{\textbf{Extrapolating species discovery} $S(n+m^*)$ and $U(n+m^*)$. An extrapolation allows to assess the trade-off between investing more resources and gaining more insight. In ecology, there exist methodologies to quantify this return on investment. In STADS, a security researcher can use these methodologies to make an informed decision whether to continue or abort a fuzzing campaign. Suppose, the client requires a statistical guarantee of $U(n+m^*)=10^{-8}$ as upper bound on the probability that the fuzzer finds a vulnerability in the program. The researcher can estimate the additional fuzzing effort $m^*$ that is required to achieve \emph{that} degree of confidence in the program's correctness. Suppose, a fuzzer has achieved a statement coverage of $G(n)=60\%$. Within STADS, the statistically well-grounded extrapolation allows to estimate the coverage $G(n+m^*)$ that would be achieved if $m^*$ more test inputs were generated.}

\subsection{Contributions}
\hl{This article addresses the fundamental challenge of statistically well-grounded extrapolation both (i)~\emph{spatially} (i.e., from behaviors observed during fuzzing to \emph{all} program behaviors) as well as (ii)~\emph{temporally} (i.e., if the campaign was continued for some more time). We provide the first general statistical model of software testing and analysis as discovery of species (STADS). For the first time, practitioners can use well-researched methodologies from ecology to make informed decisions about the fate of a fuzzing campaign and quantify what has been learned about the program. Within STADS researchers can, for the first time, formally define novel metrics, and identify or develop their estimators to investigate interesting properties of software, fuzzing campaign, and fuzzer.}
\begin{itemize}[itemsep=3pt]
  \item \hl{A fuzzer's effectiveness and efficiency may be measured and compared across other fuzzers. \emph{Effectiveness} is determined by the number of species within the fuzzer's search space. \emph{Efficiency} is determined by the number of species discovered per generated test input.}
  \item \hl{A campaign's completeness, cost-effectiveness, and residual risk may be assessed as it is ongoing. \emph{Campaign completeness} can be judged by the species coverage $G(n)$ or the sample coverage $C(n)=1-U(n)$. \emph{Cost-effectiveness}  can be assessed via extrapolation of the species discovered $S(n+m^*)$ or confidence achieved $U(n+m^*)$ if $m^*$ additional test inputs were generated. The campaign's \emph{residual risk} can be assessed via the discovery probability $U(n)$.}
  \item \hl{The difficulty to fuzz a program (i.e., \emph{software fuzzability}) can be estimated from the relative species abundance distribution. Intuitively, as the proportion of rare species increases, the difficulty to discover species increases as well.}  
\end{itemize}

\noindent
\hl{The \emph{primary contribution} of this article is the STADS model which establishes the connection with ecology to provide access to a rich statistical framework that can address the challenges in automated software testing and analysis. However, due to space limitation, we can only present and investigate some pertinent aspects of the STADS framework. Specifically, this article makes the following secondary contributions.}
\begin{itemize}[itemsep=3pt]
  \item \hl{\textbf{Hypothesis}. I hypothesize that rare species which have been discovered explain the species within the fuzzer's search space that remain undiscovered. This hypothesis underpinning the STADS framework is \emph{tested successfully} in our empirical study. Estimators and extrapolators that are based on rare species (i.e., singleton and doubleton species) demonstrate a good performance for automated software testing and analysis. Within the STADS framework, we make \emph{no assumptions} about the total number, relative abundance distribution, or location of species within the program's input space.}
  \item \hl{\textbf{STADS models}. The \emph{multinomial model}---where each input belongs to exactly one species---is integrated into the STADS framework and empirically evaluated. For instance, an input can execute only one path, exercise only one method call sequence, compute only one final output, crash only at one program location. The \emph{Bernoulli product model}---where each input belongs to one or more species---is integrated into the STADS framework. For instance, a single input can exercise multiple coverage-goals (e.g., program statements, branches, or methods), kill multiple mutants, witness multiple information flows, violate multiple assertions, expose multiple bugs, and traverse multiple program states. For both models, we provide an \emph{extensive survey} of ecological methodologies to estimate and extrapolate relevant quantities within the STADS framework, and show how these methodologies can solve hard problems that have been long-standing in automated software engineering.}
  \item \hl{\textbf{Evaluation}. In order to conduct an empirical evaluation of the multinomial model within the STADS framework, we fuzz six security-critical open-source programs for a cumulative 8.2 months using the popular, state-of-the-art fuzzer AFL }\cite{afl}\hl{. The evaluation of two estimators ($\hat G(n)$ }\cite{chao1}\hl{, $\hat U(n)$ }\cite{good}\hl{) and one extrapolator $\hat S(n+m^*)$ }\cite{shen}\hl{ demonstrate a reasonably low bias and high precision. We find that, despite the adaptive sampling bias of AFL, the methodologies are \emph{statistically consistent}, meaning that bias decreases and precision increases as more test inputs are generated. The estimate for one fuzzing campaign is fairly \emph{representative} for other fuzzing campaigns of the same length.\footnote{More specifically, an estimate is fairly representative for other fuzzing campaigns where the same program is fuzzed for the same time using the same fuzzer and seed corpus (if any).}} 
\end{itemize}

\noindent
\hl{The STADS framework exhibits some peculiar features that make the direct application of existing ecologic methodologies more challenging: One has to deal with extremely \emph{large populations} containing a \emph{huge number of species} (e.g., millions of program branches), where \emph{most species are rare}. Sampling strategies of feedback-directed fuzzers are (intentionally) subject to adaptive bias. For instance, in search-based software testing (SBST) }\cite{sbst,sbst1}\hl{ the species discovered by future test inputs depend on the ``fitness'' of past test inputs. We point out many opportunities to identify, improve, tailor, and develop novel methodologies that address the peculiarities of the STADS model and sketch solutions to correct the adaptive bias of feedback-directed fuzzers.}

\subsection{Outline}
The remainder of this article is structured as follows. \autoref{sec:motivation} illustrates the main technical challenges and contributions using a practical motivating example. \autoref{sec:model} introduces the STADS framework and multinomial model more formally and explains how the model relates to automated testing tools in practice. Sections \ref{sec:estimate} and \ref{sec:extrapolate} follow with a survey and discussion of estimation and extrapolation in multinomial model of the STADS framework, respectively. In \autoref{sec:eval}, we provide a preliminary empirical evaluation of the estimators and extrapolators within the multinomial model. In \autoref{sec:overlapping}\hl{, we extend the STADS framework to account for inputs that can belong to multiple species by introducing the Bernoulli product model.}
 In \autoref{sec:related}, we survey the relevant related literature. After an extended discussion of the peculiarities of the STADS framework and opportunities for future research in \autoref{sec:discussion}, we conclude in \autoref{sec:conclusion}.

%% file: sections/examples.tex
\section{Motivating Example}\label{sec:motivation}
We introduce the main ideas of our statistical framework of software testing and analysis as discovery of species (STADS) using the following motivating example. 
We ran the fuzzer American Fuzzy Lop (AFL) for \emph{one week} on the program libjpeg-turbo compiled with  AddressSanitizer (ASAN). \emph{AFL} \cite{afl} is the state-of-the-art fuzzer for automated vulnerability detection. \emph{Libjpeg-turbo} \cite{libjpeg} is a popular, security-critical image parsing library that is used in many browser and server frameworks. \emph{ASAN} \cite{asan} is a dynamic analyzer that identifies buffer overflows and other memory-related errors and vulnerabilities. We use that fuzzing campaign to illustrate the challenges and opportunities of automated testing and analysis in general.

\textbf{Path discovery}. %\textbf{Concrete testing objective}.
While the true objective of AFL is to discover a maximal number of errors, it is an unlikely measure of progress; errors are (thankfully) rather sparse in the program's input space. Instead, the more immediate (and measurable) goal of AFL is to explore paths.\footnote{To address path explosion, AFL clusters paths that exercise the same control-flow edges and do not yield substantially different hit counts for each edge \cite{aflfast}. Effectively, AFL reports the number of discovered path \emph{clusters} rather than the number of discovered paths. For simplicity, we stick to the AFL terminology.} AFL's compiler-wrapper \texttt{afl-gcc} instruments the program such that each path yields a different \emph{path-id}. ASAN instruments the program such that it \emph{crashes} for inputs exposing a memory-related error. Hence, AFL's \emph{concrete testing objective} is to discover a maximal number of paths and crashes. 

\textbf{Species discovery}.
In ecology, researchers sample individuals from an assemblage and identify their species to gain insights about the species richness and diversity of the assemblage.
AFL's \emph{fuzzer} \texttt{afl-fuzz} generates and executes test inputs for the instrumented program by applying random mutation operators at random points in a random seed file. In other words, AFL is a (biased) stochastic process that samples test inputs from the program's input space. \hl{Our \emph{assemblage} is the program's input space.\footnote{This is grossly simplified. Technically, our assemblage is the set of all program inputs that AFL is capable of generating using the available seed files and mutation operators. All statistical claims will hold only over AFL's search space.} Our \emph{individual} is a discrete input. Our \emph{sample} is the set of all test inputs that have been generated throughout the current campaign. In this example, our \emph{species} is the tuple $(\text{\emph{path-id}},\text{\emph{crashing}})$ where \emph{crashing} is true if the input crashes the program and false otherwise. ASAN and \texttt{afl-gcc} together form the \emph{dynamic analysis} that identifies the species for a program input. The \emph{general testing objective} is always to discover a maximal number of species.   
}

\definecolor{ballblue}{rgb}{0.13, 0.67, 0.8}
\begin{figure}
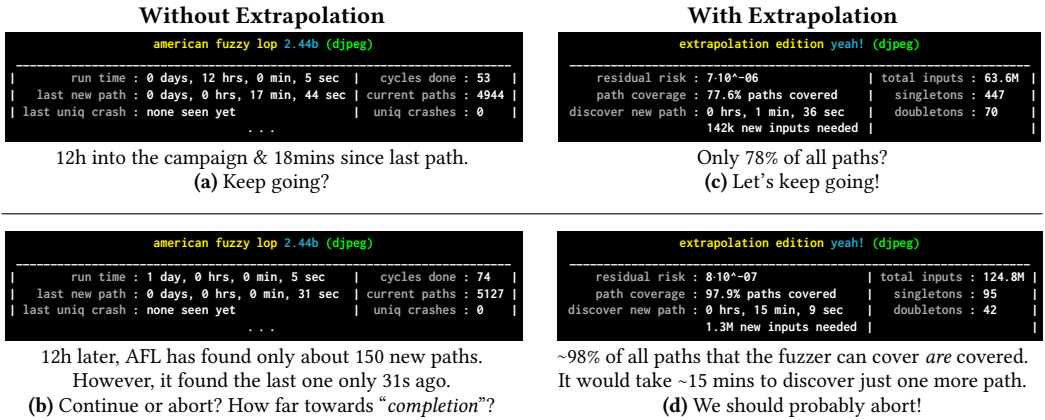
\bf
\begin{tabular}{@{}c||@{}c@{}}
\small\textbf{Without Extrapolation}&\small\textbf{With Extrapolation}\\
\scalebox{0.521}{
\colorbox{black}{\ttfamily
\begin{tabular}{@{\color{white}{|} }r@{ \color{white}{:} }l @{ \color{white}{|} }r@{ \color{white}{:} }l@{ \color{white}{|}}}
\multicolumn{4}{c}{\textbf{\color{yellow}{american fuzzy lop} \color{ballblue}{2.44b} \color{green}{(djpeg)}}}\\
\multicolumn{4}{c}{\color{white}{\_\_\_\_\_\_\_\_\_\_\_\_\_\_\_\_\_\_\_\_\_\_\_\_\_\_\_\_\_\_\_\_\_\_\_\_\_\_\_\_\_\_\_\_\_\_\_\_\_\_\_\_\_\_\_\_\_\_\_\_\_\_\_\_\_\_\_\_\_\_\_\_}}\\\hline
\textbf{\color{gray!70}{run time}} & \color{white}{0 days, 12 hrs, 0 min, 5 sec}      & \textbf{\color{gray!70}{cycles done}} & \color{white}{53}    \\
\textbf{\color{gray!70}{last new path}} & \color{white}{0 days, 0 hrs, 17 min, 44 sec}      &\textbf{\color{gray!70}{current paths}} & \color{white}{4944}\\
\textbf{\color{gray!70}{last uniq crash}} & \color{white}{none seen yet}                      & \textbf{\color{gray!70}{uniq crashes}} & \color{white}{0}      \\
\multicolumn{4}{c}{\color{white}{\textbf{\ldots}}}\\
\end{tabular}%
}} &\,\,\!
\scalebox{0.521}{
\colorbox{black}{\ttfamily
\begin{tabular}{@{ }r@{ \color{white}{:} }l @{ \color{white}{|} }r@{ \color{white}{:} }l@{ \color{white}{|}}}
\multicolumn{4}{c}{\textbf{\color{yellow}{extrapolation edition} \color{ballblue}{yeah!} \color{green}{(djpeg)}}}\\
\multicolumn{4}{c}{\color{white}{\_\_\_\_\_\_\_\_\_\_\_\_\_\_\_\_\_\_\_\_\_\_\_\_\_\_\_\_\_\_\_\_\_\_\_\_\_\_\_\_\_\_\_\_\_\_\_\_\_\_\_\_\_\_\_\_\_\_\_\_\_\_\_\_\_\_\_}}\\\hline
\textbf{\color{gray!70}{residual risk}} & \color{white}{7$\cdot$10\textasciicircum -06$$} & \textbf{\color{gray!70}{total inputs}} & \color{white}{63.6M}    \\
\textbf{\color{gray!70}{path coverage}} & \color{white}{77.6\% paths covered\quad\ \,} &\textbf{\color{gray!70}{singletons}} & \color{white}{447}\\
\textbf{\color{gray!70}{discover new path}} & \color{white}{0 hrs, 1 min, 36 sec} & \textbf{\color{gray!70}{doubletons}} & \color{white}{70}      \\
\multicolumn{2}{r@{ \color{white}{| }}}{\color{white}{142k new inputs needed}\,} &\multicolumn{2}{r@{ \color{white}{|}}}{}       \\
\end{tabular}%
}}\\ 
\Small \normalfont 12h into the campaign \& 18mins since last path. & \Small \normalfont Only 78\% of all paths?\\[-0.1cm]
\Small \textbf{(a)} \normalfont Keep going? & \Small \textbf{(c)} \normalfont Let's keep going!\\[0.2cm] \hline
\\[-0.2cm]
\scalebox{0.521}{
\colorbox{black}{\ttfamily
\begin{tabular}{@{\color{white}{|} }r@{ \color{white}{:} }l @{ \color{white}{|} }r@{ \color{white}{:} }l@{ \color{white}{|}}}
\multicolumn{4}{c}{\textbf{\color{yellow}{american fuzzy lop} \color{ballblue}{2.44b} \color{green}{(djpeg)}}}\\
\multicolumn{4}{c}{\color{white}{\_\_\_\_\_\_\_\_\_\_\_\_\_\_\_\_\_\_\_\_\_\_\_\_\_\_\_\_\_\_\_\_\_\_\_\_\_\_\_\_\_\_\_\_\_\_\_\_\_\_\_\_\_\_\_\_\_\_\_\_\_\_\_\_\_\_\_\_\_\_\_\_}}\\\hline
\textbf{\color{gray!70}{run time}} & \color{white}{1 day, 0 hrs, 0 min, 5 sec}      & \textbf{\color{gray!70}{cycles done}} & \color{white}{74}    \\
\textbf{\color{gray!70}{last new path}} & \color{white}{0 days, 0 hrs, 0 min, 31 sec\ }      &\textbf{\color{gray!70}{current paths}} & \color{white}{5127}\\
\textbf{\color{gray!70}{last uniq crash}} & \color{white}{none seen yet}                      & \textbf{\color{gray!70}{uniq crashes}} & \color{white}{0}      \\
\multicolumn{4}{c}{\color{white}{\textbf{\ldots}}}\\
\end{tabular}%
}} &\,\,\!
\scalebox{0.521}{
\colorbox{black}{\ttfamily
\begin{tabular}{@{ }r@{ \color{white}{:} }l @{ \color{white}{|} }r@{ \color{white}{:} }l@{ \color{white}{|}}}
\multicolumn{4}{c}{\textbf{\color{yellow}{extrapolation edition} \color{ballblue}{yeah!} \color{green}{(djpeg)}}}\\
\multicolumn{4}{c}{\color{white}{\_\_\_\_\_\_\_\_\_\_\_\_\_\_\_\_\_\_\_\_\_\_\_\_\_\_\_\_\_\_\_\_\_\_\_\_\_\_\_\_\_\_\_\_\_\_\_\_\_\_\_\_\_\_\_\_\_\_\_\_\_\_\_\_\_\_\_}}\\\hline
\textbf{\color{gray!70}{residual risk}} & \color{white}{8$\cdot$10\textasciicircum -07} & \textbf{\color{gray!70}{total inputs}} & \color{white}{124.8M}    \\
\textbf{\color{gray!70}{path coverage}} & \color{white}{97.9\% paths covered\quad\ \,} &\textbf{\color{gray!70}{singletons}} & \color{white}{95}\\
\textbf{\color{gray!70}{discover new path}} & \color{white}{0 hrs, 15 min, 9 sec} & \textbf{\color{gray!70}{doubletons}} & \color{white}{42}      \\
\multicolumn{2}{r@{ \color{white}{| }}}{\color{white}{1.3M new inputs needed}\,} &\multicolumn{2}{r@{ \color{white}{|}}}{}       \\
\end{tabular}%
}}\\
\Small \normalfont 12h later, AFL has found only about 150 new paths. & 
\Small \normalfont {\raise.17ex\hbox{$\scriptstyle\sim$}}98\% of all paths that the fuzzer can cover \emph{are} covered.\\[-0.1cm]
\Small \normalfont However, it found the last one only 31s ago. & \Small \normalfont It would take {\raise.17ex\hbox{$\scriptstyle\sim$}}15 mins to discover just one more path.\\[-0.1cm]
\Small \textbf{(b)} \normalfont Continue or abort? How far towards ``\emph{completion}''?  &\Small \textbf{(d)} \normalfont We should probably abort!\\
\end{tabular}
\caption{The left-hand side (``without extrapolation'') shows the first few lines of AFL's retro-style UI (AFL v2.44b). Specifically, it shows the pertinent information for the fuzzing campaign (a) at 12 hours and (b) at 24 hours. The right-hand side (``with extrapolation'') shows our extension with estimates of the residual risk (i.e., the probability to discover a (crashing) path with the next input that is generated), the path coverage (i.e., the proportion of paths discovered), and the time or test inputs needed to discover the next path---for the fuzzing campaign (c) at 12~hours and (d) at 24~hours. 
}
\label{fig:afl} 
\end{figure} 

\textbf{Challenges}. 
\autoref{fig:afl}.a) shows the progress for our fuzzing campaign after the passage of 12 hours---just like a security researcher might see it. In 12 hours, AFL has generated {\raise.17ex\hbox{$\scriptstyle\sim$}}63 million (63M) test inputs and completed 53 cycles through the seed inputs. AFL has discovered about 5 thousand (5k) paths, and about 18 minutes (18 min) have passed since the discovery of the most recent path. Since the security researcher is given only the total number of paths, she cannot make an informed decision concerning the progress of the fuzzing campaign towards completion. About 18 minutes have passed since the last discovery of a new path. So, the researcher might reckon that the probability to discover a new path is very low. However, as we will see below, the time since the last discovery is rather \emph{unreliable} and often changes several times per minute by up to \emph{four orders of magnitude}. No crashes have been found. At 12 hours, the security researcher has no handle on the progress of the fuzzing campaign towards completion or on the correctness of the program.
      
\autoref{fig:afl}.b) shows the progress for our fuzzing campaign after 24 hours. The security researcher has learned that the number of discovered paths has not increased substantially in the last 12 hours. She may (or may not) decide to discontinue the fuzzing campaign based on this observation alone. However, the most recent path was found only a few seconds ago. So, she might be swayed to continue for at least a few more hours. Still, no crashes have been found. Even after 24 hours, the security researcher has no definite handle on making an informed decision about the completeness of the fuzzing campaign or how confident she can be in the correctness of the program. 

\subsection{Assessing Residual Risk Using the Discovery Probability}\label{sec:xSample}
\epigraph{\hl{``Testing can be used to show the presence of bugs, but never to show their absence.''}}{\textit{Edsger Dijkstra (1970) \cite{djikstra}}}%

\hl{Finding no vulnerabilities in a (long-running) fuzzing campaign does not mean that none exists. A \emph{residual risk assessment} would allow us to quantify the confidence the campaign inspires in the correctness of the program. In fact, our STADS framework provides \emph{statistical guarantees} about the absence of vulnerabilities with quantifiable accuracy (e.g., 95\%-confidence intervals). In order to assess the residual risk, we suggest to estimate the probability $U$ to discover a new species with the next generated test input. If the dynamic analysis, as in our motivating example, is able to identify vulnerabilities, then undiscovered vulnerabilities correspond to undiscovered species. Hence, the discovery probability $U$ provides an upper bound on the probability to discover a new vulnerability with the next input that is generated. From this perspective, I argue that testing can be used to show that bugs are absent with a \emph{certain likelihood} $(1-U)$ that can be estimated efficiently and accurately \emph{during} a fuzzing campaign, with a likelihood that\,increases\,over\,the course of a campaign.}

\hl{In ecology, the \emph{discovery probability} $U$ gives the proportion of individuals in the assemblage whose species are \emph{not} represented in the sample. In our motivating example, 
the discovery probability gives the proportion of all inputs in the input space that exercise yet undiscovered paths. We could say, $U$ represents how much of the program behavior remains untested. The \emph{inverse of the discovery probability} $1/U$ provides the number of test inputs that we can expect to generate before discovering a new (path) species.
The \emph{sample coverage} $C=1-U$ is the complement of the discovery probability and effectively quantifies the degree of confidence that the fuzzing campaign inspires in the correctness of the program. In our example, at least $C\%$ of all inputs that AFL is capable of generating are expected to execute without crashes.}

\hl{Out of the box, AFL already reports the time since the last discovery of a new species (Fig.} \ref{fig:afl}\hl{.a+b; \texttt{last new path}). This time to last discovery can be used as an estimate of the expected time to the next discovery. However, as we will see shortly, this estimate is very unreliable.
Given the number $m$ of test inputs that have been generated in the time since the last discovery, we can compute the \emph{empirical discovery probability}  as $\hat U_\text{emp}=1/m$. However, the discovery probability thus estimated changes by orders of magnitude in a matter of seconds.}

\begin{figure}\Small
\begin{tabular}{@{}c@{}c@{}c@{}}
\quad\quad\quad \textbf{(a)} Empirical probability $(\hat U_\text{emp})$ & \textbf{(b)} Moving median $(\hat U_\text{mm})$ & \textbf{(c)} Good-Turing estimate $\hat U)$\\
\includegraphics[trim={0     0.6cm 0 0}, clip, width=0.39\columnwidth]{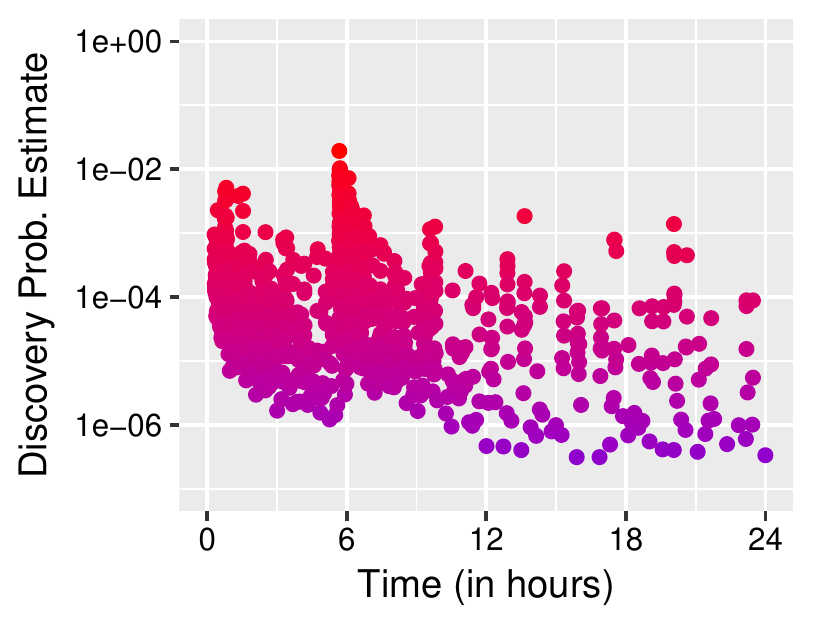} &
\includegraphics[trim={1.8cm 0.6cm 0 0}, clip, width=0.305\columnwidth]{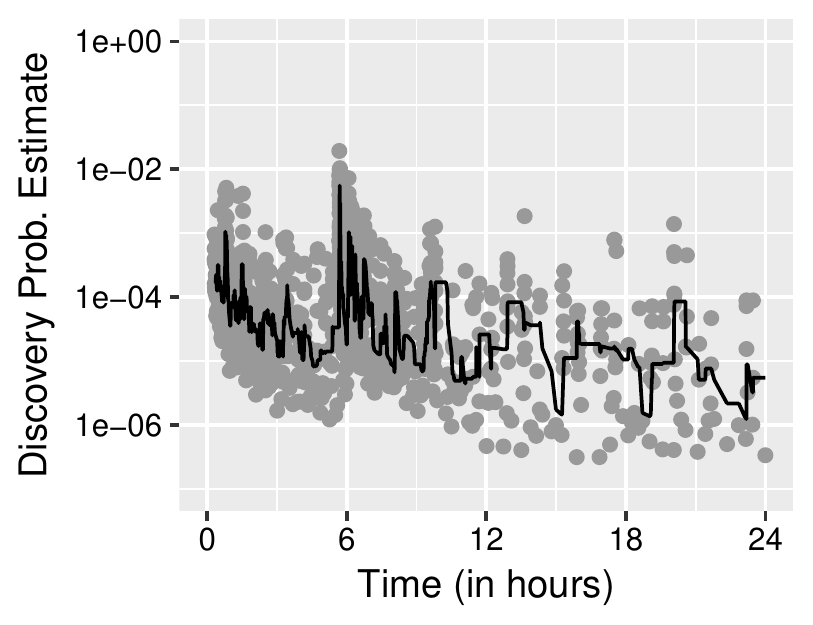} &
\includegraphics[trim={1.8cm 0.6cm 0 0}, clip, width=0.305\columnwidth]{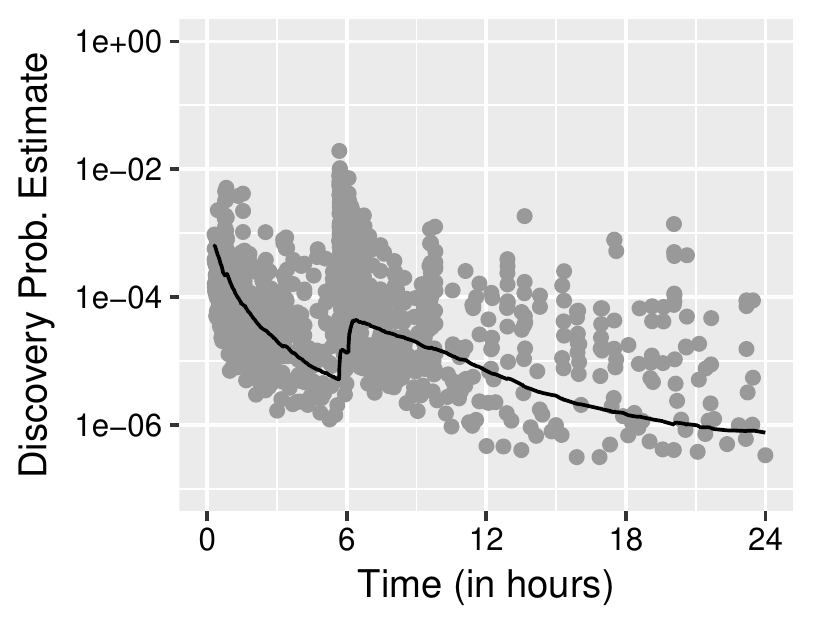} \\
\includegraphics[trim={0     0.2cm 0 0.2cm}, clip, width=0.39\columnwidth]{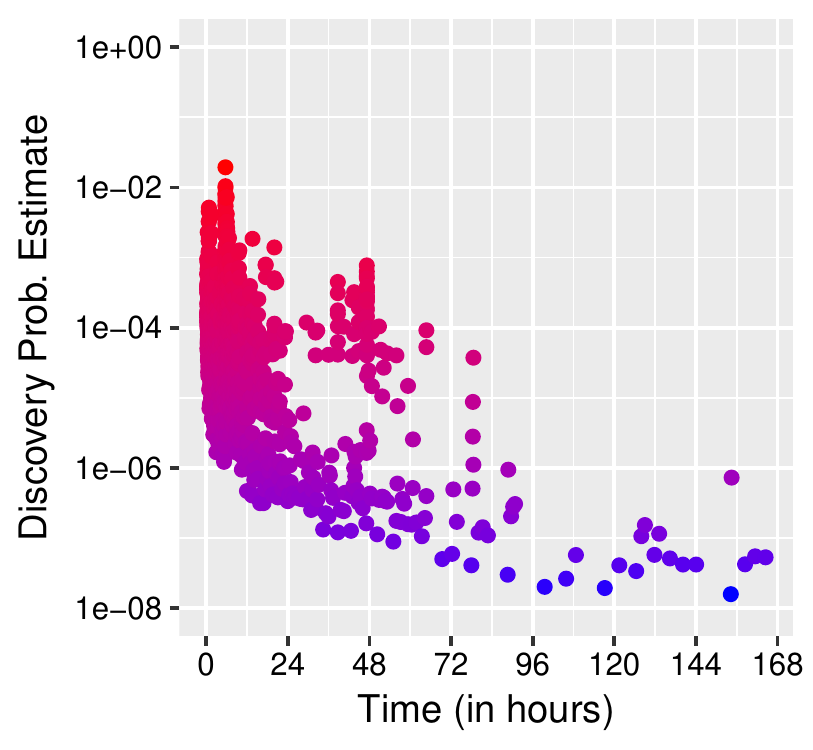} &
\includegraphics[trim={1.8cm 0.2cm 0 0.2cm}, clip, width=0.305\columnwidth]{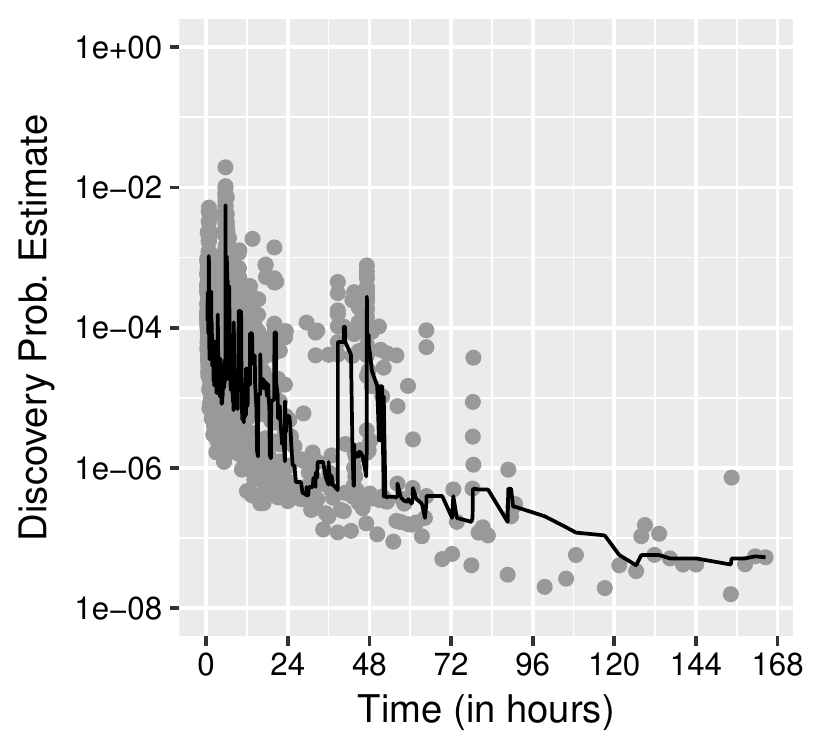} &
\includegraphics[trim={1.8cm 0.2cm 0 0.2cm}, clip, width=0.305\columnwidth]{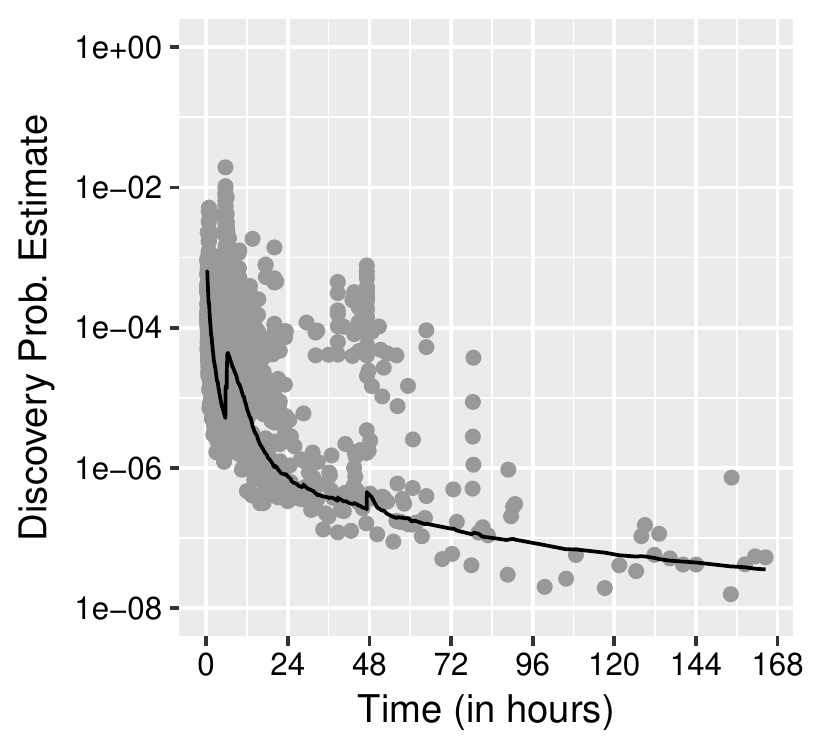} \\
\end{tabular}
\caption{Estimating the current discovery probability, i.e., the probability that a generated input discovers a previously undiscovered path over 24h (top) and 168h (bottom).}
\label{fig:xSample}
\end{figure}

In \autoref{fig:xSample}, we can see several estimators of the current discovery probability in an ongoing fuzzing campaign: \textbf{a)} the empirical  probability (i.e., $1-1/m$ where $m$ is the number of test inputs needed to discover the most recent path), \textbf{b)} the rolling median (i.e., the median empirical probability for the discovery of the $N=11$ most recent paths), and \textbf{c)} the Good-Turing estimator that is available in our STADS framework.
\autoref{fig:xSample}.a shows the \emph{empirical discovery probability} $\hat U_\text{emp}$ one day and seven days into the fuzzing campaign, respectively. Unlike the sample coverage $C=1-U$, the discovery probability $U$ can be represented on a log-scale. For instance, $t=100$ \emph{hours} into the fuzzing campaign we find the empirical probability at about $2\cdot 10^{-8}$. In other words, it took about $(2\cdot 10^{-8})^{-1}=50$ \emph{million} test inputs to discover the next path. However, the empirical probability changes quite substantially in a matter of seconds. Particularly in the first 24 hours, the change can be over four orders of magnitude (Fig.~\ref{fig:xSample}.a, top). 

In signal processing, quick but large swings are often addressed with a moving average, the mean value of a set of $N$ successive points. However, the moving average is susceptible to extreme events. Instead, the \emph{moving median is more robust}, i.e., the median value of a set of $N$ successive points. As we can see in \autoref{fig:xSample}.b, the swings of the moving median $\hat U_\text{mm}$ are still quite substantial, between one and three orders of magnitude. The moving median is right-aligned, meaning that the discovery probability estimate at time $t$ is computed as the median of the $N$ empirical values just preceding $t$. Hence, the moving median also generally \emph{over-estimates} the discovery probability. Increasing $N$ to smooth the swings would only \emph{increase the bias}. Moreover, $\hat U_\text{mm}$ is \emph{not consistent}, meaning that $\hat U_\text{mm}$ is \emph{not} guaranteed to approach the true discovery probability as sampling effort increases. So, the median (and mean) of the last $N$ empirical probabilities $\hat U_\text{mm}$ is also an unreliable estimator of the current discovery probability (Fig. \ref{fig:xSample}.b; $N=11$).

\begin{hypo*}
\hl{Almost all information about number and relative abundance of species that remain undiscovered is in the number and relative abundance of rare species that have been discovered.}
\end{hypo*}
\hl{The main hypothesis of the STADS framework applied to our motivating example is that the number and ``size'' of paths that AFL has exercised only once or twice throughout the fuzzing campaign contains almost all information about the paths that are yet to be explored.} 
Specifically, we denote as \emph{singletons} those paths that are exercised by exactly one generated test input. Similarly, we denote as \emph{doubletons} those paths that are exercised by exactly two generated test inputs. In \autoref{fig:afl}.d), we can see that one day into the fuzzing campaign after generating 125 million test inputs, there are still 95 singletons and 42 doubletons ({\raise.17ex\hbox{$\scriptstyle\sim$}}3\% of discovered paths). This is one singleton for every 1.3 million generated test inputs. Clearly, it would require at least as many new test inputs to discover the next \emph{undiscovered} path. In fact, this is the main insight of the Good-Turing estimator of the discovery probability. The \emph{Good-Turing estimator} \cite{good} is computed as the number of singletons divided by the number of samples (i.e., generated test inputs). The Good-Turing estimator is used across many disciplines of science, including rare event estimation \cite{rare}, cryptanalysis \cite{crypt}, computational linguistics \cite{smoothing}, and biology \cite{coverageSurvey}.

\hl{The \emph{main hypothesis} of the STADS model \emph{holds} for our motivating example. The proportion of generated inputs that exercise singleton paths accurately predicts the current discovery probability.} %The Good-Turing estimate $\hat U$ is a more reliable estimator of the current discovery probability $U$. 
In the bottom of \autoref{fig:xSample}.c, we can see that the Good-Turing estimate $\hat U$ is not subject to huge swings like both empirical estimators. In fact, it was formally shown that i) the estimator's accuracy strictly increases as the sample size (i.e., number of generated test inputs) increases \cite{consistent}, ii) its convergence to the true value is also reasonably fast \cite{goodNormal}, iii) its mean squared error is reasonably low \cite{goodError}, and iv) its performance is close to the best natural estimator for \emph{any} distribution \cite{goodRobust}. 

%\todo{Explain sample coverage deficit for residual risk analysis}
\autoref{fig:afl} shows the discovery probability estimate $\hat U$, just like a security researcher might see it if she uses our AFL extension. \hl{$\hat U$ is shown under \texttt{residual risk} because the discovery probability provides an upper bound on the probability of discovering a vulnerability with the next input that is generated.} Even if no crashing path has been detected in a very long running fuzzing campaign, there always exists a residual risk that an unexplored crashing path might be discovered in the future when more resources are being invested. 

Twelve hours into the fuzzing campaign the discovery probability is shown as $\hat U = 7\cdot 10^{-6}$ (Fig. \ref{fig:afl}.c). The discovery probability is estimated as $f_1/n$ where $f_1$ is number of \emph{singletons} ($f_1=447$) and $n$ is the number of \emph{total inputs} ($n=63.6\cdot 10^6$). 
Depending on the which residual risk is deemed acceptable, the security researcher can use the discovery probability to decide whether to continue or abort the fuzzing campaign. In fact, twelve hours later, one day into the fuzzing campaign, the discovery probability has decreased by one order of magnitude ($\hat U=8\cdot 10^{-7}$; Fig. \ref{fig:afl}.d).

We can use the discovery probability to compute other descriptive statistics, which the security researcher can use for her decision. For instance, the fuzzing effort has also increased by an order of magnitude: While it took only 1.5 minutes to discover a new path 12 hours into the fuzzing campaign, she can expect it takes 15 minutes to discover a new path 24 hours into the fuzzing campaign.

\subsection{Assessing the Completeness of the Fuzzing Campaign}
\epigraph{\hl{``Currently, there is no sound basis to extrapolate from tested to untested cases.''}}{\textit{Michael Whalen on the Future of V\&V \cite{whalen}}}

\hl{AFL shows the number of paths that have been discovered in the current sampling campaign (}\autoref{fig:afl}\hl{.a+b). However, without an estimate of the number of paths that remain undiscovered, a security researcher cannot judge whether this is close or far from the discovery of \emph{all} paths.

Within the STADS framework, we define \emph{species coverage} $G$ as the proportion of the asymptotic total number of species that have been discovered. In our motivating example, the \emph{path coverage}---which is one kind of species coverage---gives the proportion of paths that have been discovered in the current fuzzing campaign.} Hence, path coverage is a measure of the progress of the current fuzzing campaign towards completion. Unlike measures of code coverage, where the total number of elements is (assumed to be) known a-priori, path coverage is more difficult to measure since the total number of paths is \emph{unknown}. Currently, a security researcher has no means to compute the path coverage at any point in the fuzzing campaign.

\begin{figure}[h]\small
\begin{tabular}{@{}c l@{ : }l@{}}
\multirow{11}{*}{\includegraphics[trim={0 0 0 0.3cm}, clip, width=0.45\columnwidth]{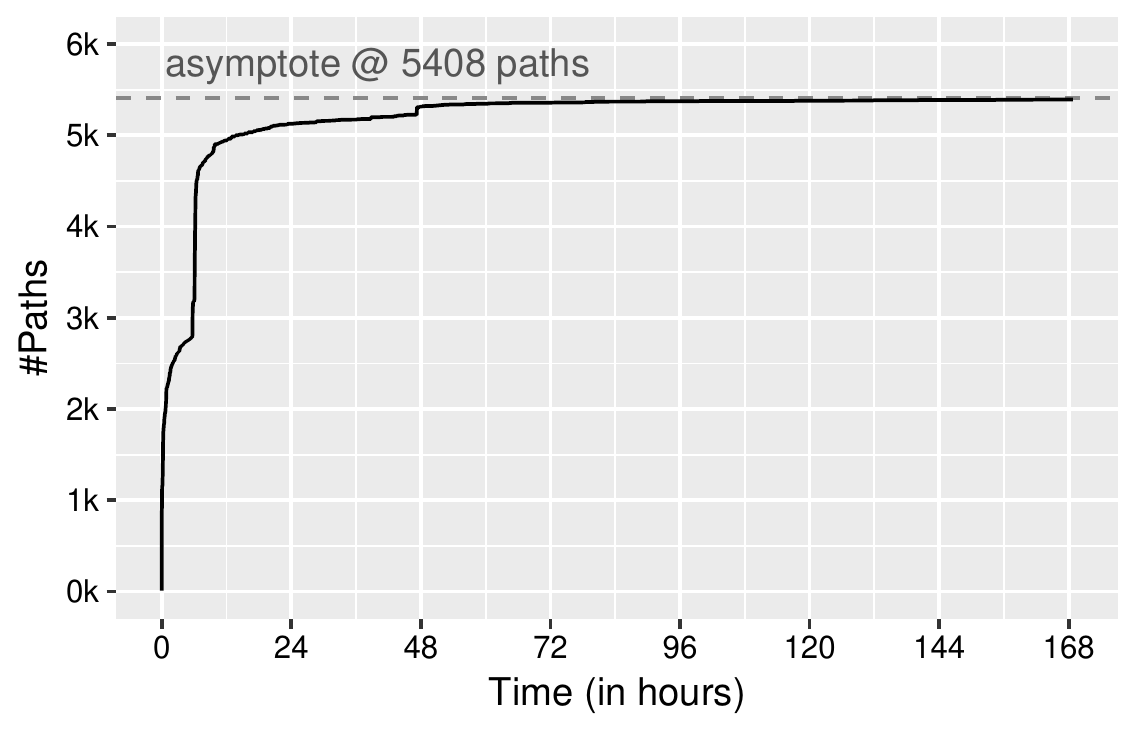}}&Subject & libjpeg-turbo \cite{libjpeg}\\
&  Test driver & djpeg\\
%&  Seeds & \texttt{testimages} folder\\
&  Fuzzer& AFL \cite{afl}\\
&  Dynamic analysis& ASAN \cite{asan}\\
&  Fuzzing time& 168 hours\\
&  Generated inputs & $973\cdot 10^6$\\
&  Discovered paths & 5392 (99.7\%)\\
&  Est. total paths & 5408\\
&  Unique crashes& 0\\
\\
\end{tabular} 
\caption{Example fuzzing campaign. (a) Number of paths discovered over time. (b) Descriptive statistics.} 
\label{fig:paths}
\end{figure}

\autoref{fig:paths} shows the number of paths $S(n)$ that AFL discovered in libjpeg-turbo as the number of generated test inputs $n$ increases.\footnote{For convenience, \autoref{fig:paths}.a actually shows $S(n)$ over \emph{time}.} We can see that the number of paths discovered approaches an asymptote, which we estimate to be at $\hat S=5408$ paths (using the \emph{Chao1}-estimator of species richness \cite{chao1}). The \emph{asymptote} represents the total number of paths that the same fuzzer can discover for the same program given unlimited time. Essentially, we estimate the y-intercept $\hat S$ of the asymptote and yield $\hat G(n)=S(n)/\hat S$. 

\autoref{fig:afl} shows the path coverage estimate, just like a security researcher might see it if she uses our AFL extension. 12 hours into the fuzzing campaign, AFL is estimated to have achieved 77.6\% path coverage for the test driver of libjpeg-turbo (Fig. \ref{fig:afl}.c). This clearly indicates that the researcher should continue the fuzzing campaign in order to explore a greater percentage of paths. 12 hours later, one day into the fuzzing campaign, the path coverage has increased to 97.7\%  (Fig. \ref{fig:afl}.d). At this point, she might decide to abort the fuzzing campaign if she feels that the time that AFL would require to explore the remaining paths is too high. In fact, two days later (i.e., 3 days into the fuzzing campaign) the path coverage has increased only to 99.1\% and six days later (i.e., 7 days into the fuzzing campaign) to 99.7\%. Basically, spending six (6) \emph{times} more hours fuzzing libjpeg-turbo only increased the path coverage by two percentage points. Clearly, the security researcher benefits tremendously from a measure such as path coverage when judging the progress of the fuzzing campaign towards completion. %Even 
 
We estimate the path coverage $\hat G$ after $n$ test inputs were generated as follows $$\hat G(n)=S(n) \left/ \left(S(n)+\frac{n-1}{n}\frac{f_1^2}{2f_2}\right)\right.$$ where the denominator is the \emph{Chao1} estimator \cite{chao1} of species richness (i.e., of the total number of paths), $S(n)$ is the current number of paths discovered, $f_1$ is the number of singletons, $f_2>0$ is the number of doubletons, and $n$ is the number of test inputs generated. Path coverage can be estimated \emph{very efficiently and scalably} (i.e., independent of the size of the fuzzed program). In fact, AFL only needs to maintain the number of singletons $f_1$ and doubletons $f_2$ (\autoref{fig:afl}). Empirically, we find that the \emph{accuracy} of the  estimate increases as the number of generated inputs increases.

\subsection{Extrapolating the Completeness of the Fuzzing Campaign}
\begin{figure}\Small
\begin{tabular}{@{}c@{}c@{}c@{}}
\quad\quad\quad (a) Coverage after & (b) Extrapolated coverage & (c) Empirical coverage \\
\quad\quad\quad \ 1 min of fuzzing & over 10 min of fuzzing & over 10 min of fuzzing\\
\includegraphics[trim={0cm   0.2cm 0 0}, clip, width=0.386\columnwidth]{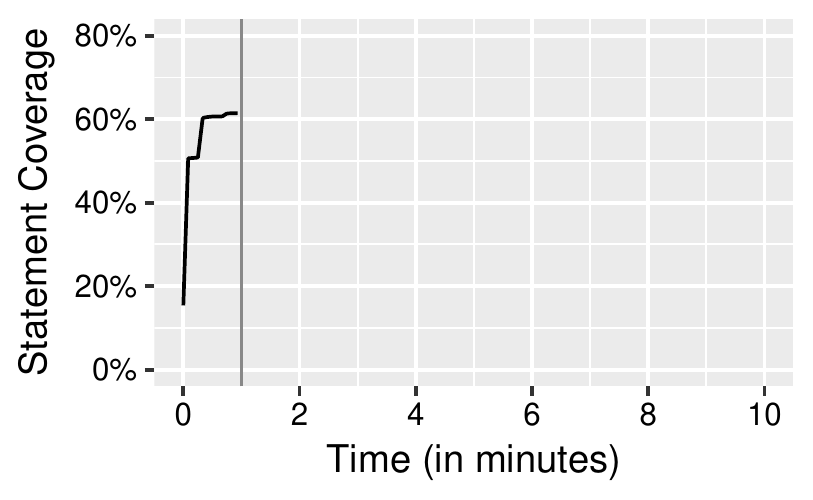} &
\includegraphics[trim={1.7cm 0.2cm 0 0}, clip, width=0.307\columnwidth]{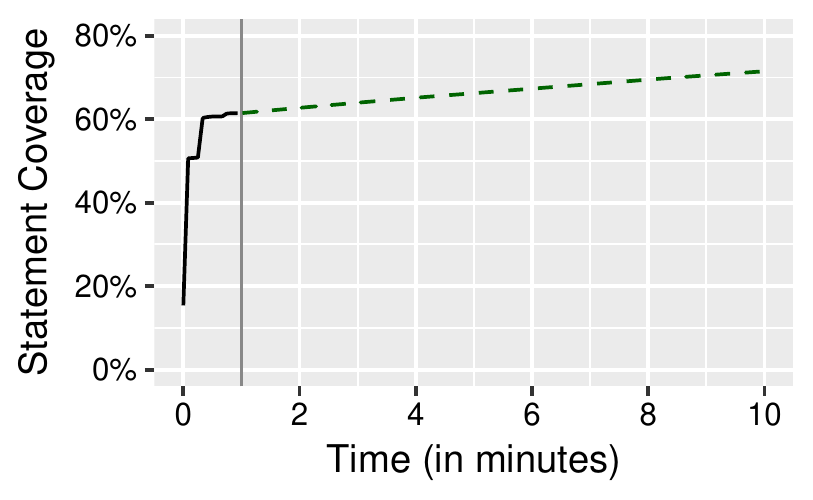} &
\includegraphics[trim={1.7cm 0.2cm 0 0}, clip, width=0.307\columnwidth]{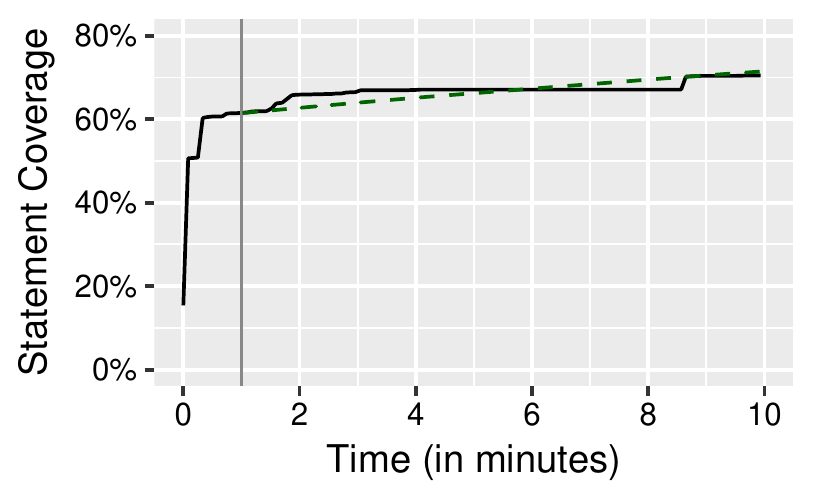}\\
\end{tabular}
\caption{Code coverage extrapolation: How much coverage is achieved if the fuzzer is run 2x, 5x, or 10x longer? We choose this particular interval because  it does not seem quite obvious how the coverage would develop. Several hours into the fuzzing campaign, the general trend appears more predictable, which is why we can increase the extrapolation intervals from minutes to hours in \autoref{fig:extrapol}.}
\label{fig:xCoverage}
\end{figure}

\hl{In automated software testing, we lack methodologies to predict how much more code coverage can be achieved if the fuzzer is run only for so much longer. 
In other words, we lack estimators of return on investment. For instance, }\autoref{fig:xCoverage}\hl{.a shows the statement coverage that AFL has achieved one minute into the fuzzing campaign.\footnote{We measured statement coverage with using \texttt{gcov} as a proportion of all \emph{executable} statements.} Even from the plot, the reader may find it difficult to estimate whether the coverage will remain at 60\% or continue to increase to 70\% within the next minute (i.e., at 2 minutes).
In the following we show how estimators from ecology can be used within our STADS framework to extrapolate statement coverage if more resources were invested.

So far, we have defined species based on the path that an input exercises, and whether it crashes or not. In the following, we allow an input to belong to multiple species where the set of species for an input $t$ is given by the program statements that $t$ covers. In our motivating example, the code coverage tool \texttt{gcov} }\cite{gcov}\hl{ forms the dynamic analysis that identifies the statements covered by an input. Notice that statement coverage is just another kind of species coverage.}

\autoref{fig:xCoverage}.b shows the extrapolation of the statement coverage within the first ten minutes of the fuzzing campaign. The statement coverage is forecasted to increase by nine (9) percentage points if the security researcher invests nine times more minutes into the fuzzing campaign. Normally, Chao and Jost \cite{coverageSurvey} would suggest to extrapolate only within twice the sample size (i.e., up to twice the length of the current fuzzing campaign). However, in this case the extrapolation is fairly accurate even within ten times the sample size as we can see by overlaying the empirical values in \autoref{fig:xCoverage}.c.
\autoref{fig:extrapol} shows more estimates of the statement coverage in the future. Despite the extrapolation up to 24 hours into the future, the coverage estimate is within $\pm 1$ percentage points of the empirical value in eight out of eleven cases. Between 10 and 15 minutes, there is a sudden coverage increase by 12 percentage points that is explained by the adaptive sampling of AFL. The extrapolation was not able to forecast this sudden increase. Otherwise, our computed estimates are all fairly accurate.

\begin{table}\Small
\caption{Extrapolating the statement coverage of libjpeg-turbo at various points $T$ into the fuzzing campaign for various times $T'$ in the future. The \emph{bias} of our estimate can be computed as the difference between the extrapolated with the empirical value of the statement coverage at time $T'$.}
\label{fig:extrapol}
\begin{tabular}{r@{ }l r| r@{ }l r r}
\multicolumn{3}{c}{}& \multicolumn{4}{c}{\textbf{Statistical Extrapolation}}\\\hline
\multicolumn{2}{c}{\emph{Current}} & \emph{Empirical} & \multicolumn{2}{c}{\emph{Future}} & \emph{Extrapolated} & \emph{Empirical} \\
\multicolumn{2}{c}{\emph{Time} $T$} &  \emph{Cov.} @ $T$  & \multicolumn{2}{c}{\emph{Time} $T'$} & \emph{Cov.} @ $T'$ & \emph{Cov.} @ $T'$\\\hline
 1 & min & 61.5\% &  2 & min & 62.7\% & 65.9\%\\
 1 & min & 61.5\% &  5 & min & 66.3\% & 67.1\%\\
 1 & min & 61.5\% &  10 & min & 71.5\% & 70.5\%\\
10 & min & 70.5\% & 15 & min & 71.7\% & 83.3\%\\
10 & min & 70.5\% & 30 & min & 75.1\% & 83.7\%\\
 1 & hour& 87.2\% & 90& min & 87.4\% & 87.5\%\\
 1 & hour& 87.2\% &  2 & hours & 87.6\% & 87.6\%\\
10 & hours& 96.6\% & 15 & hours & 96.7\% & 96.7\%\\
10 & hours& 96.6\% & 20 & hours & 96.9\% & 97.1\%\\
 1 & day & 97.1\% & 1.5 & days & 97.2\% & 97.2\%\\
 1 & day & 97.1\% & 2 & days   & 97.8\% & 98.6\%\\ \hline
\end{tabular}
\end{table}

Given that $n$ test inputs have been generated and $S(n)$ of $S$ statements have been covered in the fuzzing campaign, within the STADS framework we extrapolate the number of covered statements $S(n+m^*)$ when $m^*$ more test inputs have been generated as follows \cite{coverageSurvey}
$$\hat S(n+m^*) = S(n) + \hat Q_0\left[1-\left(1-\frac{Q_1}{n\hat Q_0+Q_1}\right)^{m^*}\right]$$
where $\hat Q_0=S-S(n)$ is the number of uncovered statements and $Q_1$ is the number of statements that are executed by exactly one generated test input. Since AFL only needs to maintain $Q_1$ (in addition to $S$, $S(n)$, and $n$), code coverage can be extrapolated \emph{very efficiently and scalably} (i.e., independent of the size of the fuzzed program). The accuracy decreases as $m^*$ increases. However, a 95\%-confidence interval that allows to assess the decrease of accuracy is available via statistical bootstrapping \cite{coverageSurvey}.

Note that the STADS statistical framework also allows us to extrapolate other quantities such as discovery probability and other kinds of species coverage.

%% file: sections/data.tex
\section{Automated Software Testing and Analysis as Discovery of Species}\label{sec:model}

In the following we present our statistical framework of automated \underline{s}oftware \underline{t}esting and \underline{a}nalysis as \underline{d}iscovery of \underline{s}pecies (STADS).
Let $\mathcal{P}$ be the program that we wish to fuzz. We call as $\mathcal{P}$'s \emph{input space} $\pmb{\mathcal{D}}$ the set of all inputs that $\mathcal{P}$ can take. \hl{As inputs, we consider command line parameters, files, event sequences, messages, data streams, data bases, but also other objects that impact program behavior that are not normally considered inputs, such as environment variables, configuration, values returned from system calls, thread schedules, and so on.}
Let $\mathcal{F}$ be a stochastic process that samples inputs $t\in \pmb{\mathcal{D}}$. We call $\mathcal{F}$ \emph{fuzzer} and the sampling of inputs \emph{test input generation}.

\subsection{Search Space of the Fuzzer}
Most fuzzers operate on a \emph{restricted search space} $\pmb{\mathcal{D}}'\subseteq \pmb{\mathcal{D}}$ such that $\mathcal{F}$ is effectively unable to generate \emph{all} inputs that $\mathcal{P}$ can take. 
For instance, a fuzzer might generate test inputs that are ``valid'' w.r.t. a pre-specified input model \cite{MWF}; yet, there might be programs that exhibit a vulnerability only for an invalid input. A fuzzer might generate inputs only up to a certain maximum size; yet, there might be programs that exhibit a vulnerability only for substantially larger test inputs. Hence, the test inputs that are generated within a fuzzing campaign are necessarily random only within the capabilities of the fuzzer. 
For instance, the search space for CSmith \cite{csmith} is a subset of all C programs (rather than a random sequence of UTF8-characters). 

Hence, estimates that are derived from methodologies in the STADS framework hold only w.r.t. the fuzzer's search space and the tested program $\mathcal{P}$.
The search space is specified either explicitly by the input model, grammar, or protocol that is used (and/or induced \cite{grammarInference}) during fuzzing to reduce the fuzzer's search space \cite{MWF,gwf,snooze}, or implicitly by the fuzzer's inherent limitations to generate certain inputs. Most fuzzers do not also fuzz a program's environment (OS, architecture, current date, etc.), further restricting the behaviors that $\mathcal{P}$ can exhibit when fuzzed by $\mathcal{F}$.

In ecology, the sampling might also operate on a restricted search space \cite{longino}. For instance, certain areas may not be accessible. A net may not trap species smaller than its mesh. A light trap may not lure light-insensitive species. 

\subsection{Species Identification}
\hl{Suppose, the fuzzer's search space $\pmb{\mathcal{D}}'$ can be subdivided into $S$ individual subdomains \{$\mathcal{D}_i\}_{i=1}^S$, called \emph{species}. All inputs belong to at least one species and multiple inputs can belong to the same species. Specifically, 
all inputs belong to the same \emph{species} that share the same discrete property of the program $\mathcal{P}$. For instance, we could consider each input that covers the same program statement to belong to the same species.} 
Depending on the specific objective (e.g., ``Cover all statements!''), we can choose a suitable species identification, and devise a sampling strategy that can discover a maximal number of species. The \emph{concrete fuzzing objective} is thus encoded by the way the species is identified for a certain input.

In our STADS framework, a \emph{dynamic analysis} identifies the specific species to which a generated test input belongs. For C programs, the \texttt{gcov} coverage-tool \cite{gcov} identifies the statements and branches an input covers; the AFL-instrumentation \cite{afl} identifies the specific path an input exercises; or AddressSanitizer \cite{asan} identifies the type of vulnerability an input exposes. \hl{Notice, some objectives require the dynamic analysis to identify a single species for each input (e.g., the path an input exercises) while others require to identify multiple species for a single input (e.g., the statements an input covers).} Notice also, we require \emph{deterministic execution}: The same input, executed an arbitrary number of times, must always belong to the same species. However, conceptually we can integrate non-deterministic programs by considering all potential non-deterministic actions as part of the program's input space. For instance, a test input for a concurrent program also specifies a specific thread-interleaving. A test input for an interactive program (e.g., an Android app) also specifies a state it must start from. 

\subsection{Fuzzing Campaigns}
The fuzzer $\mathcal{F}$ generates $n$ test inputs and is said to \emph{discover} a species $\mathcal{D}_i$ when $\mathcal{D}_i$ is sampled for the first time. The \emph{general fuzzing objective} is then to discover a maximal number of species.  
Let $p_i$ be the probability with which $\mathcal{F}$ samples species $\mathcal{D}_i$ for $i:1\le i\le S$ at any point during the fuzzing campaign.
\hl{Note that the STADS framework fully accounts for \emph{arbitrary fuzzer heuristics}, including the sampling from the operational distribution }\cite{operational}\hl{, as long as the fuzzer does not change the sampling strategy adaptively throughout the fuzzing campaign. For instance, if a fuzzer generates more ``typical'' program inputs by sampling from the program's operational distribution---because the software engineer deems the detection of bugs that could also be found by a customer as more important---then all statistical claims derived from the STADS framework strictly hold w.r.t. that fuzzer within the stipulated confidence bounds. This fuzzer is simply more likely (greater $p_i$) to discover an ``operational'' bug $\mathcal{D}_i$ than a  fuzzer without that heuristic, \emph{for all fuzzing campaigns}.}

\hl{Within our statistical framework we must assume that the \emph{relative species abundances} $\mathbf{p}=\{p_i\}^S_{i=1}$ does not substantially change during the fuzzing campaign. However, in practice this assumption might not hold. We say that (feedback-directed) fuzzers where $\mathbf{p}$ changes during the fuzzing campaign have an \emph{adaptive sampling bias}. For instance, a \emph{coverage-directed fuzzer} retains generated test inputs that previously discovered a new species, and fuzzes those in addition to the initially provided seeds }\cite{afl}\hl{. This allows to sample a ``neighboring'' species $\mathcal{D}_i$ with a greater probability $p_i$ (thus also increasing the efficiency of the fuzzer }\cite{aflfast}\hl{). Yet, the rate at which new species are discovered is consistently decelerating throughout a fuzzing campaign. Hence, the rate at which the probabilities $p_i$ change is also decelerating and the magnitude of the change decreases such that \emph{the adaptive bias reduces} as more test inputs are generated. We investigate the adaptive bias empirically in }\autoref{sec:eval}\hl{ and provide an extended discussion in} \autoref{sec:bias}.

Within our STADS statistical framework, we do \emph{not} assume that the fuzzer has any information about the species identification. Specifically, the ``location'' and relative abundance $p_i$ of a specific (undiscovered) species $\mathcal{D}_i$ as well as the total number of inputs $N$ and the total number of species $S$ are a-priori \emph{unknown}. Within a single \emph{fuzzing campaign}, the fuzzer generates $n$ test inputs and discovers $S(n)$ species. A species $\mathcal{D}_i$ is \emph{discovered} when $\mathcal{F}$ generates the first test input $t$ that belongs to $\mathcal{D}_i$, i.e., $t\in \mathcal{D}_i$. During a fuzzing campaign, the fuzzer generates $X_i$ test inputs that belong to species $\mathcal{D}_i$, $\sum_{i=1}^S X_i=n$. Only species where $X_i>0$ are marked as discovered. 

There are two pertinent measures that \emph{characterize a program}. The \emph{species richness} $S$ quantifies the total number of species, such as the number of statements, paths, vulnerabilities, information flows, et cetera, in the program. In contrast, the \emph{species evenness} $J$ quantifies how ``even'' the relative abundances $\mathbf{p}$ are distributed. Formally, we compute \emph{species evenness} $J$ using Pielou's evenness index \cite{evenness},\vspace{-0.3cm}
\begin{align}
J=\frac{H}{H_\text{max}} &\quad\quad\text{where}\ \ H = -\sum_{i=1}^S p_i\ln p_i \quad\text{is Shannon's diversity index}\\
 &\quad\quad\text{and}\quad\ \ H_\text{max}=\ln S \quad\quad \ \text{is the max. possible value of } H
\end{align}
Note that $0\le J \le 1$. Shannon's diversity index is also known as Shannon entropy. Both quantities $S$ and $J$ can be illustrated by the following example. In Case\#1, half of all inputs might exercise one path while another half might exercise another. In Case\#2, 90\% of all inputs might exercise one path and only 10\% another. Both cases have the same total number of paths ($S_1=S_2=2$) but feature a very different evenness ($J_1=1$, $J_2=0.47$). The asymptotic total number of species $S$ is important to determine how many more species we can expect to discover in a fuzzing campaign. The species evenness $J$ is important to choose the right testing tool. If $J$ is very low, symbolic execution-based fuzzers \cite{klee,s2e} might be more appropriate than random fuzzers \cite{afl,peach}.\footnote{For an extended discussion of blackbox versus whitebox testing efficiency, see B\"{o}hme and Paul \cite{efficiency}.}

There are two pertinent measures that \emph{characterize a fuzzing campaign}, species coverage and discovery probability. \hl{We define as \emph{species coverage} $G$ the proportion of species that have been discovered after generating $n$ test inputs,}\vspace{-0.1cm}
\begin{align}
G(n)=\frac{S(n)}{S}%S_\text{obs}/S
\end{align}
Examples of species coverage are path coverage in our motivating example, where the total number of species $S$ is \emph{not} known and must be estimated, and code coverage, where $S$ \emph{is} indeed known. 
\hl{We define as \emph{discovery probability} $U$ the proportion of inputs that belong to species that remain undiscovered after generating $n$ test inputs,}
\begin{align}
U(n)&=1-\dfrac{\sum_{i=1}^S p_iI(X_i>0)}{\sum_{i=1}^S p_i} = \dfrac{\sum_{i=1}^S p_iI(X_i=0)}{\sum_{i=1}^S p_i}
% &=\sum_{X_i>0}p_i \quad\text{for non-overlapping subdomains.}
\end{align}
where $I(A)$ is the indicator function, i.e., $I(A)=1$ if the event $A$ occurs, and $I(A)=0$ otherwise.
\hl{We define as \emph{sample coverage} $C$ the complement of the discovery probability. In ecology, the sample coverage is the proportion of individuals in the assemblage whose species is represented in the sample. In automated software testing, it essentially quantifies the proportion of (tested \emph{and} untested) program inputs that stress program behaviors that have already been tested before.} In \autoref{sec:estC} we show that the estimate $\hat U$ of the discovery probability provides an upper bound on the probability to expose a vulnerability, whence the sample coverage estimate measures the confidence that a fuzzing campaign inspires in the correctness of the program. The sample coverage that is achieved in a fuzzing campaign depends on the species evenness~$J$ and the number of test inputs $n$ that are generated. 
Intuitively, the lower the evenness, the more test inputs $n$ a fuzzing campaign must generate to expect a reasonably high sample coverage $C$.

\subsection{Main Hypothesis}

\hl{I hypothesize that within the STADS statistical framework the rare species which have been discovered throughout a fuzzing campaign explain the species within the fuzzer's search space that remain undiscovered. Intuitively, it is the \emph{difficulty to discover a rare species}, measured by the total number of test inputs that needed to be generated before discovering the rare species, that provides insights on the difficulty of discovering yet undetected (but detectable) species.}
\begin{hypo*}
\hl{Almost all information about number and relative abundance of undiscovered species within the fuzzer's search space is in the number and relative abundance of rare species that have already been discovered.}
\end{hypo*}
\hl{A species $\mathcal{D}_i$ is considered \emph{rare} if $1\le X_i\le\kappa$, where $\kappa$ is an arbitrary but very small constant. In fact, almost all estimators and extrapolators presented in this article are functions of the number of singleton and doubleton species (i.e., $\kappa = 2$).}

\hl{The same hypothesis is underpinning the nonparametric biostatistics in ecology. Chao and Chui argue that ``[..] abundant species (which are certain to be detected in samples) contain almost no information about the undetected species richness, whereas rare species (which are likely to be either undetected or infrequently detected) contain almost all the information about the undetected species richness'' }\cite{hypothesis}\hl{. Therefore, most nonparametric estimators and extrapolators are based on counts of rare species. In order to \emph{test the main hypothesis}, we need to establish the accuracy of these estimators and extrapolators within the STADS framework.

This hypothesis is the reason for the great scalability of the STADS framework. For most estimates, the fuzzer needs to record only the number of rare species that have been discovered. Hence, the computation of the estimates scales easily to very large programs in our experiments.}

\subsection{The Multinomial Model: One Input, Single Species}\label{sec:nonoverlapping}

\hl{Some concrete fuzzing objectives require to identify a single species for each input.} For instance, an input can execute only one path \cite{dart}, exercise only one method call sequence, compute only one final output \cite{peso}, crash only at one program location; a single input either exposes a vulnerability or does not expose a vulnerability. In ecology, one individual can also belong only to a single species. A researcher samples individuals from the assemblage at various random locations and records for each detected species the number of occurrences. When individuals are sampled, ecologists call the collected data as \emph{abundance data} and utilize the \emph{multinomial model} \cite{hurlbert,sampleSurvey}. In the multinomial model, within STADS a generated test input is considered as \emph{individual}.\footnote{In \autoref{sec:overlapping}, we extend the STADS framework to include the Bernoulli product model, where a generated test input is considered as \emph{sampling unit}. In the STADS framework, the Bernoulli model describes concrete fuzzing objectives that require to identify one or more species for a single input.} 

Define the \emph{abundance frequency count} $f_k$ as the number of species that contain exactly $k$ test inputs which were generated throughout the current fuzzing campaign, $0\le k \le n$. More formally, $f_k=\sum_{i=1}^S I(X_i=k)$, where $I(A)$ is the indicator function, i.e., $I(A)=1$ if event $A$ occurs and $I(A)=0$ otherwise. Hence, $n=\sum_{k=1}^n kf_k$ and $S(n)=\sum_{k=1}^nf_k$. The abundance frequency count $f_0$ represents the number of undiscovered species. We call $f_1$ the number of \emph{singleton species} and $f_2$ the number of \emph{doubleton species}. 
The input space contains $S$ non-overlapping subdomains, where the probability that the fuzzer generates a test input that belongs to species $\mathcal{D}_i$ is $p_i$ for $i:1\le i \le S$. Note that $\sum_{i=1}^S p_i=1$. The multinomial probability distribution has the probability mass function %\cite{sampleSurvey}
\begin{align}
P(X_1=x_1,\ldots,X_S=x_S)=\frac{n!}{x_1!\ldots x_S!}p_1^{x_1}p_2^{x_2}\ldots p_S^{x_S}\label{eq:suff}
\end{align}
From \autoref{eq:suff}, we can see that the number of generated test inputs $X_i$ that belong to species $\mathcal{D}_i$ is a \emph{sufficient statistic}, meaning that no other statistic which can be calculated from the same sample provides any additional information as to the value of the (estimated) parameter. This renders the abundance frequency counts $f_k$, which are defined from $X_i$, suitable components for the estimators and extrapolators of fuzzing progress. As Colwell et al. \cite{sampleSurvey} point out, the multinomial model assumes that the sampling procedure itself does not substantially alter the probabilities ($p_1,p_2,\ldots,p_S)$. The authors provide more details about the multinomial model and its utility in the ecologic context. The case where multiple species can be identified for a single input is explained by the Bernoulli product model \cite{incidenceSurvey} and discussed in \autoref{sec:overlapping}.

%% file: sections/progress.tex
\section{Estimating Residual Risk and Campaign Completeness}\label{sec:estimate}
\hl{Our model of software testing and analysis as discovery of species (STADS) provides access to a rich statistical framework in ecology. This unexpected connection between two otherwise unrelated fields of research provisions software testing with methodologies to accurately estimate how much we have seen and to extrapolate from the seen to the unseen. In this section, we focus on the \emph{estimation} of how much has been tested and how much more there is. We show that an estimate of the probability to discover a new species can provide a \emph{statistical guarantee} that no (detectable) vulnerability exists that has not already been discovered. Moreover, we present novel methodologies to assess \emph{campaign completeness} (i.e., the progress of an ongoing campaign towards completion).}

\subsection{Discovery Probability and Sample Completeness}\label{sec:estC}
\hl{In the STADS framework, the \emph{discovery probability} $U(n)$ measures the current probability to discover a new species with the $n+1$th generated test input where $n$ is the number of test inputs that have been generated throughout the current fuzzing campaign (i.e., $U(0)=1$).
If the dynamic analysis is able to identify vulnerabilities, then the discovery probability $U$ provides a \emph{statistical guarantee} that no detectable vulnerability exists if none has been discovered. In other words, security researchers can use the STADS statistical framework for residual risk assessment.}

The concept of discovery probability might seem to require advance knowledge of the true relative species abundance $\{p_i\}_{i=1}^S$ during a fuzzing campaign. However, the discovery probability can be \emph{very accurately and efficiently} estimated using only information contained in the single, uncompleted fuzzing campaign itself, as long as the number of generated test inputs is reasonably large \cite{good,goodError}. Hence, the concept of discovery probability finds application across many fields of science, such as rare event estimation \cite{rare}, cryptanalysis \cite{crypt}, computational linguistics \cite{smoothing}, biology \cite{coverageSurvey}, actuarial science, and so on.

In the STADS framework, the \emph{sample coverage} $C(n)=1-U(n)$ measures the probability that the $n+1$th generated test input belongs to an already discovered species. In other words, we know the species for $C\%$ of program inputs in the fuzzer's search space. Sample coverage also directly measures \emph{sample completeness}, i.e., how complete the sample is w.r.t. the remaining undiscovered species in the assemblage. Hence, in ecology sample coverage is routinely used to choose the most accurate estimator \cite{brose} and to compare attributes of species across assemblages \cite{coverageSurvey}. In software testing, sample coverage can also be used to to assess the progress of the current fuzzing campaign towards completion \emph{without} the need to estimate $\hat S$ the total number of species.
If the fuzzer has exposed no vulnerabilities, the sample coverage quantifies the \emph{degree of confidence} that the fuzzing campaign inspires in the correctness of the program. 

The \emph{inverse of the discovery probability} $1/U(n)$ gives the number of test inputs that we can expect to generate before discovering a previously undiscovered species. Given the number of test inputs generated per unit time $\Delta$, we can derive the expected time until discovery as $1/(\Delta \cdot U(n))$.

\textbf{Estimation in STADS}. 
In the multinomial model, the Good-Turing estimator estimates the probability to generate a test input that belongs to an undiscovered species. Thus, using the Good-Turing estimator \cite{good}, the estimate of the \emph{discovery probability} $\hat U(n)$ is obtained as 
\begin{align}
\hat U(n) &= \frac{f_1}{n} 
\end{align}
\hl{where $f_1$ is the number of singletons and $n$ is the total number of generated test inputs.} 
According to Good \cite{crypt}, the Good-Turing estimators were developed by Alan Turing during World War II while breaking Enigma codes. Good and Turing showed that their estimator can be accurately and efficiently computed only from the sample itself \cite{good}. Moreover, the estimator is \emph{strongly consistent}, meaning that its accuracy strictly increases as the sample size (i.e., number of generated test inputs) increases \cite{consistent}. Zhang and Zhang \cite{goodNormal} prove \emph{asymptotic normality} of the Good-Turing estimator, meaning that the convergence to the true value is also reasonably fast. Robbins \cite{goodError} showed that the \emph{mean squared error} of the Good-Turing estimator is less than $1/n$, which indicates that it is quite accurate if $n$ is large. In theory, it is assumed that the probability $p_i$ to sample a species $\mathcal{D}_i$ follows a binomial distribution. However, in practice the Good-Turing estimator seems to perform close to the \emph{best natural estimator} for \emph{any} distribution \cite{goodRobust}. 

\textbf{Statistical Guarantee}.
We show that the \emph{Good-Turing estimate $\hat U(n)$ of the discovery probability} provides an upper bound on the probability that an error in the fuzzer's search space remains undiscovered given that no error has been exposed after generating $n$ test inputs.\footnote{A vulnerability is just a special case of an error.} Depending on the dynamic analysis, the fuzzer's search space is partitioned by the species identified for each input. Inputs that belong to the same species share the same input subdomain.
Suppose, the \emph{progress-based dynamic analysis} partitions the input space according to the concrete fuzzing objective (e.g., based on the path or the statements that are exercised as in our motivating example). There are $S$ subdomains $\mathcal{A}=\{\mathcal{D}_i\}_{i=1}^S$ in the fuzzer's search space. Further suppose, an \emph{error-based dynamic analysis} partitions the same search space into $T$ subdomains $\mathcal{B}=\{\mathcal{E}_j\}_{j=1}^T$. A partitioning is \emph{error-based} if all inputs that belong to the same species homogeneously either do or do not expose an error \cite{efficiency}. One could imagine error-based partitioning as black and white regions in the restricted input space of the program, where the black regions contain inputs that expose an error. 
In practice, a dynamic analysis, such as ASAN \cite{asan}, would identify \emph{some} test input executions that expose an error. Hence, the statistical guarantees hold \emph{modulo} the dynamic analyzer's capability to identify a error-exposing input.\footnote{Similarly, in software verification the formal guarantees are valid only \emph{modulo} the provided specification. However, like a dynamic analysis in software testing, a specification may be \emph{incomplete} (i.e., does not allow to detect \emph{all} vulnerabilities; a.k.a. false negatives) or \emph{incorrect} (i.e., reports vulnerabilities when there is none; a.k.a. false positives).} 
Let a \emph{combined dynamic analysis} be derived by intersecting the progress- and error-based partitioning. The joint partitioning yields $R$ species $\mathcal{AB}=\{\mathcal{D}_i\bigcap\mathcal{E}_j\ |\ \mathcal{D}_i\in\mathcal{A}, \mathcal{E}_j\in\mathcal{B}\}/\emptyset$, where $\cdot/\cdot$ is the difference operation to remove ``empty'' species and $R \le S+T$. Notice that the number of singletons $f_1$ and doubletons $f_2$ for the progress-based analysis $\mathcal{A}$ are also the number of singletons and doubletons for the combined analysis $\mathcal{AB}$. 
Assuming that no error has been exposed throughout the fuzzing campaign, all error-\emph{exposing} species in $\mathcal{AB}$ are clearly still among the undiscovered ones. Since the estimate $U$ of the discovery probability denotes the proportion of inputs that belong to \emph{undiscovered} species for $\mathcal{AB}$ (and $\mathcal{A}$), it provides an upper bound on the proportion of inputs exposing an error. A similar argument can be constructed trivially for the Bernoulli product model. \hfill $\blacksquare$

\textbf{Quantifying Accuracy}. In the STADS framework, approximate estimators of the \emph{variance} and the associated \emph{confidence interval} can be derived with an asymptotic approach \cite{goodNormal,spade}.
We also note that one must account for the resulting \emph{missing} probability mass when estimating the relative species abundance $p_i$ for each discovered species $\mathcal{D}_i$, e.g., to estimate species evenness $J$ \cite{evenness}. In the multinomial model, the estimator $\hat p_i=X_i/n$ would evidently over-estimate $p_i$. This can be remedied with an approach called \emph{smoothing} \cite{good,smoothing}. 

\textbf{Scalability}. In practice, the computation of the discovery probability estimate $\hat U(n)$ is efficient and easily scales with program size (i.e., with \#species $S$). The fuzzer needs to store information only about doubleton and singleton species in addition to the number of generated test inputs and the number of discovered species.
In the statistical programming language $R$, the \texttt{goodTuring}-function of the \texttt{edgeR}-package \cite{edgeR} implements Good-Turing estimation while the \texttt{goodTuringProportions}-function implements the Good-Turing smoothing procedure. The \texttt{spadeR}- and \texttt{iNext}-packages \cite{spade,inext} for $R$ compute the improved discovery probability estimator. 
The \texttt{iNext}-package also provides 95\%-confidence intervals.

\subsection{Species Coverage}\label{sec:partCov}
\hl{In ecology, species richness measures the number of species in the assemblage. In the STADS model, we define species coverage as the proportion of species in the assemblage that have been discovered throughout the fuzzing campaign.} 
Hence, with an estimate of species richness $\hat S$ we can compute the current species coverage $\hat G(n)=S(n)/\hat S$ to assess the current progress of the fuzzing campaign towards completion. %For instance, humankind has recorded 1.2 million species in all kingdoms of life on Earth and the recent estimate of the total number of species is {\raise.17ex\hbox{$\scriptstyle\sim$}}8.7 million ($\pm 1.3$ million SE) \cite{species}. Hence, after 250 years of taxonomic classification only a small fraction of all species, {\raise.17ex\hbox{$\scriptstyle\sim$}}14\%, has been recorded. Clearly, this puts us at the beginning of global species discovery. 
%
%\textbf{Overview}.
At the basis of most estimators is the observation that the species discovery curve \emph{decelerates} over time as the number $n$ of generated test input increases \cite{efficiency}. At the beginning of the fuzzing campaign many species are discovered in a short time. Later, it takes more and more time to discover the next undiscovered species. For our motivating example, this deceleration can be observed in \autoref{fig:paths}. In fact, the discovery curve appears to approach an asymptote which is estimated at 5408 paths (using the \emph{Chao1}-estimator \cite{chao1}). The asymptotic total number of species is our estimation target. In the following, we review various estimators $\hat S$. %of the asymptotic total number $S$ of species in an assemblage \cite{speciesReview} using the terminology of the STADS model.
  We refer to Colwell et al. \cite{sampleSurvey} for a more extensive review of available methodologies. An empirical and simulation-based comparison of several estimators was conducted by Hortal et al. \cite{speciesEmpirical}. 

During their investigations of the species discovery curve in what we now call the STADS model, B\"{o}hme and Paul \cite{efficiency} suggest fitting an exponential curve to extrapolate how many species we can expect to discover in a given time budget. \emph{Curve fitting} would also allow us to determine the asymptotic total number of species \cite{preston,collwell2}. However, curve fitting approaches are not based on any statistical sampling model which prevents us from effectively evaluating the variance of the resulting asymptote. Moreover, different functional forms may manifest the same goodness of fit but yield vastly different estimates of the asymptote which calls into question the statistical soundness of this approach. 

\emph{Sampling-theory-based approaches} build upon a statistical foundation and can be broadly distinguished into parametric and non-parametric frameworks \cite{magurran}. In the parametric framework, it is assumed that the relative species abundances $\{p_i\}_{i=1}^S$ follow a statistical model with one or two parameters (e.g., Poisson process \cite{fisher}). However, parametric models usually require extensive numerical procedures and work well only when the correct distribution is already known \cite{chui}. Yet, in software testing and analysis, just like in ecology, the distribution is often unknown.
The most effective estimators of the total number of species are \emph{sampling-theory-based and non-parametric} \cite{sRichness}. Here, we can distinguish Jacknife, coverage-based, and Chao1/2-type estimators.

\emph{Jackknife estimators} were developed to reduce the bias of a biased estimator and allow to compute  variance and confidence intervals for the estimate \cite{palmer,walther,burnham}. The current number of discovered species $S(n)$ is obviously a negatively biased estimator of the total number of species $S$. In the multinomial model of the STADS framework, the first-order jackknife estimator $\hat S_{jk1}$ corrects this bias by assuming that the number of undiscovered species equals the number of singletons $f_1$ 
\begin{align}
\hat S_{jk1} &= S(n) + \frac{n-1}{n}f_1\\ 
&\approx S(n) + f_1
\end{align}
In the multinomial model of the STADS framework, the second-order jackknife estimator $\hat S_{jk2}$ for which the estimated number of unseen species is in terms of singletons and doubletons has the form
\begin{align}
\hat S_{jk2} &= S(n) + \frac{2n-3}{n}f_1 - \frac{(n-2)^2}{n(n-1)}f_2\\
&\approx S(n) + 2f_1 - f_2
\end{align}
Burnham and Overton \cite{burnham} provide higher orders of the jackknife estimators. All Jacknife estimators can be expressed as linear combinations of frequencies and thus variances can be obtained.

\emph{Chao1-type estimators} provide a lower bound for the total number of species rather than a point estimate \cite{chao1}. When there are a large number of undiscovered species, it will be statistically impossible to obtain a good estimate of species richness. Hence, a good lower bound is often more practical than an imprecise point estimate. Chao \cite{chao1} derived such a lower bound called \emph{Chao1} for the multinomial model:
\begin{align}
\hat S_{\text{Chao1}} &= \begin{cases}
S(n) + \frac{n-1}{n} \frac{f_1^2}{2f_2} & \text{if } f_2 > 0\label{eq:one}\\
S(n) + \frac{n-1}{n} f_1(f_1-1)/2 & \text{if } f_2 = 0
\end{cases}\\
&\approx \begin{cases}
S(n) + f_1^2 / (2f_2) & \text{if } f_2 > 0\\
S(n) + f_1(f_1-1)/2 & \text{if } f_2 = 0
\end{cases}
\end{align}
where in the current fuzzing campaign $n$ is the total number of test inputs generated, $S(n)$ is the total number of species discovered, and $f_1$ and $f_2$ are the abundance frequency counts for singleton and doubleton species, respectively.

\hl{Very recently, Chao et al. }\cite{goodtheory}\hl{ showed that $\hat S_{\text{Chao1}}$ is an \emph{unbiased point estimator} as long as very rare species (i.e., undetected and singleton species) have approximately equal relative abundance. If very rare species are unevenly distributed and the sample size is not sufficiently large, the available data do not contain sufficient information, and it is only reasonable to provide a good lower bound estimate of species richness $S$.}

An \emph{improved lower bound} can be obtained from tripleton and quadrupleton species, respectively. Chui et al. \cite{chui} derived the improved lower bound called \emph{iChao1} for the multinomial model:
\begin{align}
\hat S_{\text{iChao1}} &= \hat S_{\text{Chao1}} + \frac{n-3}{n}\frac{f_3}{4f_4} \times \max\left(f_1 - \frac{n-3}{n-1}\frac{f_2f_3}{2f_4},0\right)\\
&\approx \hat S_{\text{Chao1}} + \frac{f_3}{4f_4} \times \max\left(f_1 - \frac{f_2f_3}{2f_4},0\right)\label{eq:two}
\end{align}
where $f_3$ and $f_4$ are the frequency counts for tripleton and quadrupleton species, respectively.

\emph{Coverage-based estimators} utilize sample coverage, the proportion of inputs belonging to discovered species, to estimate the total number of species \cite{ace,chui}. As we have seen earlier, sample coverage, as the complement of the discovery probability, can be very accurately and efficiently estimated from the frequency counts alone, as long as the number of test inputs generated in the fuzzing campaign is reasonably large \cite{good}. As Chao and Chiu \cite{sRichness} point out, coverage-based estimators might be appropriate when there are many rare species, i.e., where $0\ll |\{p_i\ |\ p_i\ll \frac{1}{S}, 1\le i\le S\}|\lesssim S$ and $|\cdot|$ gives the cardinality of the set. However, for the lack of space we are adjourning to future work the discussion and evaluation of the ACE and ACE-1 estimators \cite{ace,ace1}. % for non-overlapping subdomains and ICE and ICE-1 \cite{ice,ice1} for overlapping subdomains. 

\textbf{Species coverage}. In the STADS framework, we compute the estimate $\hat G$ of the species coverage that has been achieved in the campaign by dividing the number of currently discovered species $S(n)$ by the estimated total number of species $\hat S$. If $S$ is known, then $\hat S=S$. For instance, in our motivating example path coverage is computed w.r.t. an estimated total number of species while statement coverage is computed w.r.t. the known total number of statements. Both, statement and path coverage are examples of species coverage, only that the same inputs are assigned to a different kind of species.

\textbf{Quantifying accuracy}. The \emph{variance} and 95\%-\emph{confidence intervals} for the estimators in the STADS framework can be derived by the standard statistical approximation method \cite{chao2} or using bootstrapping \cite{incidenceSurvey}.
%In practice, the Chao1-type estimators work well also as point estimates.
Hortal et al. \cite{speciesEmpirical} find that estimator accuracy strongly depends on the species evenness $J$ and on the completeness $C=1-U$ of the sample. Effectively, the accuracy of the estimate improves as the discovery probability $U$ decreases or species evenness $J$ increases. 

\textbf{Scalability}. In practice, the computation of all sampling-theoretic, non-parametric estimators of species coverage is efficient and easily scales with program size (i.e., with \#species $S$). In most cases, the fuzzer needs to store information only about doubleton and singleton species in addition to the number of generated test inputs and discovered species.%\footnote{For the improved estimators \emph{iChao}, information about quadrupleton and tripleton partitions would also need to be recorded.}
In the statistical programming language $R$, the \texttt{ChaoSpecies}-function of the \texttt{SpadeR}-package \cite{spade} implements several estimators of the total number of species. The \texttt{ChaoSpecies}-function also reports 95\%-confidence intervals. 

%% file: sections/extrapolation.tex
\section{Extrapolation of Species Discovery}\label{sec:extrapolate}
An extrapolation allows to assess the trade-off between investing more time and gaining more insight. We discuss novel methodologies from ecology to quantify this return on investment. Specifically, using extrapolation in the model of software testing and analysis as discovery of species (STADS), the security researcher can answer the following questions. %for both, overlapping and non-overlapping subdomains:
\begin{enumerate}
  \item Given in the current fuzzing campaign $n$ test inputs have been generated and the researcher has time to generate only $m^*$ more test inputs, how much species coverage $\hat G(n+m^*)$ and residual risk $\hat U(n+m^*)$ can she expect to achieve?
  \item Given in the current fuzzing campaign $n$ test inputs have been generated and the security researcher would like to achieve a specific species coverage $G^*$, how many more test inputs $m_{G^*}$ can she expect to generate before achieving $G^*$ (i.e., $m_{G^*}$ s.t. $\hat G(n+m_{G^*})=G^*$)?
\end{enumerate}
Using these extrapolators, a security researcher can make an informed decision whether to continue or abort a fuzzing campaign. Suppose, the client requires a statistical guarantee (i.e., discovery probability) of $10^{-8}$ as upper bound of the probability that the fuzzer finds a vulnerability in the program. The researcher can estimate the effort that is required to achieve that degree of confidence in the correctness of the program. 

\subsection{Estimating Progress Towards Completion within a Given Time Budget}\label{sec:extProg}
In our STADS statistical framework, there are several estimators of the expected number $\hat S(n+m^*)$ of discovered species if the reference sample of size $n$ was augmented by $m^*>0$ more individuals (i.e., if $m^*$ \emph{more} test inputs were generated) \cite{goodToulmin,collwell3,solow,shen}. Chao and Jost \cite{coverageSurvey} provide an overview.
In the multinomial model, Shen et al. \cite{shen} proposed the following sampling-theoretic extrapolator based on the asymptotic total number of species:
\begin{align}
\hat S(n + m^*) &= S(n)+\hat f_0\left[1-\left(1-\frac{f_1}{n\hat f_0 + f_1}\right)^{m^*}\right]%\\
\end{align}
where for the current fuzzing campaign, $n$ is the number of generated test inputs, $S(n)$ is the number of discovered species, $\hat f_0=\hat S - S(n)$ is the expected number of undiscovered species, and $f_1$ is the number of singletons.

\hl{The rule of thumb is to keep the extrapolation within twice the sample size (i.e., $m^*\le n$) }\cite{sampleSurvey}\hl{. However, recently Orlitzki et al. (2016) }\cite{orlitzki}\hl{ introduced the \emph{provable} extrapolation of the discovered species for $m^*$ all the way up to $n\cdot [\log(n)-1]$ additional test inputs. This shows that the number of discovered species can be estimated for a population $\log(n)$ times larger than that observed. The authors go on to show that this is also the largest possible estimation range and that the estimators' mean-square error is optimal up to constants for any $m^*$.}

In the multinomial model of the STADS framework, we can derive the expected discovery probability $\hat U(n+m^*)$ if $m^*$ more test inputs were generated by recognizing that the discovery probability is only the difference in the number of discovered species between this and the next generated test input $U(n)=S(n+1)-S(n)$. Hence,
\begin{align}
\hat U(n+m^*)&= \hat S(n+m^*+1) - \hat S(n+m^*)\\
&=\frac{f_1}{n}\left(\frac{n\hat f_0}{n\hat f_0 + f_1}\right)^{m^*+1}
\end{align}
where $\hat f_0=\hat S - S(n)$ is the expected number of undiscovered species,  $f_1$ is the number of singleton species, and $f_2$ is the number of doubleton species.

\textbf{Quantifying accuracy}. In the STADS model, confidence intervals for the estimators $\hat S(n+m^*)$ and $\hat U(n+m^*)$ can be derived using the bootstrap method \cite{sampleSurvey,coverageSurvey}. In ecology, the rule of thumb is to keep the extrapolation within twice the sample size (i.e., $m^*\le n$) \cite{sampleSurvey}. The reason can be illustrated with the following example. Intuitively, the accuracy of $\hat S(n+10)$ is better when $n=1000$ than it is when $n=1$. Firstly, the extrapolator performs better at $n=1000$ because more information is available. Secondly, the margin of error is also  reduced because the species discovery curve decelerates substantially (cf. \autoref{fig:paths}.a).

\textbf{Scalability}. In practice, the extrapolation of species coverage $\hat G=\hat S(n+m^*)/\hat S$ and of the discovery probability $\hat U(n+m^*)$ is efficient and easily scales with program size (i.e., with \#species $S$). The fuzzer needs to store information only about doubleton and singleton species in addition to the number of generated test inputs $n$ and the number of discovered species $S(n)$. 
In the statistical programming language $R$, the \texttt{iNext}-package \cite{inext,inextOnline} computes the extrapolation and also provides 95\%-confidence intervals.

\subsection{Estimating Number of Inputs Needed to Discover a Given Proportion of Species}

Chao et al. \cite{samplesRequired} developed a non-parametric method for estimating the number of further test inputs that would need to be generated in order to achieve an arbitrary species coverage $G^*$. Formally, to reach a fraction $G^*$ of estimated total number of species $\hat S$ where $\hat G(n) < G^* < 1$, in the multinomial model of the STADS framework  the required number $m_{G^*}$ of further test inputs is estimated as
\begin{align}
m_{G^*} \approx
\frac{n f_1}{2f_2}\log\left[\frac{\hat f_0}{(1-{G^*})\hat S}\right]  
\end{align}
where in the current fuzzing campaign $n$ is the number of generated test inputs, $S(n)$ is the number of discovered species, $\hat f_0=\hat S - S(n)$ is an estimate of the number of undiscovered species, and $f_1$ and $f_2$ are the abundance frequency counts for singleton and doubleton species, respectively.

\textbf{Accuracy}. In the STADS model, confidence intervals for the estimators can be derived using the bootstrap method \cite{samplesRequired}. Given the estimate $\hat G(n)$ of current species coverage, we suggest that $\hat G(n)\le G^*\le 0.5 + \frac{\hat G(n)}{2}$ to keep the accuracy within a reasonable range. This suggestion is a variant of the rule of thumb stated in \autoref{sec:extProg} that $m_{G^*}\le n$ \cite{sampleSurvey}.

\textbf{Scalability}. In practice, computing $m_{G^*}$ is efficient and easily scales with program size (i.e., with \#species $S$). The fuzzer needs to store information only about doubletons and singletons in addition to the number of generated test inputs and discovered species. The logarithm of a 32bit-floating point number in \autoref{eq:nine} can be computed efficiently with a typecast, a bit shift, and a subtraction operation \cite{logeff}. All other basic mathematical operations require one CPU step each. In the statistical programming language $R$, the number of test inputs required to discover a certain proportion of all species can be estimated with the \texttt{num.samples.required}-function in the \texttt{sprex}-package \cite{sprex}. 

%% file: sections/empirical.tex
\section{Empirical Evaluation}\label{sec:eval}

\subsection{Research Objectives}
\hl{The main objectives of this preliminary empirical evaluation are}
\begin{enumerate}[leftmargin=0.7cm,itemsep=3pt]
  \item \hl{to \emph{test my main hypothesis} that within the model of automated software testing and analysis as discovery of species (STADS) the rare species which have been discovered throughout a fuzzing campaign explain the species within the fuzzer's search space that remain undiscovered.}
  \item \hl{to \emph{evaluate ecologic estimators and predictors} for the multinomial model in STADS. Specifically, we evaluate the \emph{Chao1} estimator $\hat S$ of species richness }\cite{chao1}\hl{, the predictor $\hat S(n+m^*)$ by Shen, Chao, and Feng }\cite{shen}\hl{ of the number of species that would be discovered if $m^*$ more test inputs were generated, and the Good-Turing estimator $\hat U(n)$} \cite{good}\hl{ of the discovery probability.}
  \item \hl{to \emph{investigate the impact of the adaptive sampling bias} of a feedback-directed fuzzer. An underlying assumption of most methodologies in the STADS framework is that the probability $p_i$ to generate a test input that belongs to species $\mathcal{D}_i$ does \emph{not} change substantially during the fuzzing campaign. However, it does for feedback-directed fuzzers, such as AFL.}
\end{enumerate}
\hl{We use path coverage as one kind of species coverage (i) because path coverage is the main measure of progress for our extension of AFL }\cite{afl}\hl{ (see }\autoref{sec:motivation}\hl{), the fuzzer used for our experiments, and (ii) because path coverage satisfies the conditions of the multinomial model (one species per input).}
We employ the \emph{Chao1}-estimator \cite{chao1} to estimate the asymptotic total number of paths $\hat S$ and the \emph{Good-Turing}-estimator \cite{good} to estimate the discovery probability. We estimate path coverage as $\hat G(n)=S(n)/\hat S$.
To extrapolate the number of paths discovered in the subsequent fixed-time interval, we compute the average number of tests generated per unit time and leverage the sampling-theoretic estimator $\hat S(n+m^*)$ proposed by Shen et al. \cite{shen}.
For our evaluation, we use established measures of estimator accuracy. The \emph{bias} of an estimator measures the mean difference of the estimate to the true value of the estimation target while the \emph{precision} measures the variance of the estimates. Specifically, we ask the following research questions:
\begin{description}[leftmargin=0.95cm,itemsep=3pt]
  \item[RQ.1] Can path coverage $\hat G(n)$ be used to effectively estimate the progress of a fuzzing campaign towards completion? Do different programs achieve the same path coverage, say 6 or 48 hours into the fuzzing campaign?
  \item[RQ.2] Can discovery probability $\hat U(n)$ be used to effectively estimate the residual risk of leaving detectable vulnerabilities undetected? Is the estimate representative for different fuzzing campaigns of similar length?
  \item[RQ.3] How \emph{biased} is the \emph{Chao1}-estimator $\hat S$ of the total number of paths? Is $\hat S$ systematically positively or negatively biased? What is the bias' magnitude and how can it be corrected?
  \item[RQ.4] How \emph{precise} is the \emph{Chao1}-estimator $\hat S$ of the total number of paths? How can the precision of $\hat S$ be increased? 
  \item[RQ.5] How \emph{biased} is the extrapolation of the number of discovered paths $\hat S(n+m^*)$ if $m^*$ more inputs were generated, where $m^*$ is the number of test inputs that we can expect to generate in 30~minutes, 1~hour, 2~hours, or 4 hours? Is $\hat S(n+m^*)$ systematically positively or negatively biased? What is the magnitude of the bias and how can it be corrected? Does the rule of thumb \cite{sampleSurvey} to keep the extrapolation within twice the sampling effort apply to automated software testing and analysis?
  \item[RQ.6] How \emph{precise} is the extrapolation of the number of discovered paths $\hat S(n+m^*)$ if $m^*$ more inputs were generated, where $m^*$ is the number of test inputs that we can expect to generate in 30~minutes, 1~hour, 2~hours, or 4 hours? How can the precision of $\hat S(n+m^*)$ be increased?
\end{description}
We present a \emph{summary} of our results w.r.t. our main objectives in \autoref{sec:resultsummary}.

\subsection{Setup and Infrastructure}
\textbf{Implementation}. We implemented the pertinent estimators and extrapolators into American Fuzzy Lop (AFL) \cite{afl}; we call our tool \pythia. \pythia uses lightweight instrumentation to determine, with negligible performance overhead, a unique identifier for the path that is exercised by an input. New inputs are generated by mutating a seed input using bit flips, boundary values, and block deletion and insertion strategies. If the new input exercises a new branch, or exercises a previously exercised branch exponentially more (or less often), it is added to the fuzzer's queue. \pythia stores for each seed in the queue the path-id and the number of generated test inputs that yield the same path-id. About every five (5) seconds, \pythia writes to a file the pertinent fuzzer data, including the current unix time, the number of generated test inputs $n$, the number of discovered paths $S(n)$, and the number of singletons $f_1$ and doubletons $f_2$ (i.e., \#paths exercised once or twice).

\begin{table}[h]\small
\caption{Subjects: Four security-critical open-source C projects of different program sizes.}
\label{tab:subjects}
\begin{tabular}{rrrl}
\multicolumn{1}{c}{\textbf{Program}} & \multicolumn{1}{c}{\textbf{Size}} & \multicolumn{1}{c}{\textbf{Test Driver}} & \multicolumn{1}{l}{\textbf{Description}}\\
\hl{json}\quad\cite{json}       & \hl{44 kLOC} & \hl{\texttt{parse\_msgpack}} & \hl{JSON parser}\\
libjpeg-turbo\quad\cite{libjpeg}&     91 kLOC & \texttt{libjpeg\_turbo\_fuzzer} & JPEG image library\\
openssl\quad\cite{openssl}      &    472 kLOC & \texttt{server} & cryptography and SSL/TLS library\\
\hl{libxml2}\quad\cite{libxml2} &\hl{500 kLOC} &\hl{\texttt{xmlint -d}} & \hl{XML parser}\\ 
ffmpeg\quad\cite{ffmpeg}        &   1071 kLOC & \texttt{AV\_CODEC\_ID\_MPEG4\_fuzzer} & audio and video streaming library\\
wireshark\quad\cite{wireshark}  &   3522 kLOC & \texttt{fuzzshark\_media\_type-json} & network protocol analyzer\\
\end{tabular}
\vspace{-0.2cm}
\end{table}

\textbf{Subjects}. We chose \hl{six} subject programs from Google's OSSFuzz fuzzing infrastructure \cite{oss}. The infrastructure fully automates the fuzzing of the 50+ integrated open-source C projects. OSS-Fuzz automatically downloads the most recent version of the subject, builds the subject, compiles the test drivers, and provides the initial seed corpus. We integrated \pythia as fuzzer into the fuzzing infrastructure. The list of subject programs used for our experiments is shown in \autoref{tab:subjects}. We chose these subjects because they are all security-critical, well-fuzzed, and of different sizes. 

\textbf{Setup}. For each subject we ran ten (10) fuzzing campaigns for 100 hours. The ten-fold repetition of the fuzzing campaign allows us to discuss  bias and precision of the estimators. The fuzzer was started with the same set of seeds and targeted the same test driver (Col.~3 in \autoref{tab:subjects}). In total, we spent a cumulative 6000 hours $\approx$ 8.2 months fuzzing these six subjects.

\textbf{Estimator performance}. Bias and precision are standard performance measures for estimators and extrapolators in biostatistics and ecology \cite{estimatorEvaluation,brose}. \emph{Bias} quantifies the difference between the estimate and the true value of the estimation target. A systematically positively or negatively biased estimator consistently over- or under-estimates, respectively, the true value of the estimation target. \emph{Precision} quantifies the statistical variance of the estimator (i.e., how close repeated estimates of the same quantity are to each other). An estimator with a low variance has a high precision, and vice versa. Unlike bias, the magnitude of precision is only dependent on the estimated values and is hence completely independent of the true value. Bias and precision are \emph{scaled} w.r.t. the true value of the estimation target. For instance, an estimator with a bias of zero (0) provides exactly the true value of the estimation target, an estimator with a bias of negative one (-1) provides zero (0) as estimate, and an estimator with a bias of one (1) provides exactly twice the true value.

\subsection{Estimator Evaluation}
%\subsubsection{Measures}
We compute the \emph{mean bias} of the estimator $\hat S$ of the total number of paths $S$ as the average bias over $N=10$ runs and the \emph{imprecision} of $\hat S$ as the standard deviation of the bias \cite{brose}:
\begin{align}
\text{\emph{mean bias}} &= \sum_{i=1}^{N}\frac{\hat S_i - S_i}{NS}\\[0.2cm]
\text{\emph{imprecision}} &= \sqrt{\frac{\mathlarger{\sum}_{i=1}^N\left(\dfrac{\hat S_i - S_i}{S_i} - \dfrac{\left[\sum_{j=1}^{N}\hat S_j - S_j\right]}{NS}\right)^2}{N-1}}
\end{align}
where $S_i$ is the estimated total number of paths for the $i$th run \emph{at 100 hours of fuzzing}. Even at 100 hours, the empirical value may still substantially underestimate the true species richness. A low imprecision means that all estimates are similarly (un)biased.

\begin{table}[ht]\small
\caption{Average number of discovered paths $S_\text{obs}$ and estimated path coverage $\hat G$ over ten runs at six, forty-eight, and one-hundred hours into the fuzzing campaign, respectively.\vspace{-0.1cm}}
\label{tab:resultOverview}
\begin{tabular}{@{}rrrr@{}}
& \multicolumn{1}{l}{\ $\mathbf{S_\text{obs}\text{\quad} (\hat G)}$} 
& \multicolumn{1}{l}{\ $\mathbf{S_\text{obs}\text{\quad} (\hat G)}$}
& \multicolumn{1}{l}{\ $\mathbf{S_\text{obs}\text{\quad} (\hat G)}$}\\
\textbf{Subject} &  \multicolumn{1}{l}{\ \textbf{@ 6hrs}} 
& \multicolumn{1}{l}{\ \textbf{@ 48hrs}}
& \multicolumn{1}{l}{\ \textbf{@ 100hrs}}\\\hline
\hl{json} & 	\colorbox{green!30}{2612 (98.7\%)} & \colorbox{green!30}{2657 (99.9\%)} & \colorbox{green!30}{2665 (99.9\%)}\\
libjpeg-turbo & \colorbox{yellow!20}{2224  (95.2\%)} & \colorbox{green!30}{2547  (99.4\%)} & \colorbox{green!30}{2623  (99.6\%)} \\
openssl       & \colorbox{red!30}{1041  (86.3\%)} & \colorbox{red!30}{1356  (93.3\%)} & \colorbox{red!30}{1444  (88.6\%)} \\
\hl{libxml2}  & \colorbox{red!30}{5071  (57.3\%)}& \colorbox{red!30}{6672 (67.3\%)} & \colorbox{red!30}{7656 (66.0\%)}\\
ffmpeg        & \colorbox{red!30}{2420 (71.8\%)} & \colorbox{green!30}{2554  (99.3\%)} & \colorbox{green!30}{2568  (99.6\%)} \\  
wireshark     & \colorbox{green!30}{ \ 427  (98.0\%)} &  \colorbox{green!30}{ \ 454  (99.1\%)} &  \colorbox{green!30}{ \ 456  (99.3\%)} \\ \hline
\colorbox{red!30}{\quad $<95$\% \quad} & \multicolumn{3}{l@{}}{The campaign is considered incomplete.}\\
\colorbox{yellow!20}{\quad $<98$\% \quad} & \multicolumn{3}{l@{}}{Decide based on other factors.}\\
\colorbox{green!30}{\quad$\ge 98$\% \quad} & \multicolumn{3}{l@{}}{The campaign is considered nearly complete.}
\end{tabular}
\vspace{-0.1cm}
\end{table}

\subsubsection*{RQ.1 Completeness $\hat G(n)$}
\emph{Path coverage provides a useful indicator of the progress of a fuzzing campaign towards completion. Some programs require more time than others to achieve the same path coverage.} 
\autoref{tab:resultOverview} shows the estimated path coverage $\hat G(n)$ at 6, 48, and 100 hours into the fuzzing campaign as an average over ten runs.  
Six hours into the fuzzing campaign, we see $\hat G=99\%$ for json, meaning \pythia has discovered almost all paths that it could potentially explore. In fact, after spending seven times more time (i.e., after 48 hours), \pythia has discovered only 45 more paths. For all practical purposes the average fuzzing campaign for both json and wireshark might be considered completed shortly after the six hour mark. 

Clearly, openssl, libxml2, and ffmpeg do not appear to be completed at the six hour mark with a path coverage well below 90\%. In fact, more than 300, 1600, and 100 paths are still being discovered, respectively, until the 48 hour mark. At the 48 hour mark, the average fuzzing campaign for ffmpeg might be considered completed. In fact only 12 more paths are found until the 100 hour mark. However, the average fuzzing campaign for libxml2, and openssl remains incomplete even at the 48 hour mark. Indeed, about $1k$, and 100 more paths are being discovered, respectively, until the 100 hour mark. For libxml2, about 3200 new paths are found until the \emph{800} hour mark, on average (33 days; 10846 avg. \#paths). We explain the decrease in coverage from the 48 to the 100 hour mark for libxml2 and openssl with the lack of a discernible horizontal asymptote (2nd row in tables \ref{tab:resultEst} and \ref{tab:resultEst2}; RQ.3). %The results of RQ3 provide more insight on the estimator bias.  
We also note that AFL's existing stopping rule\footnote{If the environment variable AFL\_EXIT\_WHEN\_DONE is set, AFL automatically aborts the current fuzzing campaign at the end of a cycle in which \emph{no} new path was discovered, starting from the 100th cycle.} would have (incorrectly) aborted openssl already at the 6 hour mark while ffmpeg would (incorrectly) continue even after the 48 hour mark.

\begin{table}
\caption{Average of the path coverage estimates $\hat G(n)=S(n)/\hat S$, the number of discovered paths $S(n)$, the bias and precision of $\hat S$ over time, and the discovery probability estimate $\hat U(n)$ within the first 96 hours (4 days). The colored curves in the second and last row represent one fuzzing campaign each.} 
\label{tab:resultEst}
\begin{tabular}{@{}c@{}}
\includegraphics[width=\columnwidth,trim={0.21cm 0 0 0},clip]{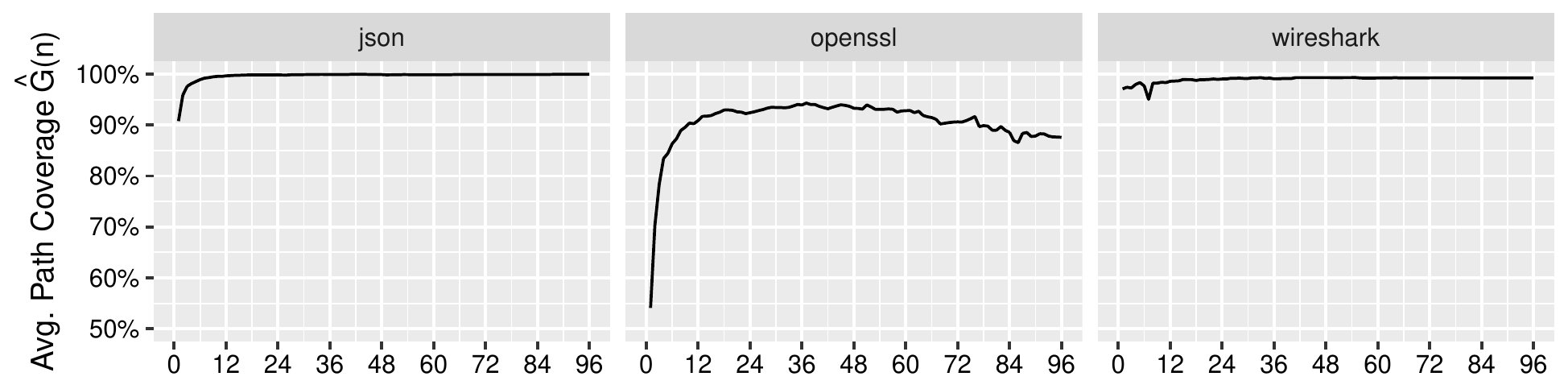}\\
\includegraphics[width=\columnwidth]{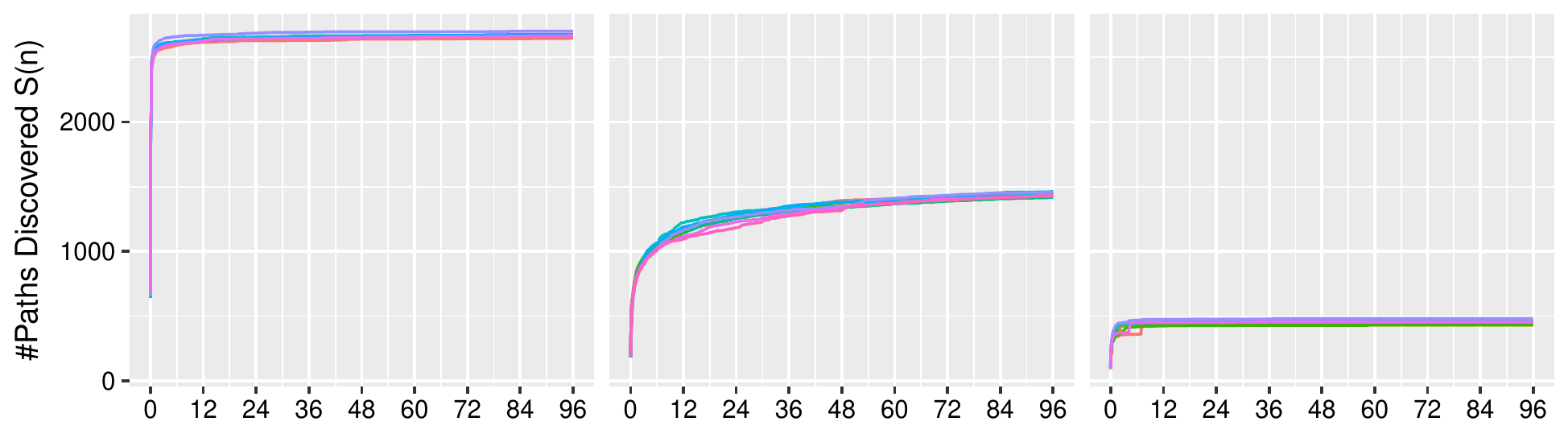}\\
\includegraphics[width=\columnwidth]{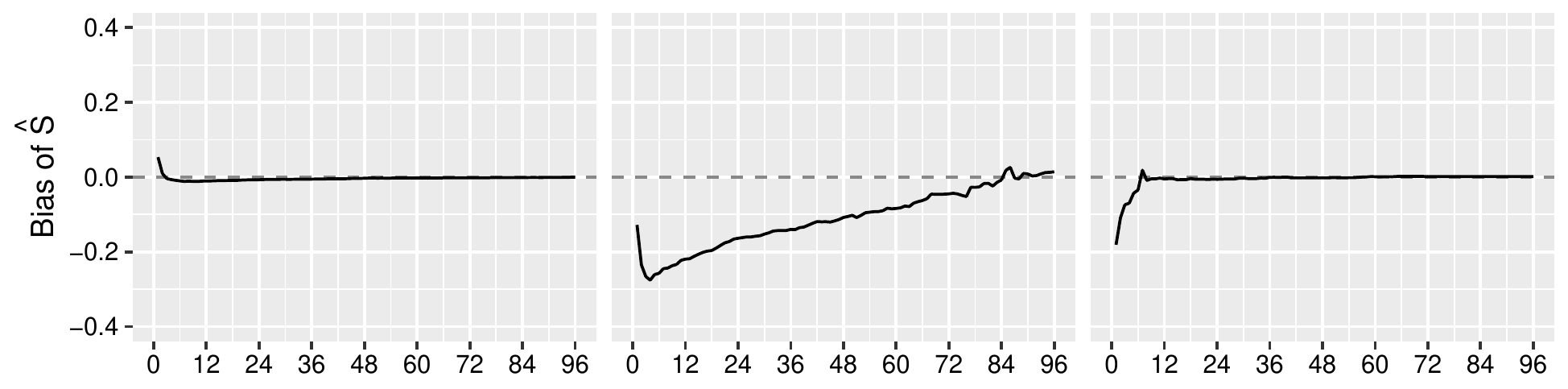}\\
\includegraphics[width=\columnwidth]{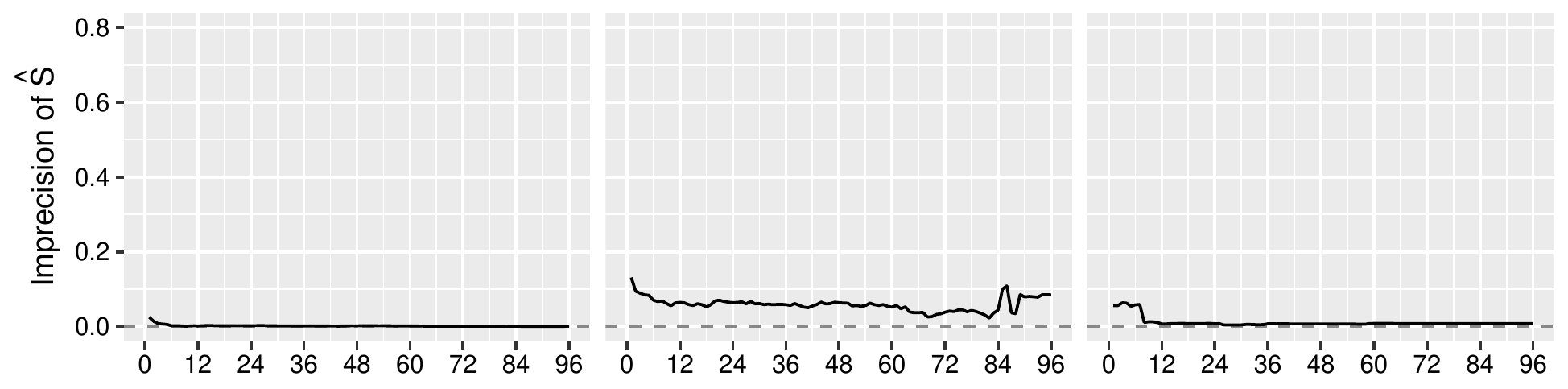}\\\hline\hline
\includegraphics[width=\columnwidth]{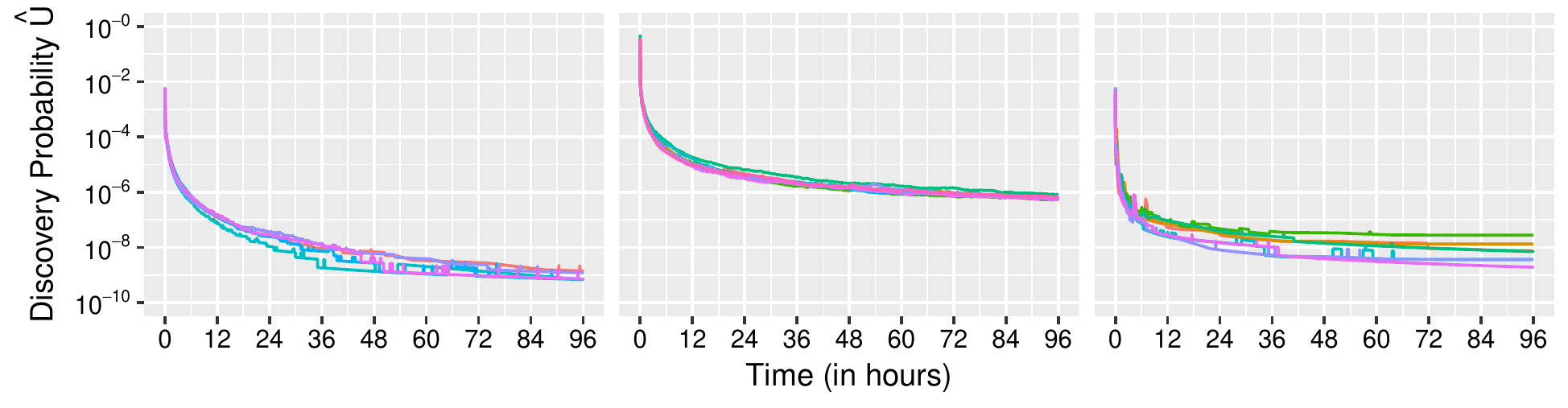}
\end{tabular}
\end{table} 

\begin{table}
\caption{Average of the path coverage estimates $\hat G(n)=S(n)/\hat S$, the number of discovered paths $S(n)$, the bias and precision of $\hat S$ over time, and the discovery probability estimate $\hat U(n)$ within the first 96 hours (4 days). The colored curves in the second and last row represent one fuzzing campaign each.} 
\label{tab:resultEst2}
\begin{tabular}{@{}c@{}}
\includegraphics[width=\columnwidth,trim={0.21cm 0 0 0},clip]{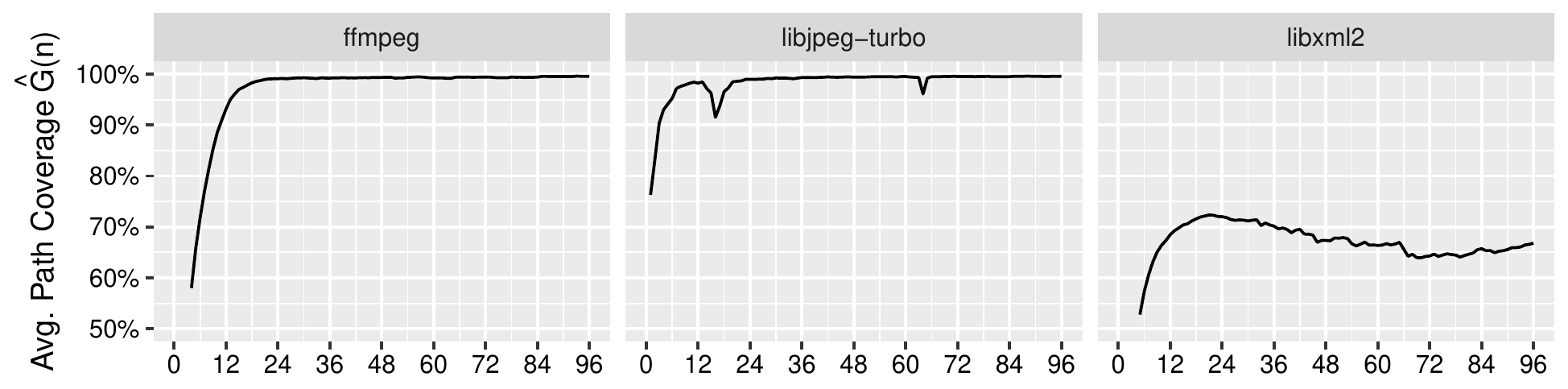}\\
\includegraphics[width=\columnwidth]{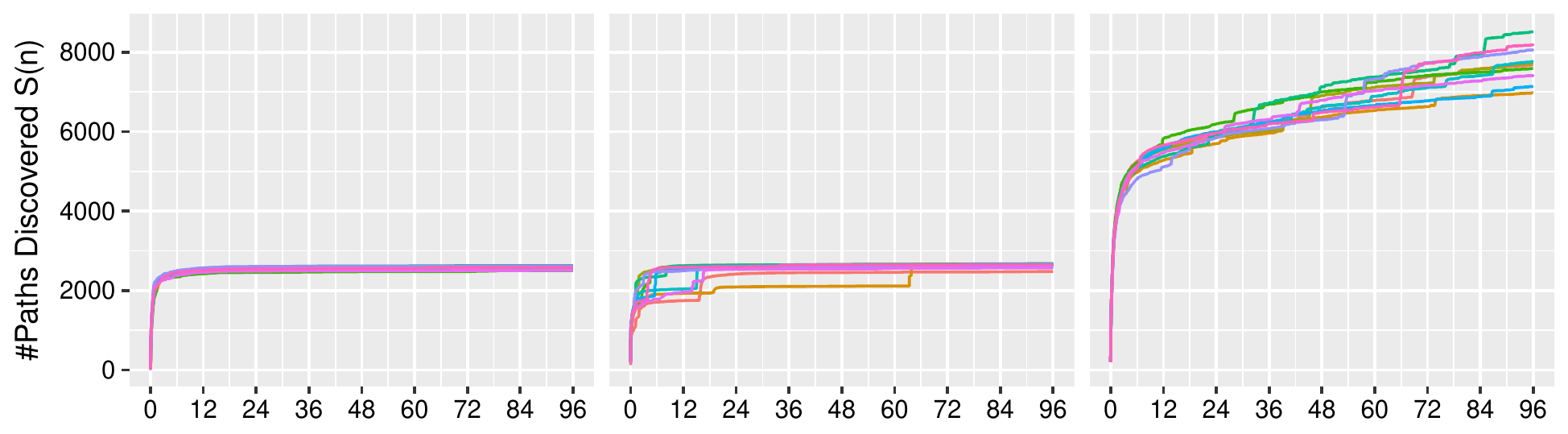}\\
\includegraphics[width=\columnwidth]{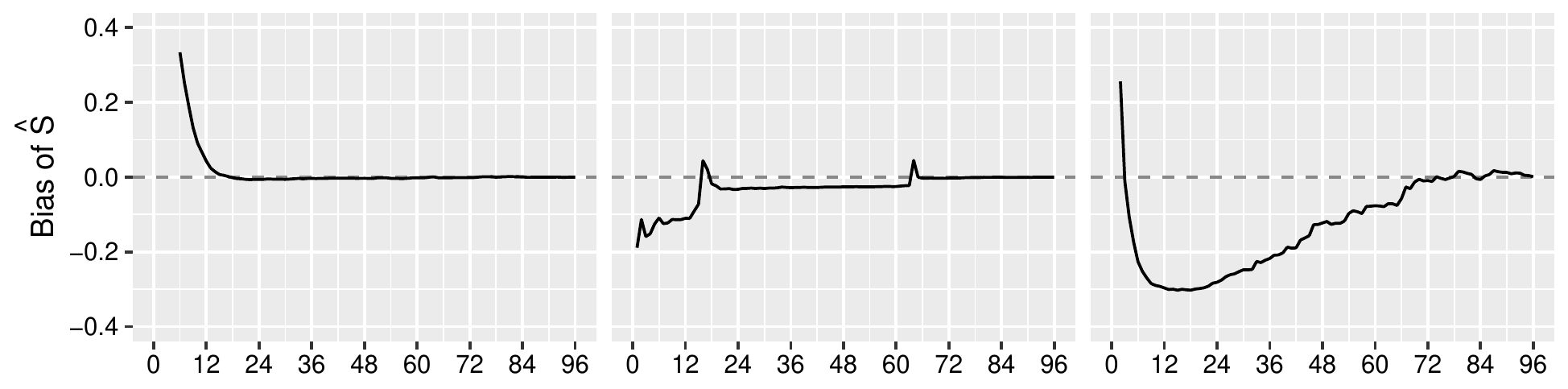}\\
\includegraphics[width=\columnwidth]{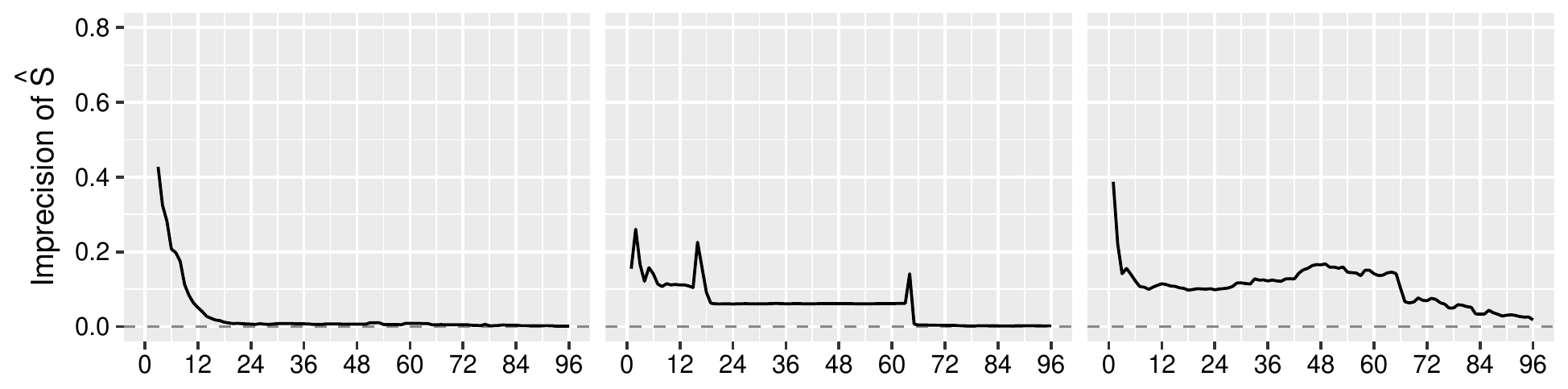}\\\hline\hline
\includegraphics[width=\columnwidth]{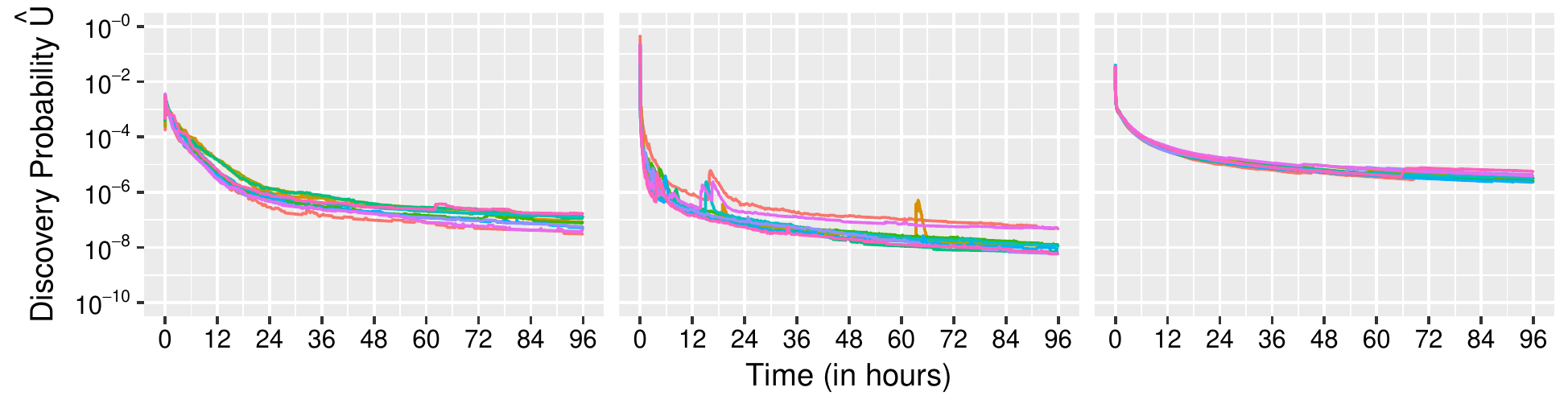}
\end{tabular}
\end{table} 

\subsubsection*{RQ.2 Residual Risk $\hat U(n)$} \emph{The current discovery probability provides a useful indicator of the current residual risk that a discoverable vulnerability exists but remains undiscovered in the ongoing fuzzing campaign. The discovery probability measured for one fuzzing campaign is fairly representative for other fuzzing campaigns of similar length, particularly later in the fuzzing campaign.} The bottom row in tables \ref{tab:resultEst} and \ref{tab:resultEst2} shows the discovery probability estimate $\hat U(n)$ over time. Notice the log-scale on the y-axis. The first observation that we can make is that fuzzing campaigns of the same length yield different degrees of residual risk for different subjects. For instance, at the 96 hour mark the discovery probability estimate across subjects ranges over four orders of magnitude (e.g., libxml2 vs. json). The second observation that we can make is that for the same subject, the variance of the estimate across fuzzing campaigns is rather small, indicating a certain degree of representativeness. The third observation is a general deceleration where each discovery probability seems to almost approach a horizontal asymptote: As the campaign continues it takes more and more test inputs to achieve the same decrease in discovery probability (and hence the same decrease of residual risk). \hl{Our fourth observation is that fuzzing campaigns with a relatively high discovery probability (libxml2, openssl) also achieve a relatively low path coverage estimate (1st row) with a relatively high estimator bias (3rd row). This provides opportunities to devise suitable adaptive bias correction strategies based on the discovery probability.}

\subsubsection*{RQ.3 Bias of $\hat S$} 
\emph{For four of six subjects, the bias reduces to within $\pm 10\%$ of the true species richness $S$ within the first 12 hours. For the other two (libxml2 and openssl), it takes 48 hours. In some cases the mean bias is systematic and negative, in other cases it is systematic and positive. The magnitude of the positive bias can be substantial in the first few hours (>5.0 for ffmpeg). Notice, a substantial over-estimation of species richness $S$ results in a (conservative) under-estimation of the path coverage $G(n)$. However, in all cases the mean bias tends to zero (0) as the number of generated test inputs increases over time.} 
%An over-estimation of $S$ results in an underestimation of the partition coverage $G(n)$, and vice versa. 
There are several sources of bias in estimating  $S$. 

\hl{\textbf{Campaign Ramp-up}. In the beginning of a fuzzing campaign, there is often substantial bias, both positive and negative. At \emph{one minute}, the total number of paths $S$ for json, openssl, and libjpeg-turbo is \emph{substantially over-estimated} with a positive mean bias of 6.2, 1.6, and 1.4, respectively. At \emph{one minute}, the total number of paths $S$ for ffmpeg, libxml2, and wireshark are \emph{under-estimated} with a considerable negative mean bias of -0.9, -0.7, and -0.2, respectively. 
Firstly, there is simply not sufficient data to extrapolate well }\cite{goodtheory}\hl{, i.e. there is no discernible asymptote that can be used for a good estimate of $S$. Secondly, when \textsc{Pythia} goes through the circular queue (i.e., the seed corpus) for the first time, the quality of the seeds varies. Meaning, the number of paths discovered by fuzzing one seed does generally not represent the number of paths discovered by fuzzing another seed. So, the estimate $\hat S$ is biased. Seed quality has less impact as more queue cycles are completed. Thirdly, \textsc{Pythia} is a coverage-based fuzzer and thus adaptively biased. Test inputs that increase coverage are added to the seed corpus. So, the fuzzer's capability to discover paths effectively improves over time }\cite{aflfast}\hl{. Hence, early estimates of $S$, particularly until the 12 hour mark, are often substantially biased. Note that an over-estimate of $S$ yields an under-estimate of path coverage $G$.}

\textbf{Adaptive bias}. \pythia (AFL) is a coverage-based fuzzer \hl{which means that the relative species abundance changes during the fuzzing campaign (see }\autoref{sec:bias}\hl{). New seeds may be added to the corpus that enable the discovery of path that might otherwise be difficult to discover }\cite{aflfast}. 
Sudden increases in the number of discovered paths can cause the current asymptote $\hat S$ to systematically \emph{under-estimate} $S$. This happens for instance when an interesting test input was generated that contains the correct value for a ``magic number'' \cite{thinair}. As we can see for libjpeg-turbo and libxml2 in the second row of \autoref{tab:resultEst2}, the magnitude of the increase can be quite substantial. The estimator $\hat S$ is negatively biased because of a ``false'' asymptote for the first 18 hours (libjpeg-turbo, 3rd row). This results in sudden drops in path coverage $\hat G$ when many new paths are discovered in short intervals (libjpeg-turbo, 1st row).

\textbf{Lower bound}. The \emph{Chao1}-estimator $\hat S$ of species richness is designed (and proved) to provide a practical lower bound rather than an imprecise point estimate \cite{chao1}. In fact, Chao1 \emph{is} an unbiased point estimator \emph{if} rare species (undetected and singleton species) have approximately equal abundance \cite{goodtheory}. If very rare species are unevenly distributed, the available data simply does not contain sufficient information.
So the estimator $\hat S$ might be negatively biased (i.e., systematically under-estimate the true number of species $S$). We can see this negative bias clearly for libxml2 and openssl (3rd row in tables \ref{tab:resultEst} and \ref{tab:resultEst2}). For openssl, this results in a path coverage that remains apparently constant between 85\% and 95\% (1st row) despite more paths being discovered (2nd row). The estimate for openssl is negatively biased because there is no discernible horizontal asymptote that could function as a less biased estimate $\hat S$ (2nd row). \hl{For libxml2, there is no discernible asymptote, either. We continued the libxml2 experiments past the 96-hour for a total of 800 hours (i.e., 33 days) to see whether the low path coverage value is warranted. Indeed, the number of paths discovered increased from an average 7.6k to an average 10.8k paths. It is interesting to note that the discovery probability (bottom row) for libxml2 and openssl is up to \emph{four order of magnitutes higher} than for the other four subjects. Hence, we attribute the negative bias for these two subjects to the large number of rare species that are unequally distributed.} We expect the negative mean bias in $\hat S$ to be smaller if the improved estimator \emph{iChao1} \cite{chui} is used. 

However, independent of the source, the bias always tends to zero (0) as the number of generated test inputs $n$ increases. In the case of libjpeg-turbo, the mean bias seems to remain negative from 18 hours onwards (row 3); the mean bias actually goes to about zero after about 65 hours due to a sudden increase from 2k to 2.5k discovered paths for one fuzzing campaign (yellow line for libjpeg-turbo in 2nd row). The tendency of the mean bias towards zero as the number of generated test inputs $n$ increases is empirical evidence of the \emph{statistical consistency} of the estimator $\hat S$ despite the adaptive sampling bias of \pythia and despite \emph{Chao1} being a biased estimator. We expect the positive and negative bias in $\hat S$ to be smaller if a coverage-based estimator, such as \emph{ACE} \cite{ace} is used.

\subsubsection*{RQ.4 Precision of $\hat S$}
\emph{For all subjects, the imprecision reduces to at most $10\%$ of $S$ within the first 12 hours and to at most $1\%$ of $S$ within the first 100 hours. The imprecision is high particularly in the beginning while the number of discovered paths increases significantly within a relatively small time interval. However, in all cases the imprecision tends to zero (0) as the number of generated test inputs increases over time.}
The precision of an estimator quantifies the variance of the provided estimates. A high precision means that the estimates are very similar across different fuzzing campaigns. The fourth row in tables \ref{tab:resultEst} and \ref{tab:resultEst2} shows the \emph{im}precision of the estimators for our subjects. When the estimator's imprecision is high, its \emph{precision} is low, and vice versa. 

As we can see for all subjects the imprecision is high when the slope of the number of paths discovered over time is steep (2nd row). This is the case in the first few hours of the fuzzing campaign when most paths are discovered (e.g., ffmpeg), and later when there are sudden increases (e.g., libjpeg-turbo). 
The single outlier run for libjpeg-turbo (2nd row, yellow line) illustrates an important challenge when computing species coverage for coverage-based greybox fuzzers, such as \pythia. The \emph{Chao1}-estimator essentially estimates the y-intercept of the horizontal asymptote of the curve describing the number of paths discovered over time (2nd row). The number of paths discovered might seem to approach a clear (but ``false'') asymptote when actually there is a sudden increase several hours later. As we can see, path coverage is bias-corrected once the sudden increase has happened (1st to 4th row in libjpeg-turbo between 12 and 18 hours). Until then, the security researcher might \emph{incorrectly} presume that the fuzzing campaign has exceeded a path coverage threshold that is required to mark the campaign as completed. However, we can also see in the second row that some programs are more prone to such sudden increases than others, and those are mostly constrained within the first few hours. 

In all cases the imprecision tends towards zero (0) as the number of generated test inputs $n$ increases. 
After 100 hours the imprecision is generally less than 0.01, indicating small variance and high precision.
Again, in the case of libjpeg-turbo, the imprecision remaining just below 0.1 from 18 hours onwards (row 3) can be explained with the outlier campaign that converges only after 65 hours. The tendency of the imprecision towards zero as the number of generated test inputs $n$ increases is empirical evidence of the \emph{representativeness} of the estimation computed for one fuzzing campaign for other fuzzing campaigns of the same length.

\subsection{Extrapolator Evaluation}
Let $t$ be the time that the fuzzer has spent generating $n$ test inputs within the fuzzing campaign. Since the time to generate a test input is fairly constant across all our fuzzing campaigns, we estimate the number $m^*$ of test inputs generated in the interval from $t$ to $t+t^*$ where $t^*\in \{30min, 1hr, 2hrs, 4hrs\}$ as $m^*=nt^*/t$. At time $t$, we compute the \emph{mean bias} for the estimate $\hat S(n+m^*)$ of the number of discovered paths $S(n+m^*)$ if $m^*$ \emph{more} test inputs were generated as the average bias over $N=10$ runs. At time $t$, we compute the \emph{imprecision} of $\hat S(n+m^*)$ as the standard deviation of the bias:
\begin{align}
\text{\emph{mean bias}} &= \sum_{i=1}^{N}\frac{\hat S_i(n+m^*) - S_i(n+m^*)}{NS_i(n+m^*)}\\
\text{\emph{imprecision}} &= \sqrt{\frac{\mathlarger{\sum}_{i=1}^N\left(\dfrac{\hat S_i(n+m^*) - S_i(n+m^*)}{S_i(n+m^*)} - \mathlarger{\sum}_{j=1}^{N}\dfrac{\hat S_j(n+m^*) - S_j(n+m^*)}{NS_j(n+m^*)}\right)^2}{N-1}}
\end{align}
where $S(n+m^*)$ is \emph{empirically} determined at time $t+t^*$ and thus provides the true value of the estimation target.

\begin{table}
\caption{Bias and imprecision for 6 subjects at 4 extrapolation intervals for  $\le 48$hrs (top) and $\le 12$hrs (bottom).}
\label{tab:resExtra}
\includegraphics[trim={0 2cm 0 0}, clip, width=\columnwidth]{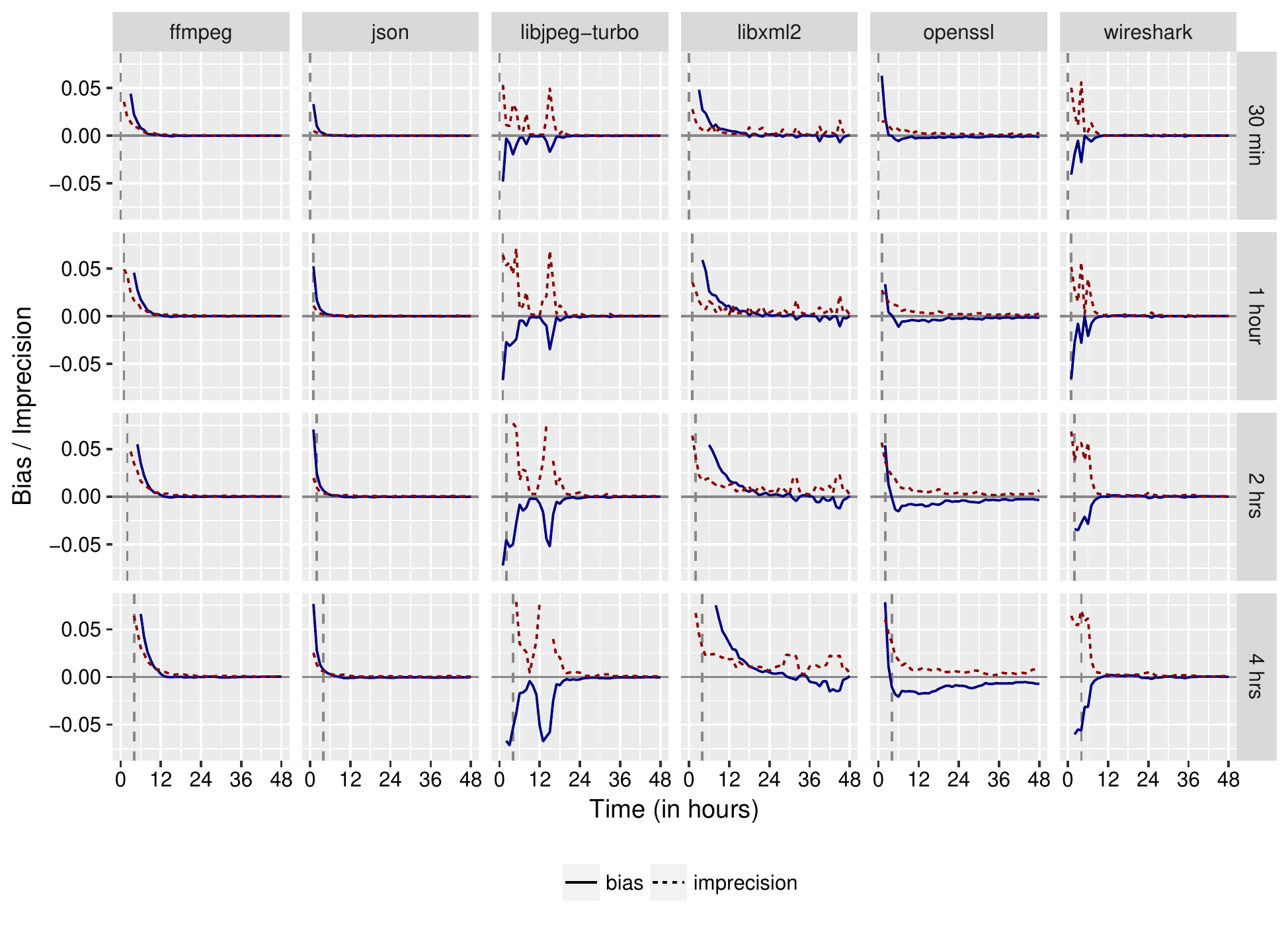}
\includegraphics[trim={0 0 0 0}, clip, width=\columnwidth]{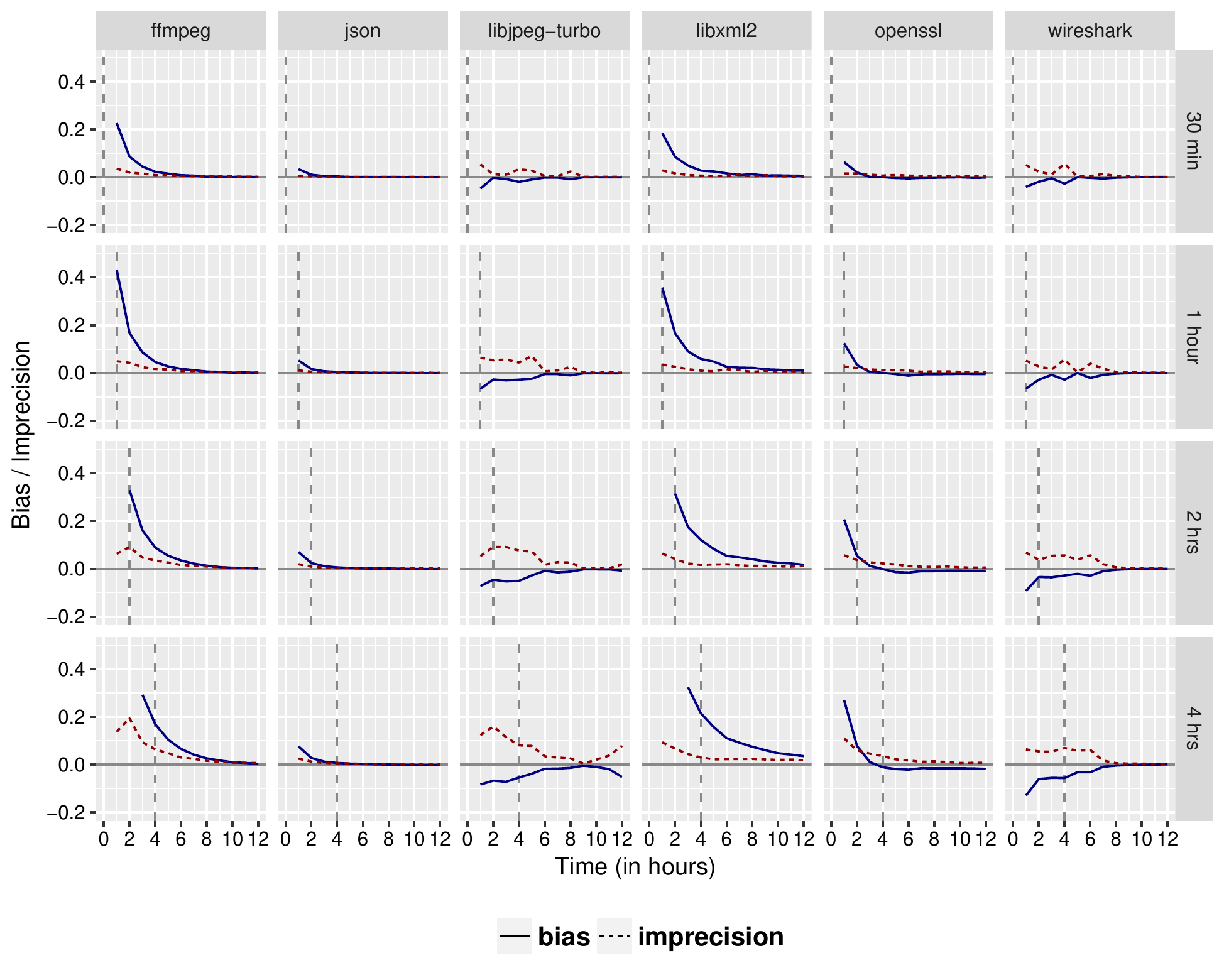}\vspace{-0.4cm}
\end{table}

\subsubsection*{RQ.5 Bias $\hat S(n+m^*)$}
\emph{The number of paths discovered and thus path coverage can be effectively extrapolated with low bias. The magnitude of the bias increases with the extrapolation interval and decreases as more test inputs $n$ are generated.} 
\autoref{tab:resExtra} shows the mean bias within the first 48 hours (top) and the first 12 hours (bottom) of the fuzzing campaign. We chose these two intervals because of the difference in the magnitude of the bias in the first few hours. In fact, the bottom four rows feature a large range of the bias between -20\% and 50\% of the true, empirical $S(n+m^*)$ while the top four rows feature a much smaller range between -8\% and 8\% of $S(n+m^*)$. 

As we can see in \autoref{tab:resExtra}, $\hat S(n+m^*)$ might substantially over-estimate the number of paths discovered in the first few hours. We explain this strong positive bias of $\hat S(n+m^*)$ with the strong positive bias of the estimate $\hat S$ of the total number of paths which forms an important component in the extrapolation methodology proposed by Shen et al. \cite{shen} (see \autoref{tab:resultEst}-3rd row).
After the initial over-estimation, $\hat S(n+m^*)$ is generally slightly (but systematically) under-estimated.
Effectively, the estimator begins to provide a \emph{conservative estimate} of the increase in path coverage. We explain this small negative bias with \pythia being a coverage-based greybox fuzzer. Indeed, we expect a blackbox fuzzer (without adaptive sampling bias) to detect less paths per unit time.
Another source of negative bias is a sudden increase in the number of paths discovered that would be difficult to anticipate (e.g., for libjpeg-turbo compare \autoref{tab:resExtra}-top and \autoref{tab:resultEst2}-2nd row).   
We also notice that the magnitude of the bias increases with the extrapolation interval. The reason is fairly obvious: The quality of the extrapolation will be worse the further we want to look into the future. However, the rule of thumb in ecology \cite{sampleSurvey} to limit the extrapolation to within twice the current sampling effort does \emph{not} find very strong empirical support for our six subjects in the domain of automated software testing and analysis.

For all six subjects and all four extrapolation intervals, the bias remains within $\pm 2\%$ of the empirical value $S(n+m^*)$ from 18 hours onwards. The tendency of the bias towards zero (0) as the number $n$ of generated test inputs increases might be explained with an expected deceleration of the number of paths $S(n)$ discovered over time approaching an asymptotic total number of paths $S$ (see 2nd row in tables \autoref{tab:resultEst} and \autoref{tab:resultEst2}).

\subsubsection*{RQ.6 Precision $\hat S(n+m^*)$}
\emph{The number of paths discovered and thus path coverage can be effectively extrapolated with high precision. The magnitude of the \emph{im}precision increases with the extrapolation interval and decreases as more test inputs $n$ are generated.}
The imprecision does not seem to be affected as substantially as the bias by the initial surge of path discoveries (see ffmpeg and openssl, bottom rows). However, the magnitude of the imprecision generally mirrors that of the bias, suggesting that bias and imprecision have the same sources. Like the bias, the imprecision tends towards zero (0) as the number $n$ of generated test inputs increases.  

\subsection{Result Summary}\label{sec:resultsummary}

\hl{The objectives of this empirical evaluation were (1)~to test my  main hypothesis within the STADS framework, (2)~to evaluate several methodologies from ecology within the STADS framework, and (3)~to investigate the impact of the adaptive sampling bias.}

\subsubsection{Main Hypothesis}
\hl{I hypothesize that within the STADS framework rare, discovered species contain almost all information about the detectable species that remain undiscovered. Rare species are those to which only a small number of generated test inputs belong. In our experiments, all methodologies used to extrapolate from the discovered to undiscovered species (i.e., from tested to untested program behaviors) are based on the number of singletons or doubleton species. The experimental results show good estimator performance for these methodologies and thus \emph{support my main hypothesis}.}

\subsubsection{Estimator Evaluation}
We find that \emph{discovery probability} provides a useful indicator of the residual risk that a discoverable vulnerability exists but remains undiscovered in the ongoing fuzzing campaign. The discovery probability estimate measured for one fuzzing campaign is fairly representative for other fuzzing campaigns of similar length.\footnote{More specifically, discovery probability is fairly representative for other fuzzing campaigns where the same program is fuzzed for the same time using the same fuzzer and seed corpus (if any).} While the estimates for 10 runs range over three orders of magnitude across subjects, they all range within the same order of magnitude for the same subject. The similarity between estimates of different runs for the same subject (i.e., the precision) increases as the number $n$ of generated inputs increases.

We find that \emph{path coverage} provides a useful indicator of the progress of a fuzzing campaign towards completion that can be used to decide effectively whether to abort or continue a fuzzing campaign. The path coverage estimate can be positively and negatively \emph{biased}. \hl{The bias is most substantial during campaign ramp-up, within the first 12 hours, when many paths are discovered. Another source of substantial bias is the existence of many rare species (see discussion in }\autoref{sec:megadiverse}\hl{), and the sudden discovery of many paths at once (see discussion in }\autoref{sec:bias}). However, for all subjects the magnitude of the bias reduces 
as the number $n$ of generated test inputs increases. Similarly, the \emph{precision} is low in the first few hours while the number of discovered paths increases significantly within a relatively small time interval. However, in all cases precision increases as $n$ increases. 

We find that path coverage can be \emph{extrapolated} with low bias and high precision. Like the path coverage estimate, the extrapolation can be positively and negatively \emph{biased}. Sudden surges in the number of discovered paths are not anticipated resulting in negative bias and some \emph{im}precision. A substantial over-estimation of the total number of paths in the first few hours might result in a substantial positive bias and some imprecision. The magnitude of bias and imprecision increases with the extrapolation interval and decreases as more test inputs $n$ are generated. We do not find very strong evidence that the rule of thumb in ecology \cite{sampleSurvey} to keep the extrapolation within twice the sampling effort applies to automated software testing and analysis. Other sources of bias and imprecision seem to have a stronger impact. 

\subsubsection{Impact of Adaptive Sampling Bias}
\hl{An underlying assumption of most methodologies in the STADS framework is that the probability $p_i$ to generate a test input that belongs to species $\mathcal{D}_i$ does \emph{not} change substantially during the fuzzing campaign. However, it does for feedback-directed fuzzers, such as \textsc{Pythia} (which is based on AFL). A \emph{feedback-directed fuzzer} continuously adapts the strategy to generate new test input based on feedback for previous test inputs. For instance, \textsc{Pythia} augments the existing seed corpus with generated test inputs that increased branch coverage. New test inputs are generated by random mutations of the seed inputs that are continuously selected from a circular queue that represents the (extended) seed corpus. As the seed corpus grows, the relative species abundance $\{p_i\}_{i=1}^S$ changes as well. This is called an \emph{adaptive sampling bias}, because the sampling strategy changes adaptively during the sampling itself.}

\hl{We find that \emph{in the beginning} of a fuzzing campaign the adaptive sampling bias has a large impact on estimator performance. In the first few hours, the total number of species $\hat S$ is often substantially under-estimated which is explained by the improving capability of \textsc{Pythia} to discover new species as new seeds are added to the seed corpus. In the beginning, a large number of new seeds are added as new species are discovered.}

\hl{We find that \emph{as more test inputs} are generated the impact reduces. For \textsc{Pythia}, the species discovery curve is strictly correlated with the adaptive sampling bias. For every new species that is discovered, a seed is added to the seed corpus. Hence, as species discovery decelerates over time, the adaptive bias reduces just as well. For some subjects, the number of discovered species seems to approach a \emph{false} asymptote, leading to a negatively biased estimate of species richness $\hat S$ (and thus a positively biased estimate of species coverage $\hat G$) when suddenly many more species are discovered, e.g., because a magic number discovered }\cite{thinair}\hl{. This is also explained by the adaptive sampling bias, and its impact reduces over time. A more general discussion on the impact of adaptive sampling bias follows in }\autoref{sec:bias}.

%% file: sections/overlapping.tex
\section{Bernoulli Product Model: One Input, Multiple Species}\label{sec:overlapping}
\begin{wrapfigure}{r}{0.44\textwidth}\small\vspace{-0.4cm}
\includegraphics[width=0.44\textwidth]{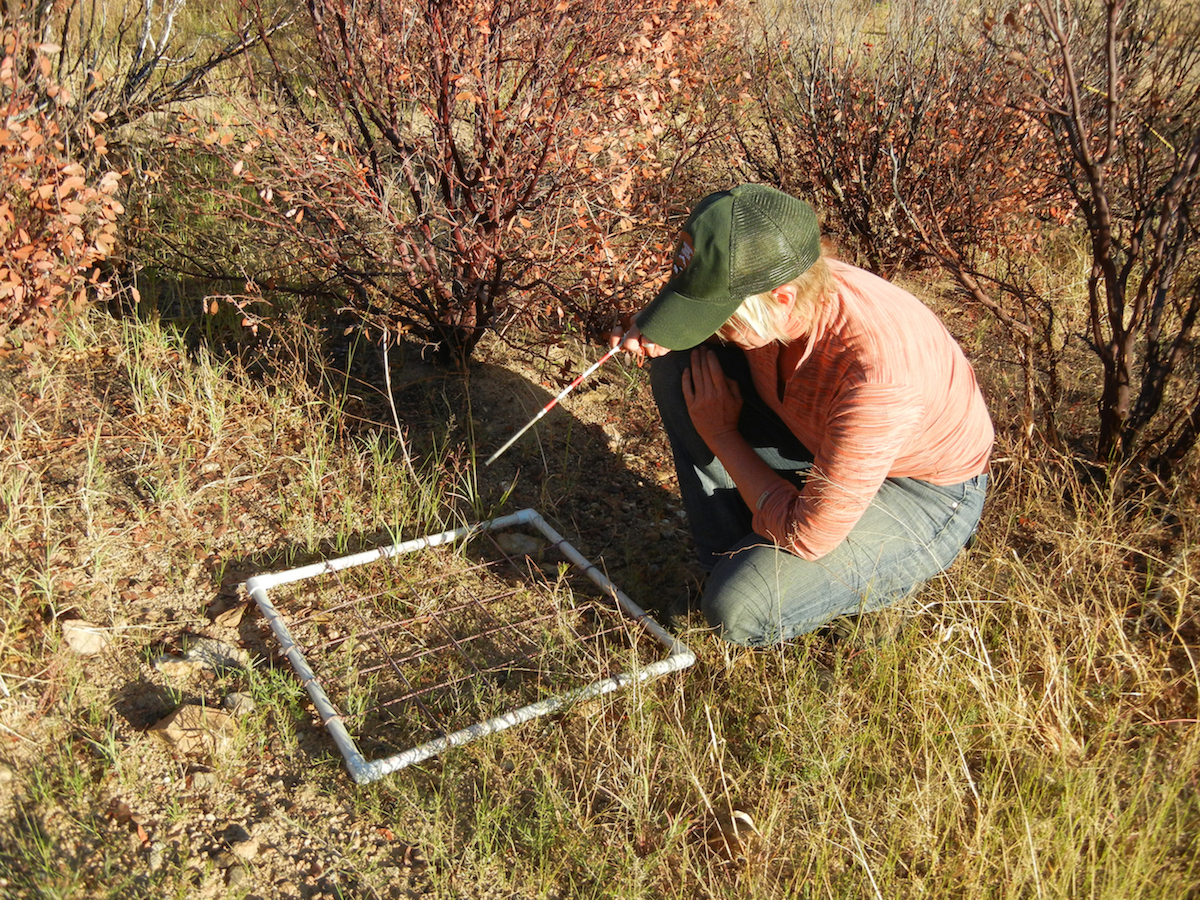}
\caption{\hl{Quadrats positioned at random locations in the assemblage. Present species are recorded for each quadrat. Image credit: NPS Sonoran Desert Network (Licence: }\href{https://creativecommons.org/licenses/by/2.0/}{CC-BY-2.0}\hl{).}}
\end{wrapfigure}

So far, we have discussed the multinomial model\footnote{The multinomial model is introduced in \autoref{sec:nonoverlapping}. Specific estimators and extrapolators for the multinomial model are discussed in sections \ref{sec:estimate} and \ref{sec:extrapolate}, respectively. Several of those are evaluated in \autoref{sec:eval}.} in the STADS
framework, where \emph{each input belongs to exactly one species}, e.g., an input can exercise only exactly one path. However, there are many other concrete testing objectives where \emph{each input belongs to one or more species}. 
For instance, a single input can exercise multiple coverage-goals, such as program statements, branches, or methods; a single input can kill multiple mutants \cite{mutation}, witness multiple information flows \cite{ase17}, violate multiple assertions, expose multiple bugs, and traverse multiple program states. In ecology, it is a  \emph{sampling unit} that can contain multiple species. A sampling unit is usually a physical trap, net, quadrat, or plot. These sampling units are distributed in the assemblage and studied exhaustively---in lieu of the assemblage itself.  
When only the presence (or absence) of species can be determined in a sampling unit, ecologist call the data as \emph{incidence data} and utilize the \emph{Bernoulli product model} \cite{incidenceSurvey}. In the Bernoulli product model, within STADS a generated test input is considered as \emph{sampling unit}.

\hl{In the following, we extend our STADS statistical framework to account for testing objectives that yield multiple species identified for a single input.}  
Let $n$ be the number of inputs that have been generated throughout the current fuzzing campaign, $S(n)$ be the number of species that have been discovered, and $S$ be the total number of species.
Define $\{W_{ij}\ |\ i = 1, 2, \ldots, S \wedge j = 1, 2, \ldots, n\}$ as the \emph{incidence matrix} where $W_{ij}=1$ if the $j$th generated test input belongs to species $\mathcal{D}_i$, and $W_{ij}=0$ otherwise. Let $Y_i$ be the number of generated test inputs that belong to species $\mathcal{D}_i$ for $i:1\le i \le S$, then $Y_i=\sum_{j=1}^n W_{ij}$. For species that exist but remain undiscovered in the current fuzzing campaign, we have $Y=0$. Define the \emph{incidence frequency count} $Q_k$ for $k:0\le k \le n$, as the number of species to which exactly $k$ test inputs belong that have been generated throughout the current fuzzing campaign. More formally, $Q_k=\sum_{i=1}^S I(Y_i=k)$. Hence, $n\le\sum_{k=1}^n kQ_k = \sum_{i=1}^S Y_i$ and $S(n)=\sum_{k=1}^n Q_k$. The incidence frequency count $Q_k$ is analogous to the abundance frequency count $f_k$ for the multinomial model. The unobservable zero frequency count $Q_0$ denotes the number of species that remain undiscovered in the current fuzzing campaign. We call $Q_1$ the number of \emph{singleton species} and $Q_2$ the number of \emph{doubleton species}.

The probability that the fuzzer generates a test input that belongs to species $\mathcal{D}_i$ is $p_i$ for $i:1\le i \le S$. Note that $\sum_{i=1}^S p_i \ge 1$. We assume that each $W_{ij}$ is a \emph{Bernoulli random variable} with probability $p_i$ that $W_{ij}=1$ (and analogously with probability $1-p_i$ that $W_{ij}=0$). Thus, the probability distribution for the incidence matrix is
\begin{align}
P(W_{ij}=w_{ij}; i=1,2,\ldots,S; j=1,2,\ldots,n) &=
\prod_{j=1}^n \prod_{i=1}^S p_i^{w_{ij}}(1-p_i)^{1-w_{ij}}\\
&=\prod_{i=1}^S p_i^{y_i}(1-p_i)^{n-y_i} & \text{where } y_i=\sum_{j=1}^n w_{ij}\label{eq:rowsums}
\end{align}

From \autoref{eq:rowsums}, we can see that the row sums $(Y_1,Y_2,\ldots,Y_S)$ are \emph{sufficient statistics}. This renders the incidence frequency counts $Q_k$ suitable components for the estimators of species richness. 
The number of generated test inputs $Y_i$ that belong to species $\mathcal{D}_i$ follows a binomial distribution
\begin{align}
P(Y_i=y_i)&=\binom{n}{y_i}p_i^{y_i}(1-p_i)^{n-y_i} \quad\quad\text{where } i=1,2,\ldots,S
\end{align}

Chao and Colwell \cite{incidenceSurvey} provide more details about the Bernoulli product model and its utility in the ecologic context.

\subsection{Estimation in the Bernoulli Product Model}

\textbf{Estimating $S$}.
In the Bernoulli product model of the STADS framework, the estimation of species richness $S$ (i.e., the asymptotic total number of species) can be done using the \emph{Chao2} and \emph{iChao2}-estimators, which were derived by Chao \cite{chao2} and Chui et al. \cite{chui}.
\begin{align}
\hat S_{\text{Chao2}} 
&\approx \begin{cases}
S(n) + Q_1^2 / (2Q_2) & \text{if } Q_2 > 0\\
S(n) + Q_1(Q_1-1)/2 & \text{if } Q_2 = 0
\end{cases}\\
\hat S_{\text{iChao2}} 
&\approx \hat S_{\text{Chao2}} + \frac{Q_3}{4Q_4} \times \max\left(Q_1 - \frac{Q_2Q_3}{2Q_4},0\right)\label{eq:two2}
\end{align}
Chao \cite{chao2} showed that the Chao2 estimator $\hat S_{\text{Chao2}}$ provides a nonparametric lower bound on the total number of species $S$ in an assemblage. Very recently, Chao and colleagues showed that the Chao2 estimator is an \emph{unbiased point estimator} as long as very rare species (specifically, undetected and singleton species) have approximately equal detection probability \cite{goodtheory}.

Alternative estimators of species richness $S$ in the Bernoulli product model include Jacknife estimators \cite{palmer,walther,burnham} and coverage-based estimators, such as ICE \cite{ice} and ICE-1 \cite{ice1}, that a particularly suitable when species diversity is high (i.e., when species evenness $J$ is low).

\textbf{Estimating $U$}.
For inputs that can belong to one or more species, the estimate $\hat U(n)$ of the discovery probability requires information not only about singleton and doubleton species but also about \emph{all discovered} species \cite{incidenceSurvey}. In the Bernoulli product model of the STADS framework, the discovery probability is estimated as
\begin{align}
\hat U(n) &= \frac{Q_1}{V} \left[\frac{n \hat Q_0}{n\hat Q_0+Q_1}\right]\\
&\approx \frac{Q_1}{V}% \quad\quad \text{ as $n$ gets larger and $n\gg Q_1$}
\end{align}
where in the current fuzzing campaign $V=\sum_{k=1}^n kQ_k = \sum_{i=1}^S \sum_{j=1}^n W_{ij}$ denotes the sum of all entries in the incidence matrix $W_{ij}$, $n$ is the number of test inputs generated, the number of undetected species can be estimated as $\hat Q_0=\hat S - S(n)$, and $Q_1$ and $Q_2$ are the incidence frequency counts of singletons and doubletons, respectively.
Note that the sum of all entries $V$ in the incidence matrix does \emph{not} require storing the complete incidence matrix $W_{ij}$. Instead, $V$ can be aggregated during the fuzzing campaign. Also notice the similarity of the approximation to the Good-Turing estimator for the multinomial model \cite{good}.

\subsection{Extrapolation in the Bernoulli Product Model}
\textbf{Extrapolating $S(n)$}. In the Bernoulli product model of the STADS framework, to estimate the expected number of discovered species $\hat S(n+m^*)$ when $n$ test inputs have already been generated and if $m^*$ more test inputs were to be generated, we have
\begin{align}
\hat S(n + m^*) &= S(n)+\hat Q_0\left[1-\left(1-\frac{Q_1}{n\hat Q_0 + Q_1}\right)^{m^*}\right]%\\
\end{align}
where for the current fuzzing campaign $n$ is the number of generated test inputs, $S(n)$ is the number of discovered species, $\hat Q_0=\hat S - S(n)$ is the expected number of undiscovered species, and $Q_1$ is the number of singletons.

\textbf{Extrapolating $U(n)$}. In the Bernoulli product model of the STADS framework, the estimate of the expected discovery probability $\hat U(n+m^*)$ if $m^*$ more test inputs were generated is computed as
\begin{align}
\hat U(n+m^*) = \dfrac{Q_1}{V}\left[\dfrac{n\hat Q_0}{n\hat Q_0+Q_1}\right]^{m^*+1}
\end{align}
where for the current fuzzing campaign $V=\sum_{k=1}^n kQ_k = \sum_{i=1}^S \sum_{j=1}^n W_{ij}$ denotes the sum of all entries in the incidence matrix $W_{ij}$, $n$ is the number of generated test inputs, $S(n)$ is the number of discovered species, $\hat Q_0=\hat S - S(n)$ is the expected number of undiscovered species, and $Q_1$ is the number of singletons. 

\textbf{Estimating $m_{G^*}$ when $G^*=S(n+m_{G^*})/\hat S$ is given}. To reach a fraction ${G^*}$ of estimated total number of species $\hat S$ where $\hat G(n) < {G^*} \le 1$, in the Bernoulli product model of the STADS framework the required number $m_{G^*}$ of further test inputs is estimated as \cite{samplesRequired}
\begin{align}
m_{G^*} \approx %\left.\log\left[1-\frac{T}{(T-1)}\frac{2Q_2}{Q_1^2}(g\hat S - S_{\text{obs}})\right] \right/ \log\left[1-\frac{2Q_2}{(T-1)Q_1+2Q_2}\right]
\dfrac{\log\left[1-\cfrac{n}{(n-1)}\cfrac{2Q_2}{Q_1^2}({G^*}\hat S - S(n))\right]}{\log\left[1-\cfrac{2Q_2}{(n-1)Q_1+2Q_2}\right]}\label{eq:nine}
\end{align}
where in the current fuzzing campaign $n$ is the number of generated test inputs, $S(n)$ is the number of discovered species, and $Q_1$ and $Q_2$ are the incidence frequency counts for singleton and doubleton species, respectively. The \emph{expected fuzzing time} can be computed by multiplying the expected number of test inputs with the average time the fuzzer takes to generate  a test input.

%% file: sections/related.tex
\section{Related Work}\label{sec:related}
\hl{To the best of our knowledge, in the domain of software testing and analysis, there exists \emph{no previous work on estimating the asymptotic total number of species, or on extrapolating the number of species discovered over time} for any definition of species. In fact, Whalen }\cite{whalen}\hl{ says about the future of verification and validation that one of the biggest problems today is that there is \emph{no sound basis to extrapolate from tested to untested cases}. We strongly believe that the STADS framework provides a statistically well-grounded basis to extrapolate from tested to untested program behaviors.}

\subsection{Residual Risk Assessment}
Finding no vulnerabilities in a (long-running) fuzzing campaign does not mean that none exists. In the STADS framework, the \emph{discovery probability} $U(n)$ measures the probability to discover a new species with the $n+1$th generated test input where $n$ is the number of test inputs that have been generated throughout the fuzzing campaign. If the dynamic analysis is able to identify vulnerabilities, then an accurate estimate of the discovery probability provides a \emph{statistical guarantee} that no detectable vulnerability exists if none has been discovered. In other words, security researchers can use the STADS statistical framework for \emph{residual risk assessment}.

There exist several \emph{systematic approaches} to quantify the reliability of a program. However, Filieri et al. \cite{jpfRelia} recently noted that most existing approaches work on the design- and architectural level rather than on the program itself. The authors present a program-level reliability estimation technique that uses \emph{probabilistic symbolic execution} \cite{pse} to compute the probability of satisfying any of the path conditions corresponding to non-error-exposing paths. In other words, the approach computes the proportion of inputs which exercise paths that do not expose an error. Since probabilistic symbolic execution leverages model counting to determine the proportion of inputs exercising a path, the approach works only for very small input spaces. In contrast, we propose a lightweight statistical technique to estimate the confidence that a fuzzing campaign inspires in the correctness of a program, and that scales to programs of arbitrary size.

There exist several \emph{statistical approaches} to quantify the reliability of a program. For instance, the problem of estimating the probability $P(n)$ to discover an error with the $n+1$th test input, given that no errors have been found after generating $n$ test inputs can be cast as a variant of the \emph{sunrise problem}\footnote{Given that we have seen the sun rise for $n$ consecutive days, what is the probability that the sun will rise tomorrow?} which is classically solved with Laplace's \emph{rule of succession} \cite{probTheory}: $P(n)=1/(n+2)$. Suppose, $s$ of $n$ generated inputs expose an error, then the probability to generate another error-exposing input follows a \emph{beta}-distribution $\text{Beta}(s+1, n-s+1)$ the postorior of which has the expected value $(s+1)/(n+2)$.

Miller et al. \cite{miller} recognized the utility of the \emph{beta}-distribution to quantify the probability of failure in the absence of failures and furthermore discuss the case where the test distribution does not overlap with the operational distribution (i.e., the fuzzer might not generate ``typical'' inputs, but we are interested in the program's reliability for typical inputs). Littlewood and Wright \cite{littlewood} also utilize the \emph{beta}-distribution but discuss how to update the previous estimate of the probability of failure \emph{after} a bug was found and fixed.
In contrast to these existing works, the STADS framework leverages information on the \emph{problem structure} by identifying the species for an input. Hence, we can provide more accurate estimates of the residual risk that a detectable vulnerability has remained undetected and of the confidence that a fuzzing campaign inspires in the correctness of the program. Moreover, the STADS framework is \emph{more general} and also provides methodologies to estimate the total number of species and to extrapolate the number of species discovered at some future point in time.

\hl{It is interesting to note that only 25 years ago, the execution of \emph{one hundred million} ($n=10^8$) test inputs was utterly \emph{unthinkable} }\cite{sleeve}\hl{. Hamlet and Voas conjecture that ``[..] direct reliability assessment by random testing of software is impractical. The levels we would like to achieve, on the order of $10^6$--$10^8$ executions without failure, cannot be established in a reasonable time. Some limitations of reliability testing can be overcome, but the ``ultrareliable'' region above $10^8$ failure-free executions is likely to remain forever untestable'' }\cite{sleeve}\hl{. Today, Google's continuous fuzzing platform OSS-Fuzz generates 10 \emph{trillion} ($10^{10}$) test inputs \emph{per day} }\cite{oss}. 

When test inputs are generated manually, a general suggestion is to increase the code coverage. The most popular measures are \emph{code coverage} metrics, such as statement, branch, or MC/DC coverage, and \emph{fault coverage} metrics, such as relative mutation adequacy \cite{mutation}. The hope is that the fault revelation of a set of test inputs increases as its coverage increases. In other words, maximal coverage should inspire maximal confidence. However, many recent empirical studies found that such coverage metrics are in fact \emph{poor indicators} of test suite effectiveness \emph{in the context of automated software test generation} \cite{tsEffectiv1,tsEffectiv2,tsEffectiv3}. The empirical results may be explained by early theoretical investigations of testing effectiveness \cite{tsEffectiv4,tsEffectiv6,tsEffectiv5}. B\"{o}hme and Paul \cite{efficiency} similar to Hamlet and Taylor \cite{tsEffectiv4} argue that a set of \emph{successful} test inputs (i.e., no input exposes an error) that achieves 100\% branch coverage, 100\% MC/DC coverage, and even 100\% relative mutation adequacy does not inspire \emph{any} degree of confidence in the correctness of the tested program. Indeed, vulnerabilities may still exist. In contrast, STADS provides a statistically well-grounded framework to assess the residual risk that a detectable vulnerability exists even if none has been found. 

\subsection{Partition Testing}
\hl{In partition testing, the program's input domain is partitioned into overlapping or non-overlapping subdomains }\cite{tsEffectiv5}\hl{. The task of a \emph{tester} is to select one or more elements from each subdomain. In the STADS framework, we would say that each and only input in the same subdomain belongs to the same species. However, unlike in the STADS framework each input subdomain in partition testing is associated with a probability $\theta_i$ that an input in this subdomain reveals an error }\cite{tsEffectiv6}\hl{. Partition testing is a probabilistic model of software testing which allows to investigate the tester's ability to detect faults.}

\hl{Analyzing the effectiveness of random testing, Duran an Ntfos }\cite{tsEffectiv6}\hl{ used the partition testing model to show that the expected number of errors $g(n)$ discovered after $n$ test inputs have been sampled \emph{uniformly at random}, for the case of non-overlapping subdomains, is given as}
$$g(n) = S - \sum_{i=1}^S (1-p_i\theta_i)^n$$
\hl{where $S$ is the total number of subdomains, and $p_i$ is the probability that the randomly sampled input lies in subdomain $\mathcal{D}_i$. Duran and Ntfos observed experimentally that a tester who samples one or more inputs from each subdomain performs only slightly better than simple random testing.}

\hl{Varying several parameters, the Hamlet and Taylor }\cite{tsEffectiv4}\hl{ repeated the experiments of Duran and Ntafos and confirm: The number of errors found by random and partition testing is very similar. In fact, the authors conclude that ``partition testing does not inspire confidence''. Weyuker and Jeng }\cite{tsEffectiv5}\hl{ found that the effectiveness of partition testing varies depending on the fault rate $\theta_i$ for each subdomain. Subsequently, several authors discussed conditions under which partition testing is generally more effective than random testing (e.g., }\cite{tsEffectiv7,tsEffectiv8}\hl{). Empirical investigations }\cite{tsEffectiv1,tsEffectiv2,tsEffectiv3}\hl{ of the effectiveness of partition testing have since confirmed Hamlet and Taylor's conclusion }\cite{tsEffectiv4}.

\hl{In our previous work }\cite{efficiency}\hl{ we leverage the partition testing model to conduct the first probabilistic analysis of the efficiency of automated software testing. We identify bounds on the time the \emph{most effective} systematic testing technique can take per test input to remain \emph{more efficient} than random testing. We develop a hypothetical hybrid testing technique that is more efficient than both, random and systematic testing. We also suggest a primitive curve fitting method to extrapolate the partitions discovered over time. However, in the present article we introduce more sophisticated sampling-theoretic extrapolation methodologies.}

\hl{While the partition testing model allows \emph{probabilistic analyses}, the STADS framework allows \emph{statistical analyses}, including estimation and extrapolation. Probabilistic and statistical analysis are inverse to each other. In a \emph{probabilistic analysis} we consider some underlying random process where the randomness is modelled by random variables, and we resolve what happens. In a \emph{statistical analysis} we observe something that has happened, and try to resolve what underlying process would explain those observations. In contrast to existing work, we present practical estimation and extrapolation methodologies. The STADS framework is the first work in automated software testing that allows to extrapolate from tested to untested program behavior with quantifiable accuracy.}

%% file: sections/future.tex
\section{Challenges and Opportunities}\label{sec:discussion}

\subsection{Programs as Megadiverse Assemblages}\label{sec:megadiverse}
\hl{The STADS framework exhibits some peculiar features that make the application of existing ecologic methodologies more challenging: specifically, one has to deal with extremely large populations containing a huge number of species (e.g., millions of program branches or exponentially more distinct paths).
Specifically, compared to common assemblages in ecology, we expect species richness $S$ to be \emph{very high} and species evenness $J$ to be \emph{very low} in the STADS model. In other words, there are a huge number of very rare species and only a few extremely abundant species.}

\subsubsection{Megadiversity}
\hl{In ecology, we call an assemblage with high richness and low evenness as \emph{megadiverse assemblage}. For instance, arthropods (i.e., bugs, millipedes, spiders, etc.) in a tropical forest would be considered a megadiverse assemblage. }\cite{tropical}\hl{. There are an estimated 6.1 \emph{million} tropical arthropod species, most of which are rare }\cite{arthopods1,arthopods2}\hl{. Such assemblages are subject to several statistical challenges during estimation and extrapolation, particularly due to the relatively small sample size }\cite{collwell2,mao}.
\hl{However, compared to species inventories common in ecology, the sample size $n$ in the STADS framework can be \emph{very large}, which should render our data precious for ecologic biostatisticians. For instance, it took 102 ecology researchers  66 person-years to sample 129,494 arthropods representing 6144 species from 0.48 ha of tropical rain forest }\cite{tropical}\hl{. In stark contrast, a fuzzer can take a million samples in only a few minutes.}

\subsubsection{Scarcity}
\hl{The main objective of fuzzing is to discover vulnerabilities in a program. Vulnerabilities are arguably \emph{very rare species} in the STADS framework. Similarly, a primary objective of many ecological surveys is to identify species that are so rare that they are close to extinction. Once identified, the necessary conservation policies are proposed and implemented to counter the diminishing biodiversity.}

\hl{For the STADS framework, we should identify, develop and employ estimators that are better suitable if many rare species are present. Colwell et al. }\cite{sampleSurvey}\hl{ suggest to employ coverage-based estimators of species richness }\cite{ace,ace1,ice,ice1}\hl{ if one expects many rare species. Mao and Collwell }\cite{mao}\hl{ propose a mixture model to compute, with confidence intervals, a lower bound on species richness when there are many rare species. In a \emph{mixture model}, species abundance or occurrence distributions are modelled as a weighted mixture of statistical distributions. Ohannessian }\cite{rare}\hl{ observes that the Good-Turing estimator of discovery probability performs well even in the presence of many rare species. Chao et al. }\cite{relAbundance}\hl{ generalize the Good-Turing estimator to develop the Good-Turing sample coverage theory.
In future, coverage-based and mixture-model-based as well as other rare event estimators }\cite{rare,rare2}\hl{ and their performance within the STADS framework can be studied. Other suitable estimators for megadiverse assemblages with many rare species can be developed that would benefit tremendously both fields of research.}

\subsubsection{Endemism}
\hl{Another challenge in fuzzing is the random generation of ``magic numbers'', such as file identifiers }\cite{thinair}\hl{. Only if the magic number is correct will the generated test input exercise interesting program behaviors. Only if the magic number is correct will many new species be discovered. A similar challenge exists in ecology. \emph{Endemism} is the ecological state of a species being unique to a defined geographic location. For instance, the rain forest of Madagascar hosts a large number of (endemic) species that can only be found in Madagascar }\cite{madagascar}\hl{. A global survey of the biodiversity in rain forests would miss many species if the ``magic island'' of Madagascar remains uninvestigated. A survey of the biodiversity in the Sahara desert would miss many species if oases remain uninvestigated }\cite{sahara}\hl{. Hence, it is sensible only to provide an improved lower bound of the total number of species $S$ }\cite{chao1,chao2}. 

\subsubsection{Opportunities} 
\hl{Strategies could be established that allow to chose the best estimator at any time during the fuzzing campaign based on estimates of species evenness $J$ and discovery probability $U$ (e.g., }\cite{brose}\hl{). Several estimators of the same quantity may be used to derive a ``best estimate'' }\cite{tropical}\hl{. In the STADS framework, the program's source code and program binary provide an additional source of information that can be used to improve estimator performance. In future, the dependence of estimator bias and precision on the sample completeness $C$ can be investigated to develop better bias-correction mechanisms.}

\subsection{Species Identification and Oracle Problem}
\hl{In the STADS framework, the species for an input $t$ is identified using a combination of dynamic analyzers which record during execution of $t$ the observed program properties of interest. For instance, to detect bugs in C programs, the compiler can be asked to inject so-called \emph{sanitizers} }\cite{asan,msan}\hl{, assertions that crash the program when a bug is detected. There are different sanitizers }\cite{llvmsan,gccsan}\hl{, e.g.,}
\begin{itemize}
  \item \hl{to detect memory related errors, such as overflows and use-after-free (AddressSanitizer),}
  \item \hl{to detect race conditions and deadlocks (ThreadSanitizer),}
  \item \hl{to detect undefined behaviors (UndefinedBehaviorSanitizer),}
  \item \hl{to detect memory leaks (LeakSanitizer), or}
  \item \hl{to check control-flow integrity (CFISanitizer).}
\end{itemize}
\hl{Whether a bug constitutes an exploitable vulnerability can be determined with excellent precision and good recall using another dynamic analysis, e.g., the CERT Triage Tools }\cite{cert}.

\subsubsection{Misidentification and Guarantees}
\hl{The correct identification of the species for an input is an important challenge in both disciplines. For instance, in ecology Austen et al. }\cite{misidentification} \hl{observed that even taxonomic experts were correct in only 60\% of cases when asked whether two images showed the same or different species of bumblebees. Similarly, in the STADS framework a dynamic analysis may misidentify the species for an input.
For instance, misidentification in software testing may lead to input incorrectly classified as \emph{not} exposing a vulnerability (when it actually does). Hence, the statistical guarantees provided by the STADS model hold \emph{modulo} the dynamic analyzer's capability to identify the correct species for an input. This motivates further research on advanced dynamic analysis techniques that are more effective at vulnerability detection.}

\hl{It is interesting to note that misidentification is also an important challenge for automated verification. The formal guarantees provided by the verifier are valid only \emph{modulo} the provided specification which may be incomplete. For instance, the specification may allow to check whether a race conditions exists (a classic model checking problem)---but not whether a buffer overflow exists (the number-one root cause of arbitrary code execution attacks).}

\subsubsection{Morphospecies and Oracle Problem}
\hl{In ecology, some individuals cannot be assigned to named species. A \emph{morphospecies} is different from previously discovered species (in its morphology) but not to a sufficient degree that it could be assigned its own species. A similar challenge is known in the software testing domain as \emph{oracle problem}. Weyuker }\cite{oracle1}\hl{ conjectures that, in general, there exists no mechanism that can accurately decide whether or not the program behavior for an input is correct---whether an observed behavior is a bug or a feature of the program. Barr et al. }\cite{oracle2}\hl{ provide an excellent survey of recent advances in tackling the oracle problem.}

\subsection{Integrating Other Models Into STADS}
The STADS framework provides opportunities to explore other topics, models, and related methodologies in ecology. For instance, a typical problem in ecology is the extrapolation of the number of species in an enlarged area of size $A+a^*$, given only a sample of a smaller area of size $A$ \cite{chaoshen}. This is modelled as \emph{continuous Poisson model} \cite{coleman}. 
In the Poisson model, the reference sample is not defined by sample size $n$ but by the area $A$ that is sampled. The $i$th species occurs at a species-specific mean rate $A\lambda_i$, so that the probability distribution is
\begin{align}
P(X_1=x_1,\ldots,X_S=x_s) = \prod_{i=1}^S(A\lambda_i)^{x_i}\frac{\exp(-A\lambda_i)}{x_i!}
\end{align} 

In fuzzing, we often restrict the size of the generated test inputs because larger inputs might take longer to execute, and species appear to be distributed more densely in the space of small inputs. Within the STADS framework, we can leverage the Poisson model to estimate the total number of species for large inputs, given a fuzzing campaign that restricted the fuzzing to only smaller inputs. For instance, in Google's continuous fuzzing platform OSS-Fuzz \cite{oss}, the main fuzzer LibFuzzer \cite{libfuzzer} is often configured with a maximum test input size. The Poisson model would allow to extrapolate the confidence such ``restricted'' fuzzing campaigns inspire in the absence of vulnerabilities from the small generated test inputs to larger ``normal-sized'' inputs. 

In future, mixture models \cite{mao} can be developed that synthesize better estimates from those provided by the multinomial and the Poisson model. In order to integrate the Poisson and Bernoulli product models successfully into the STADS framework, an empirical evaluation of estimator performance is left for future work.

\subsection{Non-Adaptive Sampling Bias}
\hl{The STADS framework fully accounts for \emph{arbitrary fuzzer heuristics}, including the sampling from the operational distribution, as long as the fuzzer does not change the sampling strategy adaptively throughout the fuzzing campaign. For instance, if a compiler fuzzer generates more programs with loops than programs without---because historically programs with loops have always found more compiler bugs---then all statistical claims derived from the STADS framework strictly hold w.r.t. that fuzzer and for that program within the stipulated confidence bounds. The main assumption upon which the (multinomial and Bernoulli product) models of the STADS framework rely is that the relative species abundance $\{p_i\}_{i=1}^S$ does not change substantially during the fuzzing campaign. The compiler fuzzer is simply more likely (greater $p_i$) to discover a loop-based bug $\mathcal{D}_i$ than a compiler fuzzer without that heuristic, \emph{for all fuzzing campaigns}.}

While there may be some bias in the test input generation, there is \emph{no adaptive bias} for blackbox fuzzers. A \emph{blackbox fuzzer} does not leverage feedback from previous test executions to adapt the test generation strategy during the fuzzing campaign. A \emph{generational blackbox fuzzer} generates test inputs either by random sampling \cite{fuzz} or by instantiating elements from an input model, grammar, or protocol \cite{peach}. A \emph{mutation-based blackbox fuzzer} generates new test inputs by random perturbations of inputs in the so-called seed corpus. The probability $p_i$ to generate a test input that belong to species $\mathcal{D}_i$ does not change \emph{at all} during the fuzzing campaign.

\hl{In ecology, the sampling is usually subject to certain biases, as well. A light trap may lure certain species more than others }\cite{samplingbias2}\hl{. An ecologist may prefer to sample certain locations in an assemblage over others }\cite{samplingbias1}. \hl{An ecologist may be more likely to sample species that are larger in body size (or perhaps prefer to sample only the smaller species in the case of the arachnida class) }\cite{samplingbias3}. 

\subsection{Adaptive Sampling Bias of Feedback-directed Fuzzers}\label{sec:bias}
\hl{The \emph{main assumption} in the STADS framework upon which the multinomial, Bernoulli product, and Poisson models rely is that the relative species abundance $\{p_i\}_{i=1}^S$ does not change substantially during the fuzzing campaign. However, feedback-directed fuzzers are based on an \emph{adaptive} sampling strategy. A \emph{feedback-directed fuzzer} leverages program feedback from previous test inputs to learn and \emph{adaptively} generate ``better'' test inputs. The probability $p_i$ to sample an input that belongs to an undiscovered species $\mathcal{D}_i$ may increase.}

\subsubsection{Search-based Software Testing}
\hl{Fuzzers developed in the field of \emph{search-based software testing} (SBST) are feedback-directed. A fitness function evaluates how close a test input or set of test inputs is towards satisfying the concrete fuzzing objective while a meta-heuristic steers the test generation adaptively towards new test inputs with improved fitness. For instance, a \emph{directed greybox fuzzer} }\cite{aflgo}\hl{ evaluates the ``distance'' of an input to a set of target locations in the program (e.g., potential buffer overflow sites) and uses simulated annealing-based power schedules to generate new test inputs that are ``closer'' to those target locations. McMinn provides an excellent survey of SBST existing techniques }\cite{sbst}\hl{ and identifies future challenges }\cite{sbst1}.

\hl{To establish the impact of the adaptive bias, we strongly suggest to evaluate estimator performance for each SBST technique. In future, customized \emph{bias-corrected estimators} can be developed that allow accurate estimation and extrapolation for SBST techniques.}

\subsubsection{Coverage-based Greybox Fuzzing}
Feedback-directed are also coverage-based greybox fuzzers \cite{aflfast,afl,libfuzzer,syzkaller}. A coverage-based greybox fuzzer is typically mutation-based and hence starts with a seed corpus. If the fuzzer generates a test input $t$ that belongs to a previously undiscovered species (e.g., by random perturbations of inputs in the corpus), then $t$ is \emph{added} to the seed corpus. Otherwise, $t$ is discarded. At the time when $t$ is added to the seed corpus, the probability $p_i$ to discover any ``neighboring'' species $\mathcal{D}_i$ slightly increases, compared to \emph{before} $t$ was added. Hence, at first a coverage-based greybox fuzzer might discover more species per unit time than a mutation-based blackbox fuzzer (which is not feedback-directed). However, \emph{in the limit} every coverage-based greybox fuzzer degenerates to a mutation-based blackbox fuzzer. Over time more and more test inputs need to be generated to discover the next species: The fuzzer cycles several times through the same set of seeds without any discoveries for hours, later for days. Hence, \emph{in the limit} the adaptive bias is non-existent. Thus, if an estimator is consistent for a fuzzer that is not feedback-directed, it is also consistent for a coverage-based greybox fuzzer. The accuracy of a \emph{consistent} estimator increases as sampling effort (i.e., the number generated test inputs) increases.

\begin{figure}\small
\begin{tabular}{@{}c@{}c@{}}
\quad\quad\quad\textbf{(a)} Estimated path coverage over time & \quad\quad \textbf{(b)} Number of paths discovered over time\\  
\includegraphics[trim={0 0 0 0}, clip, width=0.5\columnwidth]{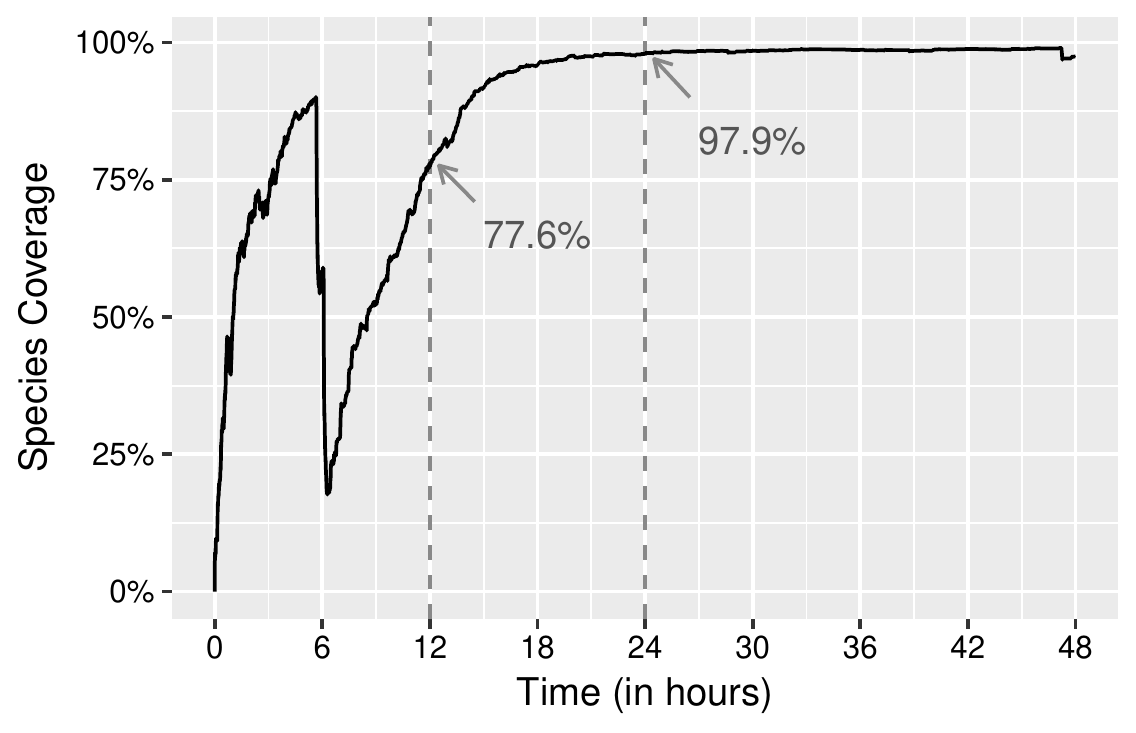} &
\includegraphics[trim={0 0 0 0}, clip, width=0.5\columnwidth]{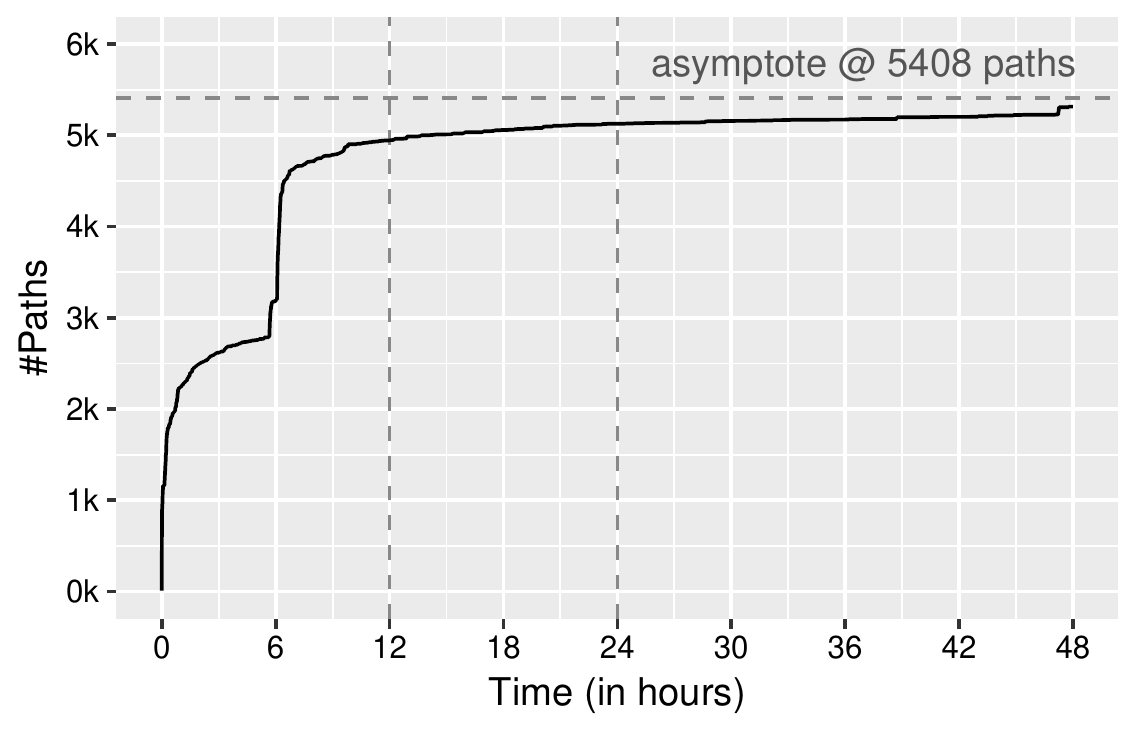}
\end{tabular}
\caption{Bias in the path coverage estimate for the AFL-fuzzing campaign in our motivating example.}
\label{fig:xBias}
\end{figure} 

The adaptive bias is obvious in \autoref{fig:xBias}.a which shows the development of the path coverage estimate over the first 48 hours of the fuzzing campaign in our motivating example (\autoref{sec:motivation}). Between five and seven hours, we see a steep drop in the path coverage estimate. The reason becomes obvious in \autoref{fig:xBias}.b which shows the number of paths discovered for the same fuzzing campaign. Just before the six-hour mark, the number of discovered paths seems to approach a different asymptote at about 3k paths when suddenly many more paths are discovered. This sudden increase is not very uncommon for AFL, particularly in the first few hours when still many new paths are discovered. However, such surges get more uncommon and their magnitude smaller as sample coverage increases. This can be explained within the Markov chain model of directed greybox fuzzing \cite{aflfast}. The path coverage estimate quickly recovers over the next 12 hours. In \autoref{fig:xBias}.b, we can see that 24 hours into the fuzzing campaign a large percentage of paths has been discovered (w.r.t. the improved estimate of the asymptote).
At this time, our path coverage estimate quite accurately puts the coverage at about 98\%. In future, we plan to investigate the correlation between the discovery probability estimate $\hat U$ of the sample and the bias/precision of the species coverage estimate $S(n)/\hat S$.

In our preliminary empirical study, we used the coverage-based greybox fuzzer AFL \cite{afl} to investigate the performance of the proposed estimators and extrapolators for \emph{the} state-of-the-art vulnerability detection tool. The results are promising (see \autoref{sec:resultsummary}). For AFL, the magnitude of the estimator bias was substantial right before and during short intervals when the number of discovered path-species increased suddenly and significantly. The extrapolation would not anticipate such sudden surges. However, the bias from adaptive sampling reduced over time. Close to the asymptotic species richness, the impact appeared negligible. 

\hl{In future work, several correction methodologies for the adaptive sampling bias may be developed.\linebreak
In software testing, we can analyze the program, e.g., to quantify the likelihood of sudden increases in species coverage. Advanced program analyses may allow for \emph{static bias-correction strategies}. 
For coverage-based greybox fuzzers, as species discovery decelerates, the impact of the adaptive bias reduces, as well. Estimates of sample completeness $C$ or species richness $S$ may be used as predictors of the adaptive bias which allow for \emph{dynamic bias-correction strategies}.
Moreover,  we anticipate \emph{empirical studies of estimator performance} for other greybox fuzzers.}

\subsubsection{Symbolic Execution}
Systematic (often symbolic execution-based) whitebox fuzzers are designed to discover previously undiscovered species with \emph{every} test input that is generated. Hence, the STADS framework explicitly \emph{does not apply} as such fuzzers violate the underlying assumptions of our statistical framework. For instance, a symbolic execution-based whitebox fuzzer is designed to systematically enumerate every (interesting) path in the program \cite{klee}. Every generated test input exercises a different path. Once a species $\mathcal{D}_i$ has been discovered, the probability $p_i$ of generating another input $t\in \mathcal{D}_i$ is $p_i=0$. This is a substantial change from before the discovery. 

However, we note that security researchers can use a blackbox or greybox fuzzer to establish the species evenness $J$ for that program, and based on its value decide whether to choose that fuzzer or the symbolic execution-based whitebox fuzzer for the actual fuzzing. As symbolic execution-based fuzzers are better suited to discover rare species, there should be a certain minimal value of $J$ below which the symbolic execution-based whitebox fuzzer performs better than the blackbox or greybox fuzzer. In future, this value can be empirically investigated.

\subsection{Adaptive Bias Correction}\label{sec:future}
Unlike sampling strategies in ecology, the STADS frameworks allows continuous estimation and extrapolation during the fuzzing campaign itself. This provides opportunities for \emph{continuous adaptive bias correction}. 
We can continuously assess the bias of our extrapolation by first predicting the value of our estimation target and later comparing it to its empirical value. The difference between predicted and empirical value describes the estimator bias. A continuous monitoring of the bias may allow to gradually control for and correct the observed bias.

Monitoring the fuzzing campaign also enables \emph{on-the-fly fuzzer selection}.
In earlier work \cite{efficiency}, we found that even the most effective systematic fuzzer would be less efficient than a random fuzzer if generating a test input takes relatively too long. For short fuzzing campaigns, a random fuzzer would always outperform a systematic fuzzer. However, at a certain time, it would be more efficient to switch to systematic fuzzing. Using the proposed extrapolators, we can make an informed decision when to switch, e.g,. from the ``biased'' random fuzzer AFL \cite{afl} to the systematic fuzzer, KLEE \cite{klee}.

%% file: conclusion.tex
\begin{table}[p]\centering
\caption{A summary of the pertinent estimators and extrapolators for the STADS model that were discussed and/or evaluated in this article.}
\label{tab:overview}
\rotatebox{90}{
\begin{tabular}{@{}l@{\quad\quad}l@{\quad\quad\quad\quad\quad} l@{\quad}}
&\textbf{Multinomial Model} & \textbf{Bernoulli Product Model}\\
&\textbf{(One Input, One Species)} & \textbf{(One Input, Multiple Species)}\\
\hline
\multicolumn{3}{@{}l@{}}{\emph{Estimating and extrapolating progress based on the total number of species} $\hat S$ \cite{sampleSurvey}}\\[0.1cm]
Total \#species &
$\hat S \approx \begin{cases}
S(n) + f_1^2 / (2f_2) & \text{if } f_2 > 0\\
S(n) + f_1(f_1-1)/2 & \text{if } f_2 = 0
\end{cases}$
&
$\hat S \approx \begin{cases}
S(n) + Q_1^2 / (2Q_2) & \text{if } Q_2 > 0\\
S(n) + Q_1(Q_1-1)/2 & \text{if } Q_2 = 0 
\end{cases}$\\[0.3cm]
(\emph{Chao1/2 estimators})& $\hat S = S$ \quad\quad\quad\quad\quad\quad\quad\quad \ \ if $S$ is known & $\hat S = S$ \quad\quad\quad\quad\quad\quad\quad\quad\quad if $S$ is known\\ 
& $\hat S = S(n) + \hat f_0$ \quad\quad\quad\quad $\rightarrow\hat f_0=\hat S - S(n)$ & $\hat S = S(n) + \hat Q_0$ \quad\quad\quad\quad \ \  $\rightarrow\hat Q_0=\hat S - S(n)$\\[0.1cm]
\emph{How many more} &  \multirow{4}{*}{$\hat S(n + m^*) = S(n)+\hat f_0\left[1-\left(1-\frac{f_1}{n\hat f_0 + f_1}\right)^{m^*}\right]$} & \multirow{4}{*}{$\hat S(n + m^*) = S(n)+\hat Q_0\left[1-\left(1-\frac{Q_1}{n\hat Q_0 + Q_1}\right)^{m^*}\right]$}\\[-.05cm]
\emph{species are}\\[-.05cm]
\emph{discovered with}\\[-.05cm]
\emph{more inputs?}\\[0.1cm]
\emph{How many more} & \multirow{5}{*}{\ $m_{G^*} \approx \dfrac{n f_1}{2f_2}\log\left[\dfrac{\hat f_0}{(1-G^*)\hat S}\right]$} & \multirow{5}{*}{$m_{G^*} \approx 
\dfrac{\log\left[1-\cfrac{n}{(n-1)}\cfrac{2Q_2}{Q_1^2}(G^*\hat S - S(n))\right]}{\log\left[1-\cfrac{2Q_2}{(n-1)Q_1+2Q_2}\right]}$}\\[-.05cm]
\emph{inputs are needed} & & \\[-.05cm]
\emph{to discover $G^*\cdot\hat S$} & &\\[-.05cm]
\emph{species where}\\[-.05cm]
\emph{$G(n)<G^*<1$?}\\[0.3cm]
\hline
\multicolumn{3}{@{}l@{}}{\emph{Estimating and extrapolating progress based on discovery probability} $\hat C$ \cite{coverageSurvey,good}}\\[0.1cm]
Discovery probability& \multirow{2}{*}{$\hat U(n) = \dfrac{f_1}{n}$} & \multirow{2}{*}{$\hat U(n) = \dfrac{Q_1}{V} \left[\dfrac{n\hat Q_0}{n\hat Q_0+Q_1}\right]\approx \dfrac{Q_1}{V}$}\\[-.05cm]
 &\\[0.4cm]
\emph{How much more}&\multirow{4}{*}{$\hat U(n+m^*) = \dfrac{f_1}{n}\left(\dfrac{n\hat f_0}{n\hat f_0 + f_1}\right)^{m^*+1}$} & \multirow{4}{*}{$\hat U(n+m^*) = \dfrac{Q_1}{V}\left[\dfrac{n\hat Q_0}{n \hat Q_0 +Q_1}\right]^{m^*+1}$}\\[-.05cm]
\emph{sample coverage}\\[-.05cm]
\emph{is achieved with}\\[-.05cm]
\emph{more inputs?}\\[0.1cm]\hline
\multicolumn{3}{@{}p{1.24\linewidth}@{}}{\textbf{Notation}: \emph{In the current fuzzing campaign $n$ is the number of generated test inputs, $S(n)$ is the number of discovered species, $f_1$ and $Q_1$ are the number of singleton species (i.e., those to which only one generated input belongs), $f_2$ and $Q_2$ are the number of doubleton species (i.e., those to which only two generated inputs belong), $\hat f_0$ and $\hat Q_0$ are estimates of the number of undiscovered species, and $V$ denotes the sum total of the number of species that each generated test input belongs to. Note that $V > n$ if multiple species can be identified for a single input.} 
}\\ 
\hline
\end{tabular}}
\end{table}
\section{Conclusion}\label{sec:conclusion}
In this article, I introduced the foundations of a general, statistical framework that models software testing and analysis as discovery of species (STADS) to address a \emph{fundamental challenge} in software testing: the statistically well-grounded \emph{extrapolation} from program behaviors observed during testing. The STADS framework draws from over three decades of research in ecological biostatistics, where the challenge is to extrapolate from properties of the species observed in a sample to properties of the species in the complete assemblage.

Based on the STADS framework, researchers can, for the first time, formally discuss, estimate and assess a \emph{fuzzer}'s effectiveness and efficiency, a \emph{campaign}'s completeness, cost-effectiveness, and residual risk, and a \emph{program}'s fuzzability. For the first time, test engineers have gained the ability to make informed decisions about whether to abort or continue a fuzzing campaign; and to quantify what has been learned about the program at any point throughout the fuzzing campaign. Beyond this initial work, I pointed to a large number of opportunities for researchers to improve and tailor the ecologic methodologies to the automated testing and analysis process.

The first empirical evidence was provided that the main hypothesis that is underpinning the STADS framework (and thus allows the usage of existing ecological methodologies in the context of automated software testing and analysis) actually holds. The \emph{multinomial model}, where each input belongs to exactly one species, was integrated and successfully evaluated. The evaluated estimators from ecology showed good performance, e.g., during estimation and extrapolation of path coverage. Thereupon, the STADS framework was extended with the \emph{Bernoulli product model}, where each input can belong to one or more species. We show that the estimators can be efficiently computed even for large programs, and are guaranteed to approach the true value as the fuzzing effort increases. An overview of the pertinent estimators for both models can be found in \autoref{tab:overview}.

The presented predictive program analysis scales to programs of arbitrary size and is general enough to work with arbitrary finite and discrete program properties. For instance, the STADS framework allows to estimate the asymptotic total number of paths, information flows, reachable target locations, unique program crashes, or the number of statements that are actually executable by the fuzzer. It allows to extrapolate code coverage, as well as mutation-adequacy efficiently and with improving accuracy. Species coverage can be used effectively to judge whether the fuzzing campaign is almost completed.

Many estimators and extrapolators are readily available as statistical analyses to try out online \cite{spadeOnline,inextOnline} or as packages in the R programming language \cite{spade,inext}. Our integration \pythia with the popular vulnerability detection tool AFL can be downloaded from Github at:
\begin{center}\bf\color{blue}{
%\vspace{0.5cm}
\url{https://github.com/mboehme/pythia}
%\vspace{0.5cm}
}
\end{center}

The STADS framework provides a large number of opportunities for future work. For instance, software can be understood as megadiverse assemblage which features a large number of very rare species. Novel estimators can be identified or developed that address the peculiarities of automated software testing as species discovery. Feedback-directed fuzzers introduce an adaptive bias that can result in sudden surges in species discovered. Adaptive bias correction strategies can be developed that leverage program analysis to anticipate and account for such surges to correct the adaptive bias dynamically.  

%% file: bibtex.bbl

%% file: main.bbl
\begin{thebibliography}{00}

%%% the Web address.

\ifx \showCODEN    \undefined \def \showCODEN     #1{\unskip}     \fi
\ifx \showDOI      \undefined \def \showDOI       #1{#1}\fi
\ifx \showISBNx    \undefined \def \showISBNx     #1{\unskip}     \fi
\ifx \showISBNxiii \undefined \def \showISBNxiii  #1{\unskip}     \fi
\ifx \showISSN     \undefined \def \showISSN      #1{\unskip}     \fi
\ifx \showLCCN     \undefined \def \showLCCN      #1{\unskip}     \fi
\ifx \shownote     \undefined \def \shownote      #1{#1}          \fi
\ifx \showarticletitle \undefined \def \showarticletitle #1{#1}   \fi
\ifx \showURL      \undefined \def \showURL       {\relax}        \fi
% invisible to TeX
\providecommand\bibfield[2]{#2}
\providecommand\bibinfo[2]{#2}
\providecommand\natexlab[1]{#1}
\providecommand\showeprint[2][]{arXiv:#2}

\bibitem[\protect\citeauthoryear{Austen, Bindemann, Griffiths, and
  Roberts}{Austen et~al\mbox{.}}{2016}]%
        {misidentification}
\bibfield{author}{\bibinfo{person}{G.~E. Austen}, \bibinfo{person}{M.
  Bindemann}, \bibinfo{person}{R.~A. Griffiths}, {and} \bibinfo{person}{D.~L.
  Roberts}.} \bibinfo{year}{2016}\natexlab{}.
\newblock \showarticletitle{Species identification by experts and non-experts:
  comparing images from field guides}.
\newblock \bibinfo{journal}{{\em Nature - Scientific Reports\/}}
  \bibinfo{volume}{6} (\bibinfo{date}{09} \bibinfo{year}{2016}).
\newblock


\bibitem[\protect\citeauthoryear{Banks, Cova, Felmetsger, Almeroth, Kemmerer,
  and Vigna}{Banks et~al\mbox{.}}{2006}]%
        {snooze}
\bibfield{author}{\bibinfo{person}{Greg Banks}, \bibinfo{person}{Marco Cova},
  \bibinfo{person}{Viktoria Felmetsger}, \bibinfo{person}{Kevin Almeroth},
  \bibinfo{person}{Richard Kemmerer}, {and} \bibinfo{person}{Giovanni Vigna}.}
  \bibinfo{year}{2006}\natexlab{}.
\newblock \showarticletitle{SNOOZE: Toward a Stateful Network Protocol fuzZEr}.
  In \bibinfo{booktitle}{{\em Proceedings of the 9th International Conference
  on Information Security}} {\em (\bibinfo{series}{ISC'06})}.
  \bibinfo{pages}{343--358}.
\newblock


\bibitem[\protect\citeauthoryear{Barr, Harman, McMinn, Shahbaz, and Yoo}{Barr
  et~al\mbox{.}}{2015}]%
        {oracle2}
\bibfield{author}{\bibinfo{person}{E.~T. Barr}, \bibinfo{person}{M. Harman},
  \bibinfo{person}{P. McMinn}, \bibinfo{person}{M. Shahbaz}, {and}
  \bibinfo{person}{S. Yoo}.} \bibinfo{year}{2015}\natexlab{}.
\newblock \showarticletitle{The Oracle Problem in Software Testing: A Survey}.
\newblock \bibinfo{journal}{{\em IEEE Transactions on Software Engineering\/}}
  \bibinfo{volume}{41}, \bibinfo{number}{5} (\bibinfo{date}{May}
  \bibinfo{year}{2015}), \bibinfo{pages}{507--525}.
\newblock


\bibitem[\protect\citeauthoryear{Basset, Cizek, Cu{\'e}noud, Didham,
  Guilhaumon, Missa, Novotny, {\O}degaard, Roslin, Schmidl, Tishechkin,
  Winchester, Roubik, Aberlenc, Bail, Barrios, Bridle, Casta{\~n}o-Meneses,
  Corbara, Curletti, Duarte~da Rocha, De~Bakker, Delabie, Dejean, Fagan,
  Floren, Kitching, Medianero, Miller, Gama~de Oliveira, Orivel, Pollet, Rapp,
  Ribeiro, Roisin, Schmidt, S{\o}rensen, and Leponce}{Basset
  et~al\mbox{.}}{2012}]%
        {tropical}
\bibfield{author}{\bibinfo{person}{Yves Basset}, \bibinfo{person}{Lukas Cizek},
  \bibinfo{person}{Philippe Cu{\'e}noud}, \bibinfo{person}{Raphael~K. Didham},
  \bibinfo{person}{Fran{\c c}ois Guilhaumon}, \bibinfo{person}{Olivier Missa},
  \bibinfo{person}{Vojtech Novotny}, \bibinfo{person}{Frode {\O}degaard},
  \bibinfo{person}{Tomas Roslin}, \bibinfo{person}{J{\"u}rgen Schmidl},
  \bibinfo{person}{Alexey~K. Tishechkin}, \bibinfo{person}{Neville~N.
  Winchester}, \bibinfo{person}{David~W. Roubik}, \bibinfo{person}{Henri-Pierre
  Aberlenc}, \bibinfo{person}{Johannes Bail}, \bibinfo{person}{H{\'e}ctor
  Barrios}, \bibinfo{person}{Jon~R. Bridle}, \bibinfo{person}{Gabriela
  Casta{\~n}o-Meneses}, \bibinfo{person}{Bruno Corbara},
  \bibinfo{person}{Gianfranco Curletti}, \bibinfo{person}{Wesley Duarte~da
  Rocha}, \bibinfo{person}{Domir De~Bakker}, \bibinfo{person}{Jacques H.~C.
  Delabie}, \bibinfo{person}{Alain Dejean}, \bibinfo{person}{Laura~L. Fagan},
  \bibinfo{person}{Andreas Floren}, \bibinfo{person}{Roger~L. Kitching},
  \bibinfo{person}{Enrique Medianero}, \bibinfo{person}{Scott~E. Miller},
  \bibinfo{person}{Evandro Gama~de Oliveira}, \bibinfo{person}{J{\'e}r{\^o}me
  Orivel}, \bibinfo{person}{Marc Pollet}, \bibinfo{person}{Mathieu Rapp},
  \bibinfo{person}{S{\'e}rvio~P. Ribeiro}, \bibinfo{person}{Yves Roisin},
  \bibinfo{person}{Jesper~B. Schmidt}, \bibinfo{person}{Line S{\o}rensen},
  {and} \bibinfo{person}{Maurice Leponce}.} \bibinfo{year}{2012}\natexlab{}.
\newblock \showarticletitle{Arthropod Diversity in a Tropical Forest}.
\newblock \bibinfo{journal}{{\em Science\/}} \bibinfo{volume}{338},
  \bibinfo{number}{6113} (\bibinfo{year}{2012}), \bibinfo{pages}{1481--1484}.
\newblock


\bibitem[\protect\citeauthoryear{B\"{o}hme, Oliveira, and
  Roychoudhury}{B\"{o}hme et~al\mbox{.}}{2013a}]%
        {prv}
\bibfield{author}{\bibinfo{person}{Marcel B\"{o}hme}, \bibinfo{person}{Bruno C.
  d.~S. Oliveira}, {and} \bibinfo{person}{Abhik Roychoudhury}.}
  \bibinfo{year}{2013}\natexlab{a}.
\newblock \showarticletitle{Partition-based Regression Verification}. In
  \bibinfo{booktitle}{{\em Proceedings of the 2013 International Conference on
  Software Engineering}} {\em (\bibinfo{series}{ICSE '13})}.
  \bibinfo{pages}{302--311}.
\newblock


\bibitem[\protect\citeauthoryear{B\"{o}hme, Oliveira, and
  Roychoudhury}{B\"{o}hme et~al\mbox{.}}{2013b}]%
        {cie}
\bibfield{author}{\bibinfo{person}{Marcel B\"{o}hme}, \bibinfo{person}{Bruno C.
  d.~S. Oliveira}, {and} \bibinfo{person}{Abhik Roychoudhury}.}
  \bibinfo{year}{2013}\natexlab{b}.
\newblock \showarticletitle{Regression Tests to Expose Change Interaction
  Errors}. In \bibinfo{booktitle}{{\em Proceedings of the 2013 9th Joint
  Meeting on Foundations of Software Engineering}} {\em
  (\bibinfo{series}{ESEC/FSE 2013})}. \bibinfo{pages}{334--344}.
\newblock


\bibitem[\protect\citeauthoryear{B\"{o}hme and Paul}{B\"{o}hme and
  Paul}{2016}]%
        {efficiency}
\bibfield{author}{\bibinfo{person}{Marcel B\"{o}hme} {and}
  \bibinfo{person}{Soumya Paul}.} \bibinfo{year}{2016}\natexlab{}.
\newblock \showarticletitle{A Probabilistic Analysis of the Efficiency of
  Automated Software Testing}.
\newblock \bibinfo{journal}{{\em IEEE Transactions on Software Engineering\/}}
  \bibinfo{volume}{42}, \bibinfo{number}{4} (\bibinfo{date}{April}
  \bibinfo{year}{2016}), \bibinfo{pages}{345--360}.
\newblock


\bibitem[\protect\citeauthoryear{B\"{o}hme, Pham, Nguyen, and
  Roychoudhury}{B\"{o}hme et~al\mbox{.}}{2017}]%
        {aflgo}
\bibfield{author}{\bibinfo{person}{Marcel B\"{o}hme},
  \bibinfo{person}{Van-Thuan Pham}, \bibinfo{person}{Manh-Dung Nguyen}, {and}
  \bibinfo{person}{Abhik Roychoudhury}.} \bibinfo{year}{2017}\natexlab{}.
\newblock \showarticletitle{Directed Greybox Fuzzing}. In
  \bibinfo{booktitle}{{\em Proceedings of the 2017 ACM SIGSAC Conference on
  Computer and Communications Security}} {\em (\bibinfo{series}{CCS '17})}.
  \bibinfo{pages}{2329--2344}.
\newblock


\bibitem[\protect\citeauthoryear{B\"{o}hme, Pham, and Roychoudhury}{B\"{o}hme
  et~al\mbox{.}}{2016}]%
        {aflfast}
\bibfield{author}{\bibinfo{person}{Marcel B\"{o}hme},
  \bibinfo{person}{Van-Thuan Pham}, {and} \bibinfo{person}{Abhik
  Roychoudhury}.} \bibinfo{year}{2016}\natexlab{}.
\newblock \showarticletitle{Coverage-based Greybox Fuzzing As Markov Chain}. In
  \bibinfo{booktitle}{{\em Proceedings of the 2016 ACM SIGSAC Conference on
  Computer and Communications Security}} {\em (\bibinfo{series}{CCS '16})}.
  \bibinfo{pages}{1032--1043}.
\newblock


\bibitem[\protect\citeauthoryear{Botev and Kroese}{Botev and Kroese}{2008}]%
        {rare2}
\bibfield{author}{\bibinfo{person}{Zdravko~I. Botev} {and}
  \bibinfo{person}{Dirk~P. Kroese}.} \bibinfo{year}{2008}\natexlab{}.
\newblock \showarticletitle{An Efficient Algorithm for Rare-event Probability
  Estimation, Combinatorial Optimization, and Counting}.
\newblock \bibinfo{journal}{{\em Methodology and Computing in Applied
  Probability\/}} \bibinfo{volume}{10}, \bibinfo{number}{4} (\bibinfo{date}{01
  Dec} \bibinfo{year}{2008}), \bibinfo{pages}{471--505}.
\newblock


\bibitem[\protect\citeauthoryear{Brose, Martinez, and Williams}{Brose
  et~al\mbox{.}}{2003}]%
        {brose}
\bibfield{author}{\bibinfo{person}{Ulrich Brose}, \bibinfo{person}{Neo~D.
  Martinez}, {and} \bibinfo{person}{Richard~J. Williams}.}
  \bibinfo{year}{2003}\natexlab{}.
\newblock \showarticletitle{Estimating Species Richness: Sensitivity to Sample
  Coverage and Insensitivity to Spatial Patterns}.
\newblock \bibinfo{journal}{{\em Ecology\/}} \bibinfo{volume}{84},
  \bibinfo{number}{9} (\bibinfo{year}{2003}), \bibinfo{pages}{2364--2377}.
\newblock


\bibitem[\protect\citeauthoryear{Bunge and Fitzpatrick}{Bunge and
  Fitzpatrick}{1993}]%
        {speciesReview}
\bibfield{author}{\bibinfo{person}{J. Bunge} {and} \bibinfo{person}{M.
  Fitzpatrick}.} \bibinfo{year}{1993}\natexlab{}.
\newblock \showarticletitle{Estimating the Number of Species: A Review}.
\newblock \bibinfo{journal}{{\it J. Amer. Statist. Assoc.}}
  \bibinfo{volume}{88}, \bibinfo{number}{421} (\bibinfo{year}{1993}),
  \bibinfo{pages}{364--373}.
\newblock


\bibitem[\protect\citeauthoryear{Burnham and Overton}{Burnham and
  Overton}{1979}]%
        {burnham}
\bibfield{author}{\bibinfo{person}{K.~P. Burnham} {and} \bibinfo{person}{W.~S.
  Overton}.} \bibinfo{year}{1979}\natexlab{}.
\newblock \showarticletitle{Robust Estimation of Population Size When Capture
  Probabilities Vary Among Animals}.
\newblock \bibinfo{journal}{{\em Ecology\/}} \bibinfo{volume}{60},
  \bibinfo{number}{5} (\bibinfo{year}{1979}), \bibinfo{pages}{927--936}.
\newblock


\bibitem[\protect\citeauthoryear{Cadar, Dunbar, and Engler}{Cadar
  et~al\mbox{.}}{2008}]%
        {klee}
\bibfield{author}{\bibinfo{person}{Cristian Cadar}, \bibinfo{person}{Daniel
  Dunbar}, {and} \bibinfo{person}{Dawson Engler}.}
  \bibinfo{year}{2008}\natexlab{}.
\newblock \showarticletitle{KLEE: Unassisted and Automatic Generation of
  High-coverage Tests for Complex Systems Programs}. In
  \bibinfo{booktitle}{{\em Proceedings of the 8th USENIX Conference on
  Operating Systems Design and Implementation}} {\em
  (\bibinfo{series}{OSDI'08})}. \bibinfo{pages}{209--224}.
\newblock


\bibitem[\protect\citeauthoryear{Chao}{Chao}{1984}]%
        {chao1}
\bibfield{author}{\bibinfo{person}{Anne Chao}.}
  \bibinfo{year}{1984}\natexlab{}.
\newblock \showarticletitle{Nonparametric Estimation of the Number of Classes
  in a Population}.
\newblock \bibinfo{journal}{{\em Scandinavian Journal of Statistics\/}}
  \bibinfo{volume}{11}, \bibinfo{number}{4} (\bibinfo{year}{1984}),
  \bibinfo{pages}{265--270}.
\newblock


\bibitem[\protect\citeauthoryear{Chao}{Chao}{1987}]%
        {chao2}
\bibfield{author}{\bibinfo{person}{Anne Chao}.}
  \bibinfo{year}{1987}\natexlab{}.
\newblock \showarticletitle{Estimating the population size for
  capture-recapture data with unequal catchability}.
\newblock \bibinfo{journal}{{\em Biometrics\/}} \bibinfo{volume}{43},
  \bibinfo{number}{4} (\bibinfo{year}{1987}), \bibinfo{pages}{783--791}.
\newblock


\bibitem[\protect\citeauthoryear{Chao and Chiu}{Chao and Chiu}{2014}]%
        {hypothesis}
\bibfield{author}{\bibinfo{person}{Anne Chao} {and} \bibinfo{person}{Chun-Huo
  Chiu}.} \bibinfo{year}{2014}\natexlab{}.
\newblock \bibinfo{booktitle}{{\em Species Richness: Estimation and
  Comparison}}.
\newblock \bibinfo{publisher}{John Wiley \& Sons, Ltd}.
\newblock


\bibitem[\protect\citeauthoryear{Chao and Chiu}{Chao and Chiu}{2016}]%
        {sRichness}
\bibfield{author}{\bibinfo{person}{Anne Chao} {and} \bibinfo{person}{Chun-Huo
  Chiu}.} \bibinfo{year}{2016}\natexlab{}.
\newblock \showarticletitle{Nonparametric estimation and comparison of species
  richness}.
\newblock \bibinfo{journal}{{\em Encyclopedia of Life Sciences (eLS)\/}}
  (\bibinfo{date}{May} \bibinfo{year}{2016}).
\newblock


\bibitem[\protect\citeauthoryear{Chao, Chiu, Colwell, Magnago, Chazdon, and
  Gotelli}{Chao et~al\mbox{.}}{2017}]%
        {goodtheory}
\bibfield{author}{\bibinfo{person}{Anne Chao}, \bibinfo{person}{Chun-Huo Chiu},
  \bibinfo{person}{Robert~K. Colwell}, \bibinfo{person}{Luiz Fernando~S.
  Magnago}, \bibinfo{person}{Robin~L. Chazdon}, {and}
  \bibinfo{person}{Nicholas~J. Gotelli}.} \bibinfo{year}{2017}\natexlab{}.
\newblock \showarticletitle{Deciphering the enigma of undetected species,
  phylogenetic, and functional diversity based on Good-Turing theory}.
\newblock \bibinfo{journal}{{\em Ecology\/}} \bibinfo{volume}{98},
  \bibinfo{number}{11} (\bibinfo{year}{2017}), \bibinfo{pages}{2914--2929}.
\newblock
\showISSN{1939-9170}


\bibitem[\protect\citeauthoryear{Chao and Colwell}{Chao and Colwell}{2017}]%
        {incidenceSurvey}
\bibfield{author}{\bibinfo{person}{Anne Chao} {and} \bibinfo{person}{Robert~K.
  Colwell}.} \bibinfo{year}{2017}\natexlab{}.
\newblock \showarticletitle{Thirty years of progeny from Chao's inequality:
  Estimating and comparing richness with incidence data and incomplete
  sampling}.
\newblock \bibinfo{journal}{{\em Statistics and Operations Research
  Transactions\/}} \bibinfo{volume}{41}, \bibinfo{number}{1}
  (\bibinfo{year}{2017}), \bibinfo{pages}{3--54}.
\newblock


\bibitem[\protect\citeauthoryear{Chao, Colwell, Lin, and Gotelli}{Chao
  et~al\mbox{.}}{2009}]%
        {samplesRequired}
\bibfield{author}{\bibinfo{person}{Anne Chao}, \bibinfo{person}{Robert~K.
  Colwell}, \bibinfo{person}{Chih-Wei Lin}, {and} \bibinfo{person}{Nicholas~J.
  Gotelli}.} \bibinfo{year}{2009}\natexlab{}.
\newblock \showarticletitle{Sufficient sampling for asymptotic minimum species
  richness estimators}.
\newblock \bibinfo{journal}{{\em Ecology\/}} \bibinfo{volume}{90},
  \bibinfo{number}{4} (\bibinfo{year}{2009}), \bibinfo{pages}{1125--1133}.
\newblock
\showISSN{1939-9170}


\bibitem[\protect\citeauthoryear{Chao, Hsieh, Chazdon, Colwell, and
  Gotelli}{Chao et~al\mbox{.}}{2015a}]%
        {relAbundance}
\bibfield{author}{\bibinfo{person}{Anne Chao}, \bibinfo{person}{T.~C. Hsieh},
  \bibinfo{person}{Robin~L. Chazdon}, \bibinfo{person}{Robert~K. Colwell},
  {and} \bibinfo{person}{Nicholas~J. Gotelli}.}
  \bibinfo{year}{2015}\natexlab{a}.
\newblock \showarticletitle{Unveiling the species-rank abundance distribution
  by generalizing the Good-Turing sample coverage theory}.
\newblock \bibinfo{journal}{{\em Ecology\/}} \bibinfo{volume}{96},
  \bibinfo{number}{5} (\bibinfo{year}{2015}), \bibinfo{pages}{1189--1201}.
\newblock


\bibitem[\protect\citeauthoryear{Chao and Jost}{Chao and Jost}{2012}]%
        {coverageSurvey}
\bibfield{author}{\bibinfo{person}{Anne Chao} {and} \bibinfo{person}{Lou
  Jost}.} \bibinfo{year}{2012}\natexlab{}.
\newblock \showarticletitle{Coverage-based rarefaction and extrapolation:
  standardizing samples by completeness rather than size}.
\newblock \bibinfo{journal}{{\em Ecology\/}} \bibinfo{volume}{93},
  \bibinfo{number}{12} (\bibinfo{year}{2012}), \bibinfo{pages}{2533--2547}.
\newblock


\bibitem[\protect\citeauthoryear{Chao and Lee}{Chao and Lee}{1992}]%
        {ace}
\bibfield{author}{\bibinfo{person}{Anne Chao} {and} \bibinfo{person}{Shen-Ming
  Lee}.} \bibinfo{year}{1992}\natexlab{}.
\newblock \showarticletitle{Estimating the Number of Classes via Sample
  Coverage}.
\newblock \bibinfo{journal}{{\it J. Amer. Statist. Assoc.}}
  \bibinfo{volume}{87}, \bibinfo{number}{417} (\bibinfo{year}{1992}),
  \bibinfo{pages}{210--217}.
\newblock


\bibitem[\protect\citeauthoryear{Chao, Ma, Hsieh, and Chiu}{Chao
  et~al\mbox{.}}{2015b}]%
        {spade}
\bibfield{author}{\bibinfo{person}{A. Chao}, \bibinfo{person}{K.~H. Ma},
  \bibinfo{person}{T.~C. Hsieh}, {and} \bibinfo{person}{C.~H. Chiu}.}
  \bibinfo{year}{2015}\natexlab{b}.
\newblock \bibinfo{title}{Online Program SpadeR (Species-richness Prediction
  And Diversity Estimation in R)}.
\newblock \bibinfo{howpublished}{Program and User's Guide published at
  \url{http://chao.stat.nthu.edu.tw/wordpress/software_download}}.
  (\bibinfo{year}{2015}).
\newblock


\bibitem[\protect\citeauthoryear{Chao and Shen}{Chao and Shen}{2004}]%
        {chaoshen}
\bibfield{author}{\bibinfo{person}{Anne Chao} {and} \bibinfo{person}{Tsung-Jen
  Shen}.} \bibinfo{year}{2004}\natexlab{}.
\newblock \showarticletitle{Nonparametric Prediction in Species Sampling}.
\newblock \bibinfo{journal}{{\em Journal of Agricultural, Biological, and
  Environmental Statistics\/}} \bibinfo{volume}{9}, \bibinfo{number}{3}
  (\bibinfo{year}{2004}), \bibinfo{pages}{253--269}.
\newblock


\bibitem[\protect\citeauthoryear{Chazdon, Colwell, Denslow, and
  Guariguata}{Chazdon et~al\mbox{.}}{1998}]%
        {ace1}
\bibfield{author}{\bibinfo{person}{R.L. Chazdon}, \bibinfo{person}{R.K.
  Colwell}, \bibinfo{person}{J.S. Denslow}, {and} \bibinfo{person}{M.R.
  Guariguata}.} \bibinfo{year}{1998}\natexlab{}.
\newblock \bibinfo{booktitle}{{\em Forest biodiversity research, monitoring and
  modeling: conceptual background and old world case studies}}.
  Vol.~\bibinfo{volume}{20}.
\newblock \bibinfo{publisher}{Man and the Biosphere Series}, Chapter
  Statistical methods for estimating species richness of woody regeneration in
  primary and secondary rain forests of Northeastern Costa Rica,
  \bibinfo{pages}{285--309}.
\newblock


\bibitem[\protect\citeauthoryear{Chekam, Papadakis, Traon, and Harman}{Chekam
  et~al\mbox{.}}{2017}]%
        {tsEffectiv3}
\bibfield{author}{\bibinfo{person}{Thierry~Titcheu Chekam},
  \bibinfo{person}{Mike Papadakis}, \bibinfo{person}{Yves~Le Traon}, {and}
  \bibinfo{person}{Mark Harman}.} \bibinfo{year}{2017}\natexlab{}.
\newblock \showarticletitle{An Empirical Study on Mutation, Statement and
  Branch Coverage Fault Revelation That Avoids the Unreliable Clean Program
  Assumption}. In \bibinfo{booktitle}{{\em Proceedings of the 39th
  International Conference on Software Engineering}} {\em
  (\bibinfo{series}{ICSE '17})}. \bibinfo{pages}{597--608}.
\newblock


\bibitem[\protect\citeauthoryear{Chen and Yu}{Chen and Yu}{1996}]%
        {tsEffectiv8}
\bibfield{author}{\bibinfo{person}{Tsong~Yueh Chen} {and}
  \bibinfo{person}{Yuen-Tak Yu}.} \bibinfo{year}{1996}\natexlab{}.
\newblock \showarticletitle{On the Expected Number of Failures Detected by
  Subdomain Testing and Random Testing.}
\newblock \bibinfo{journal}{{\em IEEE Transactions on Software Engineering\/}}
  \bibinfo{volume}{22}, \bibinfo{number}{2} (\bibinfo{year}{1996}),
  \bibinfo{pages}{109--119}.
\newblock


\bibitem[\protect\citeauthoryear{Chipounov, Kuznetsov, and Candea}{Chipounov
  et~al\mbox{.}}{2011}]%
        {s2e}
\bibfield{author}{\bibinfo{person}{Vitaly Chipounov},
  \bibinfo{person}{Volodymyr Kuznetsov}, {and} \bibinfo{person}{George
  Candea}.} \bibinfo{year}{2011}\natexlab{}.
\newblock \showarticletitle{S2E: A Platform for In-vivo Multi-path Analysis of
  Software Systems}. In \bibinfo{booktitle}{{\em ASPLOS XVI}}.
  \bibinfo{pages}{265--278}.
\newblock


\bibitem[\protect\citeauthoryear{Chiu, Wang, Walther, and Chao}{Chiu
  et~al\mbox{.}}{2014}]%
        {chui}
\bibfield{author}{\bibinfo{person}{C.H. Chiu}, \bibinfo{person}{Y.T. Wang},
  \bibinfo{person}{B.A. Walther}, {and} \bibinfo{person}{A. Chao}.}
  \bibinfo{year}{2014}\natexlab{}.
\newblock \showarticletitle{An improved nonparametric lower bound of species
  richness via a modified good-turing frequency formula}.
\newblock \bibinfo{journal}{{\em Biometrics\/}} \bibinfo{volume}{70},
  \bibinfo{number}{3} (\bibinfo{year}{2014}), \bibinfo{pages}{671--682}.
\newblock


\bibitem[\protect\citeauthoryear{Coleman}{Coleman}{1981}]%
        {coleman}
\bibfield{author}{\bibinfo{person}{Bernard~D. Coleman}.}
  \bibinfo{year}{1981}\natexlab{}.
\newblock \showarticletitle{On random placement and species-area relations}.
\newblock \bibinfo{journal}{{\em Mathematical Biosciences\/}}
  \bibinfo{volume}{54}, \bibinfo{number}{3} (\bibinfo{year}{1981}),
  \bibinfo{pages}{191 -- 215}.
\newblock


\bibitem[\protect\citeauthoryear{Colwell, Chao, Gotelli, Lin, Mao, Chazdon, and
  Longino}{Colwell et~al\mbox{.}}{2012}]%
        {sampleSurvey}
\bibfield{author}{\bibinfo{person}{Robert~K. Colwell}, \bibinfo{person}{Anne
  Chao}, \bibinfo{person}{Nicholas~J. Gotelli}, \bibinfo{person}{Shang-Yi Lin},
  \bibinfo{person}{Chang~Xuan Mao}, \bibinfo{person}{Robin~L. Chazdon}, {and}
  \bibinfo{person}{John~T. Longino}.} \bibinfo{year}{2012}\natexlab{}.
\newblock \showarticletitle{Models and estimators linking individual-based and
  sample-based rarefaction, extrapolation and comparison of assemblages}.
\newblock \bibinfo{journal}{{\em Journal of Plant Ecology\/}}
  \bibinfo{volume}{5}, \bibinfo{number}{1} (\bibinfo{year}{2012}),
  \bibinfo{pages}{3}.
\newblock


\bibitem[\protect\citeauthoryear{Colwell and Coddington}{Colwell and
  Coddington}{1994}]%
        {collwell2}
\bibfield{author}{\bibinfo{person}{Robert~K. Colwell} {and}
  \bibinfo{person}{Jonathan~A. Coddington}.} \bibinfo{year}{1994}\natexlab{}.
\newblock \showarticletitle{Estimating Terrestrial Biodiversity through
  Extrapolation}.
\newblock \bibinfo{journal}{{\em Philosophical Transactions of the Royal
  Society of London B: Biological Sciences\/}} \bibinfo{volume}{345},
  \bibinfo{number}{1311} (\bibinfo{year}{1994}), \bibinfo{pages}{101--118}.
\newblock


\bibitem[\protect\citeauthoryear{Colwell and Elsensohn}{Colwell and
  Elsensohn}{2014}]%
        {collwell20}
\bibfield{author}{\bibinfo{person}{Robert~K. Colwell} {and}
  \bibinfo{person}{Johanna~E. Elsensohn}.} \bibinfo{year}{2014}\natexlab{}.
\newblock \showarticletitle{EstimateS turns 20: statistical estimation of
  species richness and shared species from samples, with non-parametric
  extrapolation}.
\newblock \bibinfo{journal}{{\em Ecography\/}} \bibinfo{volume}{37},
  \bibinfo{number}{6} (\bibinfo{year}{2014}), \bibinfo{pages}{609--613}.
\newblock


\bibitem[\protect\citeauthoryear{Colwell, Mao, and Chang}{Colwell
  et~al\mbox{.}}{2004}]%
        {collwell3}
\bibfield{author}{\bibinfo{person}{Robert~K. Colwell},
  \bibinfo{person}{Chang~Xuan Mao}, {and} \bibinfo{person}{Jing Chang}.}
  \bibinfo{year}{2004}\natexlab{}.
\newblock \showarticletitle{Interpolating, Extrapolating, and Comparing
  Incidence-based Species Accumulation Curves}.
\newblock \bibinfo{journal}{{\em Ecology\/}} \bibinfo{volume}{85},
  \bibinfo{number}{10} (\bibinfo{year}{2004}), \bibinfo{pages}{2717--2727}.
\newblock


\bibitem[\protect\citeauthoryear{Dijkstra}{Dijkstra}{1970}]%
        {djikstra}
\bibfield{author}{\bibinfo{person}{Edsger~W. Dijkstra}.}
  \bibinfo{year}{1970}\natexlab{}.
\newblock \bibinfo{title}{Notes on {S}tructured {P}rogramming}.
  (\bibinfo{year}{1970}).
\newblock


\bibitem[\protect\citeauthoryear{Duran and Ntafos}{Duran and Ntafos}{1984}]%
        {tsEffectiv6}
\bibfield{author}{\bibinfo{person}{Joe~W. Duran} {and}
  \bibinfo{person}{Simeon~C. Ntafos}.} \bibinfo{year}{1984}\natexlab{}.
\newblock \showarticletitle{An Evaluation of Random Testing}.
\newblock \bibinfo{journal}{{\em IEEE Transactions on Software Engineering\/}}
  \bibinfo{volume}{10}, \bibinfo{number}{4} (\bibinfo{date}{July}
  \bibinfo{year}{1984}), \bibinfo{pages}{438--444}.
\newblock


\bibitem[\protect\citeauthoryear{Durant, Wacher, Bashir, Woodroffe,
  De~Ornellas, Ransom, Newby, Abáigar, Abdelgadir, El~Alqamy, Baillie,
  Beddiaf, Belbachir, Belbachir-Bazi, Berbash, Bemadjim, Beudels-Jamar,
  Boitani, Breitenmoser, Cano, Chardonnet, Collen, Cornforth, Cuzin, Gerngross,
  Haddane, Hadjeloum, Jacobson, Jebali, Lamarque, Mallon, Minkowski, Monfort,
  Ndoassal, Niagate, Purchase, Samaïla, Samna, Sillero-Zubiri, Soultan,
  Stanley~Price, and Pettorelli}{Durant et~al\mbox{.}}{2014}]%
        {sahara}
\bibfield{author}{\bibinfo{person}{S.~M. Durant}, \bibinfo{person}{T. Wacher},
  \bibinfo{person}{S. Bashir}, \bibinfo{person}{R. Woodroffe},
  \bibinfo{person}{P. De~Ornellas}, \bibinfo{person}{C. Ransom},
  \bibinfo{person}{J. Newby}, \bibinfo{person}{T. Abáigar},
  \bibinfo{person}{M. Abdelgadir}, \bibinfo{person}{H. El~Alqamy},
  \bibinfo{person}{J. Baillie}, \bibinfo{person}{M. Beddiaf},
  \bibinfo{person}{F. Belbachir}, \bibinfo{person}{A. Belbachir-Bazi},
  \bibinfo{person}{A.~A. Berbash}, \bibinfo{person}{N.~E. Bemadjim},
  \bibinfo{person}{R. Beudels-Jamar}, \bibinfo{person}{L. Boitani},
  \bibinfo{person}{C. Breitenmoser}, \bibinfo{person}{M. Cano},
  \bibinfo{person}{P. Chardonnet}, \bibinfo{person}{B. Collen},
  \bibinfo{person}{W.~A. Cornforth}, \bibinfo{person}{F. Cuzin},
  \bibinfo{person}{P. Gerngross}, \bibinfo{person}{B. Haddane},
  \bibinfo{person}{M. Hadjeloum}, \bibinfo{person}{A. Jacobson},
  \bibinfo{person}{A. Jebali}, \bibinfo{person}{F. Lamarque},
  \bibinfo{person}{D. Mallon}, \bibinfo{person}{K. Minkowski},
  \bibinfo{person}{S. Monfort}, \bibinfo{person}{B. Ndoassal},
  \bibinfo{person}{B. Niagate}, \bibinfo{person}{G. Purchase},
  \bibinfo{person}{S. Samaïla}, \bibinfo{person}{A.~K. Samna},
  \bibinfo{person}{C. Sillero-Zubiri}, \bibinfo{person}{A.~E. Soultan},
  \bibinfo{person}{M.~R. Stanley~Price}, {and} \bibinfo{person}{N.
  Pettorelli}.} \bibinfo{year}{2014}\natexlab{}.
\newblock \showarticletitle{Fiddling in biodiversity hotspots while deserts
  burn? Collapse of the Sahara's megafauna}.
\newblock \bibinfo{journal}{{\em Diversity and Distributions\/}}
  \bibinfo{volume}{20}, \bibinfo{number}{1} (\bibinfo{year}{2014}),
  \bibinfo{pages}{114--122}.
\newblock
\showISSN{1472-4642}


\bibitem[\protect\citeauthoryear{Filieri, P\u{a}s\u{a}reanu, and
  Visser}{Filieri et~al\mbox{.}}{2013}]%
        {jpfRelia}
\bibfield{author}{\bibinfo{person}{Antonio Filieri}, \bibinfo{person}{Corina~S.
  P\u{a}s\u{a}reanu}, {and} \bibinfo{person}{Willem Visser}.}
  \bibinfo{year}{2013}\natexlab{}.
\newblock \showarticletitle{Reliability Analysis in Symbolic Pathfinder}. In
  \bibinfo{booktitle}{{\em Proceedings of the 2013 International Conference on
  Software Engineering}} {\em (\bibinfo{series}{ICSE '13})}.
  \bibinfo{pages}{622--631}.
\newblock


\bibitem[\protect\citeauthoryear{Fisher, Corbet, and Williams}{Fisher
  et~al\mbox{.}}{1943}]%
        {fisher}
\bibfield{author}{\bibinfo{person}{R.~A. Fisher}, \bibinfo{person}{A.~S.
  Corbet}, {and} \bibinfo{person}{C.~B. Williams}.}
  \bibinfo{year}{1943}\natexlab{}.
\newblock \showarticletitle{The relation between the number of species and the
  number of individuals in a random sample of an animal population}.
\newblock \bibinfo{journal}{{\em Journal of Animal Ecology\/}}
  \bibinfo{volume}{12} (\bibinfo{year}{1943}), \bibinfo{pages}{42--58}.
\newblock


\bibitem[\protect\citeauthoryear{Gale and Sampson}{Gale and Sampson}{1995}]%
        {smoothing}
\bibfield{author}{\bibinfo{person}{William~A. Gale} {and}
  \bibinfo{person}{Geoffrey Sampson}.} \bibinfo{year}{1995}\natexlab{}.
\newblock \showarticletitle{Good-Turing smoothing without tears}.
\newblock \bibinfo{journal}{{\em Journal of Quantitative Linguistics\/}}
  \bibinfo{volume}{2} (\bibinfo{year}{1995}), \bibinfo{pages}{217--237}.
\newblock


\bibitem[\protect\citeauthoryear{Gay, Staats, Whalen, and Heimdahl}{Gay
  et~al\mbox{.}}{2015}]%
        {tsEffectiv2}
\bibfield{author}{\bibinfo{person}{G. Gay}, \bibinfo{person}{M. Staats},
  \bibinfo{person}{M. Whalen}, {and} \bibinfo{person}{M.~P.~E. Heimdahl}.}
  \bibinfo{year}{2015}\natexlab{}.
\newblock \showarticletitle{The Risks of Coverage-Directed Test Case
  Generation}.
\newblock \bibinfo{journal}{{\em IEEE Transactions on Software Engineering\/}}
  \bibinfo{volume}{41}, \bibinfo{number}{8} (\bibinfo{date}{Aug}
  \bibinfo{year}{2015}), \bibinfo{pages}{803--819}.
\newblock


\bibitem[\protect\citeauthoryear{Geldenhuys, Dwyer, and Visser}{Geldenhuys
  et~al\mbox{.}}{2012}]%
        {pse}
\bibfield{author}{\bibinfo{person}{Jaco Geldenhuys},
  \bibinfo{person}{Matthew~B. Dwyer}, {and} \bibinfo{person}{Willem Visser}.}
  \bibinfo{year}{2012}\natexlab{}.
\newblock \showarticletitle{Probabilistic Symbolic Execution}. In
  \bibinfo{booktitle}{{\em Proceedings of the 2012 International Symposium on
  Software Testing and Analysis}} {\em (\bibinfo{series}{ISSTA 2012})}.
  \bibinfo{pages}{166--176}.
\newblock


\bibitem[\protect\citeauthoryear{Godefroid, Kiezun, and Levin}{Godefroid
  et~al\mbox{.}}{2008}]%
        {gwf}
\bibfield{author}{\bibinfo{person}{Patrice Godefroid}, \bibinfo{person}{Adam
  Kiezun}, {and} \bibinfo{person}{Michael~Y. Levin}.}
  \bibinfo{year}{2008}\natexlab{}.
\newblock \showarticletitle{Grammar-based Whitebox Fuzzing}. In
  \bibinfo{booktitle}{{\em Proceedings of the 29th ACM SIGPLAN Conference on
  Programming Language Design and Implementation}} {\em (\bibinfo{series}{PLDI
  '08})}. \bibinfo{pages}{206--215}.
\newblock


\bibitem[\protect\citeauthoryear{Godefroid, Klarlund, and Sen}{Godefroid
  et~al\mbox{.}}{2005}]%
        {dart}
\bibfield{author}{\bibinfo{person}{Patrice Godefroid}, \bibinfo{person}{Nils
  Klarlund}, {and} \bibinfo{person}{Koushik Sen}.}
  \bibinfo{year}{2005}\natexlab{}.
\newblock \showarticletitle{DART: Directed Automated Random Testing}. In
  \bibinfo{booktitle}{{\em Proceedings of the 2005 ACM SIGPLAN Conference on
  Programming Language Design and Implementation}} {\em (\bibinfo{series}{PLDI
  '05})}. \bibinfo{pages}{213--223}.
\newblock


\bibitem[\protect\citeauthoryear{Good}{Good}{2000}]%
        {crypt}
\bibfield{author}{\bibinfo{person}{I.J. Good}.}
  \bibinfo{year}{2000}\natexlab{}.
\newblock \showarticletitle{Turing's anticipation of empirical bayes in
  connection with the cryptanalysis of the naval enigma}.
\newblock \bibinfo{journal}{{\em Journal of Statistical Computation and
  Simulation\/}} \bibinfo{volume}{66}, \bibinfo{number}{2}
  (\bibinfo{year}{2000}), \bibinfo{pages}{101--111}.
\newblock


\bibitem[\protect\citeauthoryear{Good}{Good}{1953}]%
        {good}
\bibfield{author}{\bibinfo{person}{Irving~John Good}.}
  \bibinfo{year}{1953}\natexlab{}.
\newblock \showarticletitle{The population frequencies of species and the
  estimation of population parameters}.
\newblock \bibinfo{journal}{{\em Biometrika\/}}  \bibinfo{volume}{40}
  (\bibinfo{year}{1953}), \bibinfo{pages}{237--264}.
\newblock


\bibitem[\protect\citeauthoryear{Good and Toulmin}{Good and Toulmin}{1956}]%
        {goodToulmin}
\bibfield{author}{\bibinfo{person}{I.~J. Good} {and} \bibinfo{person}{G.~H.
  Toulmin}.} \bibinfo{year}{1956}\natexlab{}.
\newblock \showarticletitle{The Number of New Species, and the Increase in
  Population Coverage, when a Sample is Increased}.
\newblock \bibinfo{journal}{{\em Biometrika\/}} \bibinfo{volume}{43},
  \bibinfo{number}{1/2} (\bibinfo{year}{1956}), \bibinfo{pages}{45--63}.
\newblock


\bibitem[\protect\citeauthoryear{Goodman and Benstead}{Goodman and
  Benstead}{2005}]%
        {madagascar}
\bibfield{author}{\bibinfo{person}{Steven~M. Goodman} {and}
  \bibinfo{person}{Jonathan~P. Benstead}.} \bibinfo{year}{2005}\natexlab{}.
\newblock \showarticletitle{Updated estimates of biotic diversity and endemism
  for Madagascar}.
\newblock \bibinfo{journal}{{\em Oryx\/}} \bibinfo{volume}{39},
  \bibinfo{number}{1} (\bibinfo{year}{2005}), \bibinfo{pages}{73–77}.
\newblock


\bibitem[\protect\citeauthoryear{Gotelli and Chao}{Gotelli and Chao}{2013}]%
        {ice1}
\bibfield{author}{\bibinfo{person}{Nicholas~J. Gotelli} {and}
  \bibinfo{person}{Anne Chao}.} \bibinfo{year}{2013}\natexlab{}.
\newblock \bibinfo{booktitle}{{\em Encyclopedia of Biodiversity\/}
  (\bibinfo{edition}{2} ed.)}. Vol.~\bibinfo{volume}{5}.
\newblock \bibinfo{publisher}{Academic Press}, Chapter Measuring and Estimating
  Species Richness, Species Diversity, and Biotic Similarity from Sampling
  Data, \bibinfo{pages}{195--211}.
\newblock


\bibitem[\protect\citeauthoryear{Gutjahr}{Gutjahr}{1999}]%
        {tsEffectiv7}
\bibfield{author}{\bibinfo{person}{Walter~J. Gutjahr}.}
  \bibinfo{year}{1999}\natexlab{}.
\newblock \showarticletitle{Partition Testing vs. Random Testing: The Influence
  of Uncertainty}.
\newblock \bibinfo{journal}{{\em IEEE Transactions on Software Engineering\/}}
  \bibinfo{volume}{25}, \bibinfo{number}{5} (\bibinfo{date}{Sept.}
  \bibinfo{year}{1999}), \bibinfo{pages}{661--674}.
\newblock


\bibitem[\protect\citeauthoryear{Hamilton, Basset, Benke, Grimbacher, Miller,
  Novotn{\`y}, Samuelson, Stork, Weiblen, and Yen}{Hamilton
  et~al\mbox{.}}{2010}]%
        {arthopods1}
\bibfield{author}{\bibinfo{person}{Andrew~J Hamilton}, \bibinfo{person}{Yves
  Basset}, \bibinfo{person}{Kurt~K Benke}, \bibinfo{person}{Peter~S
  Grimbacher}, \bibinfo{person}{Scott~E Miller}, \bibinfo{person}{Vojtech
  Novotn{\`y}}, \bibinfo{person}{G~Allan Samuelson}, \bibinfo{person}{Nigel~E
  Stork}, \bibinfo{person}{George~D Weiblen}, {and} \bibinfo{person}{Jian~DL
  Yen}.} \bibinfo{year}{2010}\natexlab{}.
\newblock \showarticletitle{Quantifying uncertainty in estimation of tropical
  arthropod species richness}.
\newblock \bibinfo{journal}{{\em The American Naturalist\/}}
  \bibinfo{volume}{176}, \bibinfo{number}{1} (\bibinfo{year}{2010}),
  \bibinfo{pages}{90--95}.
\newblock


\bibitem[\protect\citeauthoryear{Hamilton, Basset, Benke, Grimbacher, Miller,
  NovotnÃœ, Samuelson, Stork, Weiblen, and Yen}{Hamilton
  et~al\mbox{.}}{2011}]%
        {arthopods2}
\bibfield{author}{\bibinfo{person}{Andrew~J. Hamilton}, \bibinfo{person}{Yves
  Basset}, \bibinfo{person}{Kurt~K. Benke}, \bibinfo{person}{Peter~S.
  Grimbacher}, \bibinfo{person}{Scott~E. Miller}, \bibinfo{person}{Vojtech
  NovotnÃœ}, \bibinfo{person}{G.~Allan Samuelson}, \bibinfo{person}{Nigel~E.
  Stork}, \bibinfo{person}{George~D. Weiblen}, {and} \bibinfo{person}{Jian
  D.~L. Yen}.} \bibinfo{year}{2011}\natexlab{}.
\newblock \showarticletitle{Correction}.
\newblock \bibinfo{journal}{{\em The American Naturalist\/}}
  \bibinfo{volume}{177}, \bibinfo{number}{4} (\bibinfo{year}{2011}),
  \bibinfo{pages}{544--545}.
\newblock


\bibitem[\protect\citeauthoryear{Hamlet and Taylor}{Hamlet and Taylor}{1990}]%
        {tsEffectiv4}
\bibfield{author}{\bibinfo{person}{D. Hamlet} {and} \bibinfo{person}{R.
  Taylor}.} \bibinfo{year}{1990}\natexlab{}.
\newblock \showarticletitle{Partition testing does not inspire confidence
  [program testing]}.
\newblock \bibinfo{journal}{{\em IEEE Transactions on Software Engineering\/}}
  \bibinfo{volume}{16}, \bibinfo{number}{12} (\bibinfo{date}{Dec}
  \bibinfo{year}{1990}), \bibinfo{pages}{1402--1411}.
\newblock


\bibitem[\protect\citeauthoryear{Hamlet and Voas}{Hamlet and Voas}{1993}]%
        {sleeve}
\bibfield{author}{\bibinfo{person}{Dick Hamlet} {and} \bibinfo{person}{Jeff
  Voas}.} \bibinfo{year}{1993}\natexlab{}.
\newblock \showarticletitle{Faults on Its Sleeve: Amplifying Software
  Reliability Testing}. In \bibinfo{booktitle}{{\em Proceedings of the 1993 ACM
  SIGSOFT International Symposium on Software Testing and Analysis}} {\em
  (\bibinfo{series}{ISSTA '93})}. \bibinfo{pages}{89--98}.
\newblock


\bibitem[\protect\citeauthoryear{Hernandez and Lindquist}{Hernandez and
  Lindquist}{1999}]%
        {samplingbias2}
\bibfield{author}{\bibinfo{person}{Frank~J. Hernandez} {and}
  \bibinfo{person}{David~G. Lindquist}.} \bibinfo{year}{1999}\natexlab{}.
\newblock \showarticletitle{A comparison of two light-trap designs for sampling
  larval and presettlement juvenile fish above a reef in Onslow Bay, North
  Carolina}.
\newblock \bibinfo{journal}{{\em Bulletin of Marine Science\/}}
  \bibinfo{volume}{64}, \bibinfo{number}{1} (\bibinfo{year}{1999}),
  \bibinfo{pages}{173--184}.
\newblock


\bibitem[\protect\citeauthoryear{Hortal, Borges, and Gaspar}{Hortal
  et~al\mbox{.}}{2006}]%
        {speciesEmpirical}
\bibfield{author}{\bibinfo{person}{Joaquin Hortal}, \bibinfo{person}{Paulo
  A.~V. Borges}, {and} \bibinfo{person}{Clara Gaspar}.}
  \bibinfo{year}{2006}\natexlab{}.
\newblock \showarticletitle{Evaluating the performance of species richness
  estimators: sensitivity to sample grain size}.
\newblock \bibinfo{journal}{{\em Journal of Animal Ecology\/}}
  \bibinfo{volume}{75}, \bibinfo{number}{1} (\bibinfo{year}{2006}),
  \bibinfo{pages}{274--287}.
\newblock
\showISSN{1365-2656}


\bibitem[\protect\citeauthoryear{H\"{o}schele and Zeller}{H\"{o}schele and
  Zeller}{2016}]%
        {grammarInference}
\bibfield{author}{\bibinfo{person}{Matthias H\"{o}schele} {and}
  \bibinfo{person}{Andreas Zeller}.} \bibinfo{year}{2016}\natexlab{}.
\newblock \showarticletitle{Mining Input Grammars from Dynamic Taints}. In
  \bibinfo{booktitle}{{\em Proceedings of the 31st IEEE/ACM International
  Conference on Automated Software Engineering}} {\em (\bibinfo{series}{ASE
  2016})}. \bibinfo{pages}{720--725}.
\newblock


\bibitem[\protect\citeauthoryear{Hsieh, Ma, and Chao}{Hsieh
  et~al\mbox{.}}{2016}]%
        {inext}
\bibfield{author}{\bibinfo{person}{T.~C. Hsieh}, \bibinfo{person}{K.~H. Ma},
  {and} \bibinfo{person}{Anne Chao}.} \bibinfo{year}{2016}\natexlab{}.
\newblock \showarticletitle{iNEXT: an R package for rarefaction and
  extrapolation of species diversity (Hill numbers)}.
\newblock \bibinfo{journal}{{\em Methods in Ecology and Evolution\/}}
  \bibinfo{volume}{7}, \bibinfo{number}{12} (\bibinfo{year}{2016}),
  \bibinfo{pages}{1451--1456}.
\newblock
\showDOI{%
\url{https://doi.org/10.1111/2041-210X.12613}}


\bibitem[\protect\citeauthoryear{Hurlbert}{Hurlbert}{1971}]%
        {hurlbert}
\bibfield{author}{\bibinfo{person}{Stuart~H. Hurlbert}.}
  \bibinfo{year}{1971}\natexlab{}.
\newblock \showarticletitle{The Nonconcept of Species Diversity: A Critique and
  Alternative Parameters}.
\newblock \bibinfo{journal}{{\em Ecology\/}} \bibinfo{volume}{52},
  \bibinfo{number}{4} (\bibinfo{year}{1971}), \bibinfo{pages}{577--586}.
\newblock


\bibitem[\protect\citeauthoryear{Inozemtseva and Holmes}{Inozemtseva and
  Holmes}{2014}]%
        {tsEffectiv1}
\bibfield{author}{\bibinfo{person}{Laura Inozemtseva} {and}
  \bibinfo{person}{Reid Holmes}.} \bibinfo{year}{2014}\natexlab{}.
\newblock \showarticletitle{Coverage is Not Strongly Correlated with Test Suite
  Effectiveness}. In \bibinfo{booktitle}{{\em Proceedings of the 36th
  International Conference on Software Engineering}} {\em
  (\bibinfo{series}{ICSE 2014})}. \bibinfo{pages}{435--445}.
\newblock


\bibitem[\protect\citeauthoryear{Jaynes}{Jaynes}{2003}]%
        {probTheory}
\bibfield{author}{\bibinfo{person}{E.~T. Jaynes}.}
  \bibinfo{year}{2003}\natexlab{}.
\newblock \bibinfo{booktitle}{{\em Probability Theory: The Logic of Science}}.
\newblock \bibinfo{publisher}{Cambridge University Press}.
\newblock


\bibitem[\protect\citeauthoryear{Jia and Harman}{Jia and Harman}{2011}]%
        {mutation}
\bibfield{author}{\bibinfo{person}{Y. Jia} {and} \bibinfo{person}{M. Harman}.}
  \bibinfo{year}{2011}\natexlab{}.
\newblock \showarticletitle{An Analysis and Survey of the Development of
  Mutation Testing}.
\newblock \bibinfo{journal}{{\em IEEE Transactions on Software Engineering\/}}
  \bibinfo{volume}{37}, \bibinfo{number}{5} (\bibinfo{date}{Sept}
  \bibinfo{year}{2011}), \bibinfo{pages}{649--678}.
\newblock


\bibitem[\protect\citeauthoryear{Lee and Chao}{Lee and Chao}{1994}]%
        {ice}
\bibfield{author}{\bibinfo{person}{Shen-Ming Lee} {and} \bibinfo{person}{Anne
  Chao}.} \bibinfo{year}{1994}\natexlab{}.
\newblock \showarticletitle{Estimating Population Size Via Sample Coverage for
  Closed Capture-Recapture Models}.
\newblock \bibinfo{journal}{{\em Biometrics\/}} \bibinfo{volume}{50},
  \bibinfo{number}{1} (\bibinfo{year}{1994}), \bibinfo{pages}{88--97}.
\newblock


\bibitem[\protect\citeauthoryear{Littlewood and Wright}{Littlewood and
  Wright}{1997}]%
        {littlewood}
\bibfield{author}{\bibinfo{person}{B. Littlewood} {and} \bibinfo{person}{D.
  Wright}.} \bibinfo{year}{1997}\natexlab{}.
\newblock \showarticletitle{Some conservative stopping rules for the
  operational testing of safety critical software}.
\newblock \bibinfo{journal}{{\em IEEE Transactions on Software Engineering\/}}
  \bibinfo{volume}{23}, \bibinfo{number}{11} (\bibinfo{date}{Nov}
  \bibinfo{year}{1997}), \bibinfo{pages}{673--683}.
\newblock


\bibitem[\protect\citeauthoryear{Longino and Colwell}{Longino and
  Colwell}{1997}]%
        {longino}
\bibfield{author}{\bibinfo{person}{John~T. Longino} {and}
  \bibinfo{person}{Robert~K. Colwell}.} \bibinfo{year}{1997}\natexlab{}.
\newblock \showarticletitle{Biodiversity assessment using structured inventory:
  Capturing the ant fauna of a tropical rain forest}.
\newblock \bibinfo{journal}{{\em Ecological Applications\/}}
  \bibinfo{volume}{7}, \bibinfo{number}{4} (\bibinfo{year}{1997}),
  \bibinfo{pages}{1263--1277}.
\newblock


\bibitem[\protect\citeauthoryear{Magurran and McGill}{Magurran and
  McGill}{2011}]%
        {magurran}
\bibfield{author}{\bibinfo{person}{Anne~E. Magurran} {and}
  \bibinfo{person}{Brian~J. McGill}.} \bibinfo{year}{2011}\natexlab{}.
\newblock \bibinfo{booktitle}{{\em Biological diversity: frontiers in
  measurement and assessment}}.
\newblock \bibinfo{publisher}{Oxford University Press}.
\newblock


\bibitem[\protect\citeauthoryear{Mao and Colwell}{Mao and Colwell}{2005}]%
        {mao}
\bibfield{author}{\bibinfo{person}{Chang~Xuan Mao} {and}
  \bibinfo{person}{Robert~K. Colwell}.} \bibinfo{year}{2005}\natexlab{}.
\newblock \showarticletitle{Estimation of Species Richness: Mixture Models, the
  Role of Rare Species, and Inferential Challenges}.
\newblock \bibinfo{journal}{{\em Ecology\/}} \bibinfo{volume}{86},
  \bibinfo{number}{5} (\bibinfo{year}{2005}), \bibinfo{pages}{1143--1153}.
\newblock
\showISSN{1939-9170}


\bibitem[\protect\citeauthoryear{Mao, Harman, and Jia}{Mao
  et~al\mbox{.}}{2016}]%
        {sapienz}
\bibfield{author}{\bibinfo{person}{Ke Mao}, \bibinfo{person}{Mark Harman},
  {and} \bibinfo{person}{Yue Jia}.} \bibinfo{year}{2016}\natexlab{}.
\newblock \showarticletitle{Sapienz: Multi-objective Automated Testing for
  Android Applications}. In \bibinfo{booktitle}{{\em Proceedings of the 25th
  International Symposium on Software Testing and Analysis}} {\em
  (\bibinfo{series}{ISSTA 2016})}. \bibinfo{pages}{94--105}.
\newblock


\bibitem[\protect\citeauthoryear{Matthis, Avdiienko, Soremekun, B\"{o}hme, and
  Zeller}{Matthis et~al\mbox{.}}{2017}]%
        {ase17}
\bibfield{author}{\bibinfo{person}{Bj\"{o}rn Matthis}, \bibinfo{person}{Vitalii
  Avdiienko}, \bibinfo{person}{Ezekiel Soremekun}, \bibinfo{person}{Marcel
  B\"{o}hme}, {and} \bibinfo{person}{Andreas Zeller}.}
  \bibinfo{year}{2017}\natexlab{}.
\newblock \showarticletitle{Detecting Information Flow by Mutating Input Data}.
  In \bibinfo{booktitle}{{\em Proceedings of the 32nd IEEE/ACM International
  Conference on Automated Software Engineering}} {\em (\bibinfo{series}{ASE
  '17})}. \bibinfo{pages}{1--11}.
\newblock


\bibitem[\protect\citeauthoryear{Maurer, Brown, and Rusler}{Maurer
  et~al\mbox{.}}{1992}]%
        {samplingbias3}
\bibfield{author}{\bibinfo{person}{Brian~A. Maurer}, \bibinfo{person}{James~H.
  Brown}, {and} \bibinfo{person}{Renee~D. Rusler}.}
  \bibinfo{year}{1992}\natexlab{}.
\newblock \showarticletitle{The Micro and Macro in Body Size Evolution}.
\newblock \bibinfo{journal}{{\em Evolution\/}} \bibinfo{volume}{46},
  \bibinfo{number}{4} (\bibinfo{year}{1992}), \bibinfo{pages}{939--953}.
\newblock


\bibitem[\protect\citeauthoryear{McMinn}{McMinn}{2004}]%
        {sbst}
\bibfield{author}{\bibinfo{person}{Phil McMinn}.}
  \bibinfo{year}{2004}\natexlab{}.
\newblock \showarticletitle{Search-based Software Test Data Generation: A
  Survey: Research Articles}.
\newblock \bibinfo{journal}{{\em Journal of Software Testing, Verification and
  Reliability\/}} \bibinfo{volume}{14}, \bibinfo{number}{2}
  (\bibinfo{date}{June} \bibinfo{year}{2004}), \bibinfo{pages}{105--156}.
\newblock


\bibitem[\protect\citeauthoryear{McMinn}{McMinn}{2011}]%
        {sbst1}
\bibfield{author}{\bibinfo{person}{P. McMinn}.}
  \bibinfo{year}{2011}\natexlab{}.
\newblock \showarticletitle{Search-Based Software Testing: Past, Present and
  Future}. In \bibinfo{booktitle}{{\em Proceedings of the 4th IEEE
  International Conference on Software Testing, Verification and Validation
  Workshops}} {\em (\bibinfo{series}{ICSTW '11})}. \bibinfo{pages}{153--163}.
\newblock


\bibitem[\protect\citeauthoryear{Miller, Fredriksen, and So}{Miller
  et~al\mbox{.}}{1990}]%
        {fuzz}
\bibfield{author}{\bibinfo{person}{Barton~P. Miller}, \bibinfo{person}{Louis
  Fredriksen}, {and} \bibinfo{person}{Bryan So}.}
  \bibinfo{year}{1990}\natexlab{}.
\newblock \showarticletitle{An Empirical Study of the Reliability of UNIX
  Utilities}.
\newblock \bibinfo{journal}{{\em Commun. ACM\/}} \bibinfo{volume}{33},
  \bibinfo{number}{12} (\bibinfo{date}{Dec.} \bibinfo{year}{1990}),
  \bibinfo{pages}{32--44}.
\newblock


\bibitem[\protect\citeauthoryear{Miller, Morell, Noonan, Park, Nicol, Murrill,
  and Voas}{Miller et~al\mbox{.}}{1992}]%
        {miller}
\bibfield{author}{\bibinfo{person}{K.~W. Miller}, \bibinfo{person}{L.~J.
  Morell}, \bibinfo{person}{R.~E. Noonan}, \bibinfo{person}{S.~K. Park},
  \bibinfo{person}{D.~M. Nicol}, \bibinfo{person}{B.~W. Murrill}, {and}
  \bibinfo{person}{M. Voas}.} \bibinfo{year}{1992}\natexlab{}.
\newblock \showarticletitle{Estimating the probability of failure when testing
  reveals no failures}.
\newblock \bibinfo{journal}{{\em IEEE Transactions on Software Engineering\/}}
  \bibinfo{volume}{18}, \bibinfo{number}{1} (\bibinfo{date}{Jan}
  \bibinfo{year}{1992}), \bibinfo{pages}{33--43}.
\newblock


\bibitem[\protect\citeauthoryear{Mora, Tittensor, Adl, Simpson, and Worm}{Mora
  et~al\mbox{.}}{2011}]%
        {species}
\bibfield{author}{\bibinfo{person}{Camilo Mora}, \bibinfo{person}{Derek~P.
  Tittensor}, \bibinfo{person}{Sina Adl}, \bibinfo{person}{Alastair G.~B.
  Simpson}, {and} \bibinfo{person}{Boris Worm}.}
  \bibinfo{year}{2011}\natexlab{}.
\newblock \showarticletitle{How Many Species Are There on Earth and in the
  Ocean?}
\newblock \bibinfo{journal}{{\em PLOS Biology\/}} \bibinfo{volume}{9},
  \bibinfo{number}{8} (\bibinfo{date}{08} \bibinfo{year}{2011}),
  \bibinfo{pages}{1--8}.
\newblock


\bibitem[\protect\citeauthoryear{Ohannessian}{Ohannessian}{2012}]%
        {rare}
\bibfield{author}{\bibinfo{person}{Mesrob~I. Ohannessian}.}
  \bibinfo{year}{2012}\natexlab{}.
\newblock {\em \bibinfo{title}{On Inference about Rare Events}}.
\newblock {PhD} dissertation. \bibinfo{school}{Massachusetts Institute of
  Technology}.
\newblock


\bibitem[\protect\citeauthoryear{Orlitsky and Suresh}{Orlitsky and
  Suresh}{2015}]%
        {goodRobust}
\bibfield{author}{\bibinfo{person}{Alon Orlitsky} {and}
  \bibinfo{person}{Ananda~Theertha Suresh}.} \bibinfo{year}{2015}\natexlab{}.
\newblock \showarticletitle{Competitive Distribution Estimation: Why is
  Good-Turing Good}. In \bibinfo{booktitle}{{\em Proceedings of the 28th
  International Conference on Neural Information Processing Systems}} {\em
  (\bibinfo{series}{NIPS'15})}. \bibinfo{pages}{2143--2151}.
\newblock


\bibitem[\protect\citeauthoryear{Orlitsky, Suresh, and Wu}{Orlitsky
  et~al\mbox{.}}{2016}]%
        {orlitzki}
\bibfield{author}{\bibinfo{person}{Alon Orlitsky},
  \bibinfo{person}{Ananda~Theertha Suresh}, {and} \bibinfo{person}{Yihong Wu}.}
  \bibinfo{year}{2016}\natexlab{}.
\newblock \showarticletitle{Optimal prediction of the number of unseen
  species}.
\newblock \bibinfo{journal}{{\em Proceedings of the National Academy of
  Sciences\/}} \bibinfo{volume}{113}, \bibinfo{number}{47}
  (\bibinfo{year}{2016}), \bibinfo{pages}{13283--13288}.
\newblock


\bibitem[\protect\citeauthoryear{Palmer}{Palmer}{1991}]%
        {palmer}
\bibfield{author}{\bibinfo{person}{Michael~W. Palmer}.}
  \bibinfo{year}{1991}\natexlab{}.
\newblock \showarticletitle{Estimating Species Richness: The Second-Order
  Jackknife Reconsidered}.
\newblock \bibinfo{journal}{{\em Ecology\/}} \bibinfo{volume}{72},
  \bibinfo{number}{4} (\bibinfo{year}{1991}), \bibinfo{pages}{1512--1513}.
\newblock


\bibitem[\protect\citeauthoryear{Pham, B\"{o}hme, and Roychoudhury}{Pham
  et~al\mbox{.}}{2016}]%
        {MWF}
\bibfield{author}{\bibinfo{person}{Van-Thuan Pham}, \bibinfo{person}{Marcel
  B\"{o}hme}, {and} \bibinfo{person}{Abhik Roychoudhury}.}
  \bibinfo{year}{2016}\natexlab{}.
\newblock \showarticletitle{Model-based whitebox fuzzing for program binaries}.
  In \bibinfo{booktitle}{{\em Proceedings of the 2016 31st IEEE/ACM
  International Conference on Automated Software Engineering}} {\em
  (\bibinfo{series}{ASE '16})}. \bibinfo{pages}{543--553}.
\newblock


\bibitem[\protect\citeauthoryear{Pielou}{Pielou}{1966}]%
        {evenness}
\bibfield{author}{\bibinfo{person}{E.C. Pielou}.}
  \bibinfo{year}{1966}\natexlab{}.
\newblock \showarticletitle{Species-diversity and pattern-diversity in the
  study of ecological succession}.
\newblock \bibinfo{journal}{{\em Journal of Theoretical Biology\/}}
  \bibinfo{volume}{10}, \bibinfo{number}{2} (\bibinfo{year}{1966}),
  \bibinfo{pages}{370 -- 383}.
\newblock


\bibitem[\protect\citeauthoryear{Preston}{Preston}{1948}]%
        {preston}
\bibfield{author}{\bibinfo{person}{F.~W. Preston}.}
  \bibinfo{year}{1948}\natexlab{}.
\newblock \showarticletitle{The Commonness, And Rarity, of Species}.
\newblock \bibinfo{journal}{{\em Ecology\/}} \bibinfo{volume}{29},
  \bibinfo{number}{3} (\bibinfo{year}{1948}), \bibinfo{pages}{254--283}.
\newblock


\bibitem[\protect\citeauthoryear{Qi, Nguyen, and Roychoudhury}{Qi
  et~al\mbox{.}}{2013}]%
        {peso}
\bibfield{author}{\bibinfo{person}{Dawei Qi}, \bibinfo{person}{Hoang D.~T.
  Nguyen}, {and} \bibinfo{person}{Abhik Roychoudhury}.}
  \bibinfo{year}{2013}\natexlab{}.
\newblock \showarticletitle{Path Exploration Based on Symbolic Output}.
\newblock \bibinfo{journal}{{\em ACM Transactions on Software Engineering and
  Methodology\/}} \bibinfo{volume}{22}, \bibinfo{number}{4}
  (\bibinfo{date}{Oct.} \bibinfo{year}{2013}), \bibinfo{pages}{32:1--32:41}.
\newblock


\bibitem[\protect\citeauthoryear{Reddy and Dávalos}{Reddy and
  Dávalos}{2003}]%
        {samplingbias1}
\bibfield{author}{\bibinfo{person}{Sushma Reddy} {and}
  \bibinfo{person}{Liliana~M. Dávalos}.} \bibinfo{year}{2003}\natexlab{}.
\newblock \showarticletitle{Geographical sampling bias and its implications for
  conservation priorities in Africa}.
\newblock \bibinfo{journal}{{\em Journal of Biogeography\/}}
  \bibinfo{volume}{30}, \bibinfo{number}{11} (\bibinfo{year}{2003}),
  \bibinfo{pages}{1719--1727}.
\newblock


\bibitem[\protect\citeauthoryear{Robbins}{Robbins}{1968}]%
        {goodError}
\bibfield{author}{\bibinfo{person}{Herbert~E. Robbins}.}
  \bibinfo{year}{1968}\natexlab{}.
\newblock \showarticletitle{Estimating the Total Probability of the Unobserved
  Outcomes of an Experiment}.
\newblock \bibinfo{journal}{{\em The Annals of Mathematical Statistics\/}}
  \bibinfo{volume}{39}, \bibinfo{number}{1} (\bibinfo{date}{02}
  \bibinfo{year}{1968}), \bibinfo{pages}{256--257}.
\newblock


\bibitem[\protect\citeauthoryear{Robinson, McCarthy, and Smyth}{Robinson
  et~al\mbox{.}}{2010}]%
        {edgeR}
\bibfield{author}{\bibinfo{person}{Mark~D. Robinson}, \bibinfo{person}{Davis~J.
  McCarthy}, {and} \bibinfo{person}{Gordon~K. Smyth}.}
  \bibinfo{year}{2010}\natexlab{}.
\newblock \showarticletitle{edgeR: a Bioconductor package for differential
  expression analysis of digital gene expression data.}
\newblock \bibinfo{journal}{{\em Bioinformatics\/}} \bibinfo{volume}{26},
  \bibinfo{number}{1} (\bibinfo{year}{2010}), \bibinfo{pages}{139--140}.
\newblock


\bibitem[\protect\citeauthoryear{Serebryany, Bruening, Potapenko, and
  Vyukov}{Serebryany et~al\mbox{.}}{2012}]%
        {asan}
\bibfield{author}{\bibinfo{person}{Konstantin Serebryany},
  \bibinfo{person}{Derek Bruening}, \bibinfo{person}{Alexander Potapenko},
  {and} \bibinfo{person}{Dmitry Vyukov}.} \bibinfo{year}{2012}\natexlab{}.
\newblock \showarticletitle{AddressSanitizer: A Fast Address Sanity Checker}.
  In \bibinfo{booktitle}{{\em Proceedings of the 2012 USENIX Conference on
  Annual Technical Conference}} {\em (\bibinfo{series}{USENIX ATC'12})}.
  \bibinfo{pages}{28--28}.
\newblock


\bibitem[\protect\citeauthoryear{Shen, Chao, and Lin}{Shen
  et~al\mbox{.}}{2003}]%
        {shen}
\bibfield{author}{\bibinfo{person}{Tsung-Jen Shen}, \bibinfo{person}{Anne
  Chao}, {and} \bibinfo{person}{Chih-Feng Lin}.}
  \bibinfo{year}{2003}\natexlab{}.
\newblock \showarticletitle{Predicting the Number of New Species in Further
  Taxonomic Sampling}.
\newblock \bibinfo{journal}{{\em Ecology\/}} \bibinfo{volume}{84},
  \bibinfo{number}{3} (\bibinfo{year}{2003}), \bibinfo{pages}{798--804}.
\newblock


\bibitem[\protect\citeauthoryear{Solow and Polasky}{Solow and Polasky}{1999}]%
        {solow}
\bibfield{author}{\bibinfo{person}{Andrew~R. Solow} {and}
  \bibinfo{person}{Stephen Polasky}.} \bibinfo{year}{1999}\natexlab{}.
\newblock \showarticletitle{A Quick Estimator for Taxonomic Surveys}.
\newblock \bibinfo{journal}{{\em Ecology\/}} \bibinfo{volume}{80},
  \bibinfo{number}{8} (\bibinfo{year}{1999}), \bibinfo{pages}{2799--2803}.
\newblock


\bibitem[\protect\citeauthoryear{Stepanov and Serebryany}{Stepanov and
  Serebryany}{2015}]%
        {msan}
\bibfield{author}{\bibinfo{person}{Evgeniy Stepanov} {and}
  \bibinfo{person}{Konstantin Serebryany}.} \bibinfo{year}{2015}\natexlab{}.
\newblock \showarticletitle{MemorySanitizer: fast detector of uninitialized
  memory use in C++}. In \bibinfo{booktitle}{{\em Proceedings of the 2015
  IEEE/ACM International Symposium on Code Generation and Optimization}} {\em
  (\bibinfo{series}{CGO'15})}. \bibinfo{pages}{46--55}.
\newblock


\bibitem[\protect\citeauthoryear{Wagner, Viswanath, and Kulkarni}{Wagner
  et~al\mbox{.}}{2006}]%
        {consistent}
\bibfield{author}{\bibinfo{person}{A.~B. Wagner}, \bibinfo{person}{P.
  Viswanath}, {and} \bibinfo{person}{S.~R. Kulkarni}.}
  \bibinfo{year}{2006}\natexlab{}.
\newblock \showarticletitle{Strong Consistency of the Good-Turing Estimator}.
  In \bibinfo{booktitle}{{\em Proceedings of the 2006 IEEE International
  Symposium on Information Theory}}. \bibinfo{pages}{2526--2530}.
\newblock


\bibitem[\protect\citeauthoryear{Walther and Moore}{Walther and Moore}{2005a}]%
        {estimatorEvaluation}
\bibfield{author}{\bibinfo{person}{{Bruno A.} Walther} {and}
  \bibinfo{person}{{Joslin L.} Moore}.} \bibinfo{year}{2005}\natexlab{a}.
\newblock \showarticletitle{The concepts of bias, precision and accuracy, and
  their use in testing the performance of species richness estimators, with a
  literature review of estimator performance}.
\newblock \bibinfo{journal}{{\em Ecography\/}} \bibinfo{volume}{28},
  \bibinfo{number}{6} (\bibinfo{date}{12} \bibinfo{year}{2005}),
  \bibinfo{pages}{815--829}.
\newblock


\bibitem[\protect\citeauthoryear{Walther and Moore}{Walther and Moore}{2005b}]%
        {walther}
\bibfield{author}{\bibinfo{person}{Bruno~A. Walther} {and}
  \bibinfo{person}{Joslin~L. Moore}.} \bibinfo{year}{2005}\natexlab{b}.
\newblock \showarticletitle{The concepts of bias, precision and accuracy, and
  their use in testing the performance of species richness estimators, with a
  literature review of estimator performance}.
\newblock \bibinfo{journal}{{\em Ecography\/}} \bibinfo{volume}{28},
  \bibinfo{number}{6} (\bibinfo{year}{2005}), \bibinfo{pages}{815--829}.
\newblock


\bibitem[\protect\citeauthoryear{Website}{Website}{}]%
        {ponemon1}
\bibfield{author}{\bibinfo{person}{Website}.}
\newblock \bibinfo{title}{2017 Ponemon Cost of Cyber Crime Study}.
\newblock
  \bibinfo{howpublished}{\url{https://www.accenture.com/us-en/insight-cost-of-cybercrime-2017}}.
    (\bibinfo{year}{????}).
\newblock
\newblock
\shownote{Accessed: 2017-11-13.}


\bibitem[\protect\citeauthoryear{Website}{Website}{2013}]%
        {whalen}
\bibfield{author}{\bibinfo{person}{Website}.} \bibinfo{year}{2013}\natexlab{}.
\newblock \bibinfo{title}{Lockheed Martin Webinar Series: Michael Whalen on the
  Future of Verification and Validation.}
\newblock
  \bibinfo{howpublished}{\url{https://www.computer.org/cms/Computer.org/webinars/lmco/012413Slides-Whalen.pdf}}.
    (\bibinfo{date}{Jan.} \bibinfo{year}{2013}).
\newblock
\newblock
\shownote{Accessed: 2017-05-13.}


\bibitem[\protect\citeauthoryear{Website}{Website}{2017a}]%
        {afl}
\bibfield{author}{\bibinfo{person}{Website}.} \bibinfo{year}{2017}\natexlab{a}.
\newblock \bibinfo{title}{AFL: American Fuzzy Lop Fuzzer}.
\newblock
  \bibinfo{howpublished}{\url{http://lcamtuf.coredump.cx/afl/technical_details.txt}}.
    (\bibinfo{year}{2017}).
\newblock
\newblock
\shownote{Accessed: 2017-05-13.}


\bibitem[\protect\citeauthoryear{Website}{Website}{2017b}]%
        {thinair}
\bibfield{author}{\bibinfo{person}{Website}.} \bibinfo{year}{2017}\natexlab{b}.
\newblock \bibinfo{title}{{AFL: Pulling Jpegs out of Thin Air, Michael
  Zalewski}}.
\newblock
  \bibinfo{howpublished}{\url{https://lcamtuf.blogspot.com/2014/11/pulling-jpegs-out-of-thin-air.html}}.
    (\bibinfo{year}{2017}).
\newblock
\newblock
\shownote{Accessed: 2017-05-13.}


\bibitem[\protect\citeauthoryear{Website}{Website}{2017c}]%
        {cert}
\bibfield{author}{\bibinfo{person}{Website}.} \bibinfo{year}{2017}\natexlab{c}.
\newblock \bibinfo{title}{CERT Triage Tools}.
\newblock
  \bibinfo{howpublished}{\url{https://www.cert.org/vulnerability-analysis/tools/triage.cfm?}}.
    (\bibinfo{year}{2017}).
\newblock
\newblock
\shownote{Accessed: 2017-05-13.}


\bibitem[\protect\citeauthoryear{Website}{Website}{2017d}]%
        {llvmsan}
\bibfield{author}{\bibinfo{person}{Website}.} \bibinfo{year}{2017}\natexlab{d}.
\newblock \bibinfo{title}{Clang compiler documentation}.
\newblock \bibinfo{howpublished}{\url{https://clang.llvm.org/docs/index.html}}.
    (\bibinfo{year}{2017}).
\newblock
\newblock
\shownote{Accessed: 2017-05-13.}


\bibitem[\protect\citeauthoryear{Website}{Website}{2017e}]%
        {cgc}
\bibfield{author}{\bibinfo{person}{Website}.} \bibinfo{year}{2017}\natexlab{e}.
\newblock \bibinfo{title}{DARPA Cyber Grand Challenge}.
\newblock
  \bibinfo{howpublished}{\url{http://www.darpa.mil/news-events/2016-08-04}}.
  (\bibinfo{year}{2017}).
\newblock
\newblock
\shownote{Accessed: 2017-05-13.}


\bibitem[\protect\citeauthoryear{Website}{Website}{2017f}]%
        {harmanFacebook}
\bibfield{author}{\bibinfo{person}{Website}.} \bibinfo{year}{2017}\natexlab{f}.
\newblock \bibinfo{title}{Facebook: Mark Harman on software engineering at
  Facebook scale.}
\newblock
  \bibinfo{howpublished}{\url{https://research.fb.com/mark-harmon-on-software-engineering-at-facebook-scale/}}.
    (\bibinfo{year}{2017}).
\newblock
\newblock
\shownote{Accessed: 2017-05-13.}


\bibitem[\protect\citeauthoryear{Website}{Website}{2017g}]%
        {ffmpeg}
\bibfield{author}{\bibinfo{person}{Website}.} \bibinfo{year}{2017}\natexlab{g}.
\newblock \bibinfo{title}{FFMPEG: A complete, cross-platform solution to
  record, convert and stream audio and video.}
\newblock \bibinfo{howpublished}{\url{https://www.ffmpeg.org/}}.
  (\bibinfo{year}{2017}).
\newblock
\newblock
\shownote{Accessed: 2017-05-13.}


\bibitem[\protect\citeauthoryear{Website}{Website}{2017h}]%
        {gcov}
\bibfield{author}{\bibinfo{person}{Website}.} \bibinfo{year}{2017}\natexlab{h}.
\newblock \bibinfo{title}{GCov: coverage testing tool}.
\newblock \bibinfo{howpublished}{\url{https://linux.die.net/man/1/gcov}}.
  (\bibinfo{year}{2017}).
\newblock
\newblock
\shownote{Accessed: 2017-11-13.}


\bibitem[\protect\citeauthoryear{Website}{Website}{2017i}]%
        {gccsan}
\bibfield{author}{\bibinfo{person}{Website}.} \bibinfo{year}{2017}\natexlab{i}.
\newblock \bibinfo{title}{GNU GCC sanitizer options}.
\newblock
  \bibinfo{howpublished}{\url{https://gcc.gnu.org/onlinedocs/gcc-6.3.0/gcc/Instrumentation-Options.html\#index-fsanitize_003daddress-947}}.
    (\bibinfo{year}{2017}).
\newblock
\newblock
\shownote{Accessed: 2017-05-13.}


\bibitem[\protect\citeauthoryear{Website}{Website}{2017j}]%
        {inextOnline}
\bibfield{author}{\bibinfo{person}{Website}.} \bibinfo{year}{2017}\natexlab{j}.
\newblock \bibinfo{title}{iNext Online: Species iNterpolation and
  EXTrapolation}.
\newblock \bibinfo{howpublished}{\url{https://chao.shinyapps.io/iNEXTOnline/}}.
    (\bibinfo{year}{2017}).
\newblock
\newblock
\shownote{Accessed: 2017-05-13.}


\bibitem[\protect\citeauthoryear{Website}{Website}{2017k}]%
        {json}
\bibfield{author}{\bibinfo{person}{Website}.} \bibinfo{year}{2017}\natexlab{k}.
\newblock \bibinfo{title}{JSON for Modern C++}.
\newblock \bibinfo{howpublished}{\url{https://github.com/nlohmann/json}}.
  (\bibinfo{year}{2017}).
\newblock
\newblock
\shownote{Accessed: 2017-11-13.}


\bibitem[\protect\citeauthoryear{Website}{Website}{2017l}]%
        {libfuzzer}
\bibfield{author}{\bibinfo{person}{Website}.} \bibinfo{year}{2017}\natexlab{l}.
\newblock \bibinfo{title}{LibFuzzer: A library for coverage-guided fuzz
  testing}.
\newblock \bibinfo{howpublished}{\url{http://llvm.org/docs/LibFuzzer.html}}.
  (\bibinfo{year}{2017}).
\newblock
\newblock
\shownote{Accessed: 2017-05-13.}


\bibitem[\protect\citeauthoryear{Website}{Website}{2017m}]%
        {libjpeg}
\bibfield{author}{\bibinfo{person}{Website}.} \bibinfo{year}{2017}\natexlab{m}.
\newblock \bibinfo{title}{libjpeg-turbo is a JPEG image codec to accelerate
  baseline JPEG compression and decompression}.
\newblock \bibinfo{howpublished}{\url{http://libjpeg-turbo.virtualgl.org/}}.
  (\bibinfo{year}{2017}).
\newblock
\newblock
\shownote{Accessed: 2017-05-13.}


\bibitem[\protect\citeauthoryear{Website}{Website}{2017n}]%
        {libxml2}
\bibfield{author}{\bibinfo{person}{Website}.} \bibinfo{year}{2017}\natexlab{n}.
\newblock \bibinfo{title}{LibXML2: The XML C parser and toolkit of Gnome}.
\newblock \bibinfo{howpublished}{\url{http://xmlsoft.org/}}.
  (\bibinfo{year}{2017}).
\newblock
\newblock
\shownote{Accessed: 2017-11-13.}


\bibitem[\protect\citeauthoryear{Website}{Website}{2017o}]%
        {springfield}
\bibfield{author}{\bibinfo{person}{Website}.} \bibinfo{year}{2017}\natexlab{o}.
\newblock \bibinfo{title}{Microsoft: Project Springfield}.
\newblock \bibinfo{howpublished}{\url{https://www.microsoft.com/Springfield/}}.
    (\bibinfo{year}{2017}).
\newblock
\newblock
\shownote{Accessed: 2017-05-13.}


\bibitem[\protect\citeauthoryear{Website}{Website}{2017p}]%
        {monkey}
\bibfield{author}{\bibinfo{person}{Website}.} \bibinfo{year}{2017}\natexlab{p}.
\newblock \bibinfo{title}{Monkey: Android Random Testing.}
\newblock
  \bibinfo{howpublished}{\url{http://developer.android.com/tools/help/monkey.html}}.
    (\bibinfo{year}{2017}).
\newblock
\newblock
\shownote{Accessed: 2017-05-13.}


\bibitem[\protect\citeauthoryear{Website}{Website}{2017q}]%
        {mozillaFuzzing}
\bibfield{author}{\bibinfo{person}{Website}.} \bibinfo{year}{2017}\natexlab{q}.
\newblock \bibinfo{title}{Mozilla: Fuzzing Firefox with Peach.}
\newblock
  \bibinfo{howpublished}{\url{https://wiki.mozilla.org/Security/Fuzzing/Peach}}.
    (\bibinfo{year}{2017}).
\newblock
\newblock
\shownote{Accessed: 2017-05-13.}


\bibitem[\protect\citeauthoryear{Website}{Website}{2017r}]%
        {openssl}
\bibfield{author}{\bibinfo{person}{Website}.} \bibinfo{year}{2017}\natexlab{r}.
\newblock \bibinfo{title}{OpenSSL: A toolkit for the Transport Layer Security
  (TLS) and Secure Sockets Layer (SSL) protocols.}
\newblock \bibinfo{howpublished}{\url{https://www.openssl.org/}}.
  (\bibinfo{year}{2017}).
\newblock
\newblock
\shownote{Accessed: 2017-05-13.}


\bibitem[\protect\citeauthoryear{Website}{Website}{2017s}]%
        {oss}
\bibfield{author}{\bibinfo{person}{Website}.} \bibinfo{year}{2017}\natexlab{s}.
\newblock \bibinfo{title}{{OSS-Fuzz: Five Months Later}}.
\newblock
  \bibinfo{howpublished}{\url{https://testing.googleblog.com/2017/05/oss-fuzz-five-months-later-and.html}}.
    (\bibinfo{year}{2017}).
\newblock
\newblock
\shownote{Accessed: 2017-05-13.}


\bibitem[\protect\citeauthoryear{Website}{Website}{2017t}]%
        {peach}
\bibfield{author}{\bibinfo{person}{Website}.} \bibinfo{year}{2017}\natexlab{t}.
\newblock \bibinfo{title}{{Peach Fuzzer Platform}}.
\newblock
  \bibinfo{howpublished}{\url{http://www.peachfuzzer.com/products/peach-platform/}}.
    (\bibinfo{year}{2017}).
\newblock
\newblock
\shownote{Accessed: 2017-05-13.}


\bibitem[\protect\citeauthoryear{Website}{Website}{2017u}]%
        {spadeOnline}
\bibfield{author}{\bibinfo{person}{Website}.} \bibinfo{year}{2017}\natexlab{u}.
\newblock \bibinfo{title}{SpadeR Online: Species-richness Prediction And
  Diversity Estimation in R}.
\newblock \bibinfo{howpublished}{\url{https://chao.shinyapps.io/SpadeR/}}.
  (\bibinfo{year}{2017}).
\newblock
\newblock
\shownote{Accessed: 2017-05-13.}


\bibitem[\protect\citeauthoryear{Website}{Website}{2017v}]%
        {sprex}
\bibfield{author}{\bibinfo{person}{Website}.} \bibinfo{year}{2017}\natexlab{v}.
\newblock \bibinfo{title}{sprex: Calculate Species Richness and Extrapolation
  Metrics}.
\newblock
  \bibinfo{howpublished}{\url{https://cran.r-project.org/web/packages/sprex/}}.
    (\bibinfo{year}{2017}).
\newblock
\newblock
\shownote{Accessed: 2017-05-13.}


\bibitem[\protect\citeauthoryear{Website}{Website}{2017w}]%
        {syzkaller}
\bibfield{author}{\bibinfo{person}{Website}.} \bibinfo{year}{2017}\natexlab{w}.
\newblock \bibinfo{title}{Syzkaller: Coverage-guided kernel fuzzing.}
\newblock \bibinfo{howpublished}{\url{https://github.com/google/syzkaller}}.
  (\bibinfo{year}{2017}).
\newblock
\newblock
\shownote{Accessed: 2017-05-13.}


\bibitem[\protect\citeauthoryear{Website}{Website}{2017x}]%
        {wireshark}
\bibfield{author}{\bibinfo{person}{Website}.} \bibinfo{year}{2017}\natexlab{x}.
\newblock \bibinfo{title}{Wireshark is the world's foremost and widely-used
  network protocol analyzer.}
\newblock \bibinfo{howpublished}{\url{https://www.wireshark.org/}}.
  (\bibinfo{year}{2017}).
\newblock
\newblock
\shownote{Accessed: 2017-05-13.}


\bibitem[\protect\citeauthoryear{Website\quad}{Website\quad}{2017a}]%
        {logeff}
\bibfield{author}{\bibinfo{person}{Website\quad}.}
  \bibinfo{year}{2017}\natexlab{a}.
\newblock \bibinfo{title}{Bithacks: Implementing logarithm efficiently}.
\newblock
  \bibinfo{howpublished}{\url{https://graphics.stanford.edu/~seander/bithacks.html\#IntegerLogFloat}}.
    (\bibinfo{year}{2017}).
\newblock
\newblock
\shownote{Accessed: 2017-11-13.}


\bibitem[\protect\citeauthoryear{Website\quad}{Website\quad}{2017b}]%
        {cloudbleed}
\bibfield{author}{\bibinfo{person}{Website\quad}.}
  \bibinfo{year}{2017}\natexlab{b}.
\newblock \bibinfo{title}{{Incident report on memory leak caused by Cloudflare
  parser bug}}.
\newblock
  \bibinfo{howpublished}{\url{https://blog.cloudflare.com/incident-report-on-memory-leak-caused-by-cloudflare-parser-bug/}}.
    (\bibinfo{year}{2017}).
\newblock
\newblock
\shownote{Accessed: 2017-11-13.}


\bibitem[\protect\citeauthoryear{Website\quad}{Website\quad}{2017c}]%
        {ether}
\bibfield{author}{\bibinfo{person}{Website\quad}.}
  \bibinfo{year}{2017}\natexlab{c}.
\newblock \bibinfo{title}{Medium: A hacker stole USD 31M of Ether.}
\newblock
  \bibinfo{howpublished}{\url{https://medium.freecodecamp.org/a-hacker-stole-31m-of-ether-how-it-happened-and-what-it-means-for-ethereum-9e5dc29e33ce}}.
    (\bibinfo{year}{2017}).
\newblock
\newblock
\shownote{Accessed: 2017-05-13.}


\bibitem[\protect\citeauthoryear{Website\quad}{Website\quad}{2017d}]%
        {guardian}
\bibfield{author}{\bibinfo{person}{Website\quad}.}
  \bibinfo{year}{2017}\natexlab{d}.
\newblock \bibinfo{title}{NHS seeks to recover from global cyber-attack as
  security concerns resurface}.
\newblock
  \bibinfo{howpublished}{\url{https://www.theguardian.com/society/2017/may/12/hospitals-across-england-hit-by-large-scale-cyber-attack}}.
    (\bibinfo{year}{2017}).
\newblock
\newblock
\shownote{Accessed: 2017-11-13.}


\bibitem[\protect\citeauthoryear{Weyuker}{Weyuker}{1982}]%
        {oracle1}
\bibfield{author}{\bibinfo{person}{Elaine~J. Weyuker}.}
  \bibinfo{year}{1982}\natexlab{}.
\newblock \showarticletitle{On Testing Non-Testable Programs}.
\newblock \bibinfo{journal}{{\it Comput. J.}} \bibinfo{volume}{25},
  \bibinfo{number}{4} (\bibinfo{year}{1982}), \bibinfo{pages}{465--470}.
\newblock


\bibitem[\protect\citeauthoryear{Weyuker and Jeng}{Weyuker and Jeng}{1991}]%
        {tsEffectiv5}
\bibfield{author}{\bibinfo{person}{E.~J. Weyuker} {and} \bibinfo{person}{B.
  Jeng}.} \bibinfo{year}{1991}\natexlab{}.
\newblock \showarticletitle{Analyzing partition testing strategies}.
\newblock \bibinfo{journal}{{\em IEEE Transactions on Software Engineering\/}}
  \bibinfo{volume}{17}, \bibinfo{number}{7} (\bibinfo{date}{Jul}
  \bibinfo{year}{1991}), \bibinfo{pages}{703--711}.
\newblock


\bibitem[\protect\citeauthoryear{Yang and Xie}{Yang and Xie}{2000}]%
        {operational}
\bibfield{author}{\bibinfo{person}{B. Yang} {and} \bibinfo{person}{M. Xie}.}
  \bibinfo{year}{2000}\natexlab{}.
\newblock \showarticletitle{A study of operational and testing reliability in
  software reliability analysis}.
\newblock \bibinfo{journal}{{\em Reliability Engineering \& System Safety\/}}
  \bibinfo{volume}{70}, \bibinfo{number}{3} (\bibinfo{year}{2000}),
  \bibinfo{pages}{323 -- 329}.
\newblock


\bibitem[\protect\citeauthoryear{Yang, Chen, Eide, and Regehr}{Yang
  et~al\mbox{.}}{2011}]%
        {csmith}
\bibfield{author}{\bibinfo{person}{Xuejun Yang}, \bibinfo{person}{Yang Chen},
  \bibinfo{person}{Eric Eide}, {and} \bibinfo{person}{John Regehr}.}
  \bibinfo{year}{2011}\natexlab{}.
\newblock \showarticletitle{Finding and Understanding Bugs in C Compilers}. In
  \bibinfo{booktitle}{{\em Proceedings of the 32Nd ACM SIGPLAN Conference on
  Programming Language Design and Implementation}} {\em (\bibinfo{series}{PLDI
  '11})}. \bibinfo{pages}{283--294}.
\newblock


\bibitem[\protect\citeauthoryear{Zhang and Zhang}{Zhang and Zhang}{2009}]%
        {goodNormal}
\bibfield{author}{\bibinfo{person}{Cun-Hui Zhang} {and} \bibinfo{person}{Zhiyi
  Zhang}.} \bibinfo{year}{2009}\natexlab{}.
\newblock \showarticletitle{Asymptotic normality of a nonparametric estimator
  of sample coverage}.
\newblock \bibinfo{journal}{{\em The Annals of Statistics\/}}
  \bibinfo{volume}{37}, \bibinfo{number}{5A} (\bibinfo{date}{10}
  \bibinfo{year}{2009}), \bibinfo{pages}{2582--2595}.
\newblock


\end{thebibliography}
